\documentclass[12pt,doublespace]{report2}
\usepackage{epsf}
\usepackage{graphics}
\usepackage{axodraw}
\input epsf
\renewcommand{\baselinestretch}{2}

\newcommand {\bsone}{\renewcommand{\baselinestretch}{1}\Large{}\normalsize}
\newcommand {\bstwo}{\renewcommand{\baselinestretch}{2}\Large{}\normalsize}

\begin{document}
\epsfclipon
\pagenumbering{roman}
\begin{center}

{\large\bf  ANGULAR DISTRIBUTION OF J/PSI DECAYS}\\ 
{\large\bf  IN DIMUON CHANNEL IN 800 GEV}\\
{\large\bf  PROTON-COPPER COLLISIONS}

\vspace*{.4cm}

{\large  BY }

\vspace*{.4cm}

{\large  TING-HUA CHANG}

\vspace{20mm}

A Dissertation submitted to the Graduate School \\
in partial fulfillment of the requirements \\
for the Degree   \\
Doctor of Philosophy  \\

\vspace{20mm}

Subject: Physics \\

\vspace{25mm}

New Mexico State University \\
Las Cruces, New Mexico  \\
July 1999
\end{center}

\thispagestyle{empty}

\newpage
\setcounter{page}{2}
\noindent ``Angular distribution of J/psi decays in dimuon channel
 in 800 GeV proton copper collisions,'' a dissertation
prepared by
Ting-Hua Chang in partial fulfillment of the requirements for the
degree, Doctor of Philosophy, has been approved and accepted by the
following:

{\begin{flushleft}
{\bsone
\vspace{.5in}
\noindent
\hrulefill

\noindent Timothy J. Pettibone

\noindent Dean of the Graduate School

\vspace*{.75in}
\noindent 
\hrulefill

\noindent Vassili Papavassiliou

\noindent Chair of the Examining Committee
\vspace*{.75in}

\noindent 
\hrulefill

\noindent Date

}%

\vspace*{.5in}
\noindent Committee in charge:

\hspace*{.5in}
Dr. Vassili Papavassiliou, Chair

\hspace*{.5in}
Dr. Sidney A. Coon

\hspace*{.5in}
Dr. Gary S. Kyle

\hspace*{.5in}
Dr. Robert J. Liefield

\hspace*{.5in}
Dr. Richard L. Long
\end{flushleft}

\newpage
%dedication
\begin{center}
{\large
DEDICATION}
\end{center}

\vspace{.25in}

\indent
\begin{center}
To all of you who read this thesis.
\end{center}

\newpage
%Acknowledgements
\begin{center}
{\large
ACKNOWLEDGEMENTS}
\end{center}

\vspace{.25in}

\indent I would like to thank all the members of the FNAL E866/NuSea 
Collaboration who made this experiment, and hence this thesis, possible. 

I would like to thank my thesis advisor Prof. Vassili Papavassiliou for his
guidance on my thesis work. Everything I know about QDC is from him. Without
his constant help and knowledge, I would have never understood this subject.

I would like to thank Prof. Gary Kyle for his non-stop supporting during 
the last seven years, financially and personally. I never had to worry about any 
budget problems on travelling or attending conferences during the time of study.
He was also my formal advisor, who introduced me into the career of high energy
physics. Without his support, I don't know to where I would have ended up now.

I would like to thank Dr. Thomas Carey. We worked together on E866 DAQ upgrade 
project at LANL for more than a year. He has taught me so many special skills
about hardware electronics and software coding which are nowhere to learn in
campus. He is a great teacher and I enjoyed the time we spent together very
much.

I would like to thank Dr. Patrick McGaughey and Dr. Chuck Brown, for their work 
on the angular distribution analysis. They have provided me with many valuable 
suggestions and ideas. Especially for Dr. Patrick McGaughey for his deep
physics insight on this subject, I have gained a lot. 

I would like to thank Dr. Mike Leitch and Dr. Paul Reimer, who maintained the 
analysis and Monte Carlo code. Without their efforts, the data analysis would
be impossible.

I would like to thank Prof. Carl Gagliardi and Dr. Don Geesaman, for their 
careful reading of my thesis draft and giving critical comments.

I would like to give thanks to the following list of people, for their help in 
many ways during the past days: Jim Bread, Siguud and Ingrid Smitz, Bob and
Rona Paz, Ho-Fu and Lily Dai, Paul and Lily Horng, Job and Grace Lee, 
John Powell, Ron and May Guo, Prof. Twan-Wu Chen, and Jason Webb. 

I would like to give special thanks to my parents Zhe-Yu and Pi-Tao Chang for
their spiritual and financial support. 

Last but not least, I would like to thank my loving wife Yan, for her patient 
and constant encouragement. She is the behind-unseen author of this thesis.

\newpage
%VITA

\begin{center}

{\large VITA}

\end{center}

\vspace*{.25in}

{\bsone
\noindent September 28th, 1967--Born in Changhua, Taiwan

\vspace{.15in}

\noindent 1989--B.S., Tunghai University, Taichung, Taiwan 

\vspace{.15in}

\noindent 1991--M.S., New Mexico State University, Las Cruces, New Mexico

%\vspace*{.25in}

%\begin{center}

%{\large\bf PROFESSIONAL AND HONORARY SOCIETIES}

%\end{center}

%\vspace*{.25in}

%\noindent American Physical Society

\vspace*{.25in}

\begin{center}

{\large PUBLICATIONS}

\end{center}

\vspace{.25in}

\noindent {\it Pion-induced nucleon knockout from a polarized $^{7}Li$ target},
 M.G. Khayat {\it et al.}, Bull. Am. Phys. Soc. {\bf 39}, 1146 (1994). 

\vspace{.15in}

\noindent {\it $\pi^{-}$-induced single charge exchange on polarized $^{3}He$},
 Q. Zhao {\it et al.}, Bull. Am. Phys. Soc. {\bf 40}, 963 (1995).

\vspace{.15in}

\noindent {\it Analyzing powers for polarized $^{1}H(\pi^{+}, \pi^{-}p)$ at
$T_{\pi}$ = 165 MeV and 240 MeV}, B. A. Raue {\it et al.}, Phys. Rev. {\bf C53},
1005 (1996).

\vspace{.15in}

\noindent {\it Anti-D/Anti-U asymmetry and the origin of the nucleon sea}, 
FNAL E866/NuSea Collaboration, J. C. Peng {\it et al.}, Phys. Rev. {\bf D58},
2004 (1998).

\vspace{.15in}

\noindent {\it Measurement of the light anti-quark flavor asymmetry in the 
nucleon sea}, FNAL E866/NuSea Collaboration, E. A. Hawker {\it et al.}, Phys.
Rev. Lett. {\bf 80}, 3715 (1998).

\vspace{.15in}

\noindent {\it Measurement of $J/\psi$ decay angular distribution in dilepton
channel in 800 GeV p+Cu collisions}, to be published. 

\vspace{.25in}

\begin{center}

{\large FIELD OF STUDY}

\end{center}

\vspace{.25in}

\noindent Major Field:  Physics

\vspace{.25in}

\noindent Minor Field:  Chemical Engineering

}%

\newpage
\begin{center}

{\large ABSTRACT}

\vspace*{.25in}

{\large ANGULAR DISTRIBUTION OF J/PSI DECAYS}

\vspace*{.3cm}

{\large IN DIMUON CHANNEL IN 800 GEV}

\vspace*{.3cm}

{\large PROTON-COPPER COLLISIONS}

\vspace*{.3cm}

{\large BY}

\vspace*{.3cm}

{\large TING-HUA CHANG}

\vspace*{.25in}

Doctor of Philosophy in Physics

New Mexico State University

Las Cruces, New Mexico, 1999

Dr. Vassili Papavassiliou, Chair

\end{center}

\vspace{.25in}

The angular distribution of $J/\psi$ decays in the $\mu^{+} \mu^{-}$ channel in 
800 GeV proton-copper collisions has been measured for $x_{F} > 0.25$. The 
polarization parameter $\lambda$ is extracted in 1 GeV of $p_T$ and 0.1 of 
$x_F$ bins for two magnet configurations with different acceptances. The data
indicate that the $J/\psi$'s are produced with a slight transverse polarization
at small $x_F$, which turns to longitudinal at $x_{f} > 0.6$. No $p_T$ 
dependence of $\lambda$ is observed. Theoretical calculations are needed in 
order to interpret the measurements.

\newpage

\renewcommand{\baselinestretch}{1}
\tableofcontents
\newpage
{\bstwo
\addcontentsline{toc}{chapter}{LIST OF TABLES}
\addtocontents{toc}{\vspace*{+0.2in}}
\listoftables
\newpage
\addcontentsline{toc}{chapter}{LIST OF FIGURES}
\addtocontents{toc}{\vspace*{+0.2in}}
\listoffigures
}
\newpage

\pagenumbering{arabic}

%%%%%%%%%%%%%%%%%%%%%%%%%%%%%%%%%%%%%%%%%%%%%%%%%%%%%%%%%%%%%%%%%%%%%%%%%%%%%%%

\chapter{\small INTRODUCTION}
\indent    
    Since the 1970s the Standard Model has provided a satisfactory description 
of the interactions of all known elementary particles.  The underlying theory
for describing the electromagnetic force in the sub-atomic world is known as
Quantum Electrodynamics (QED), while Quantum Chromodynamics (QCD) describes
the strong force. QED has been tested to be a valid theory by its amazing 
predictions of the lepton magnetic moments and atomic energy spectra. But unlike 
QED, even though QCD was developed following the same fundamental idea of 
gauge invariance and seems to be a straightforward extension of QED, QCD is facing 
the difficulties associated with non-perturbative calculations in the low-energy 
regime. After decades of effort physicists have developed techniques, such as 
renormalization and resummation, as well as non-perturbative ones, such as 
effective theories, to help solve some of the mathematical 
difficulties.  And nowadays we are able to compare many experimental results with 
QCD predictions, and indeed QCD has proved to be the best candidate theory for 
describing the strong interaction.
   
There are many successful examples of QCD. The earliest and most profound one
is the prediction of the evolution of the structure functions in Deep Inelastic 
Scattering (DIS). Given the quark and gluon distributions at a fixed 
energy-momentum transfer $Q_0^2$, QCD can actually predict the nucleon structure 
functions at arbitrary $Q^2$ using the Evolution Equations \cite{Pic 95}. Later in 
collider experiments, Next-to-Leading Order (NLO) QCD calculations 
predicted the inclusive jet-production cross section
over several orders of magnitude and over a wide range of center-of-mass (COM)
energy and jet transverse momentum \cite{Arn 86, Ali 91, Abe 93} using the Parton
Distribution Functions (PDFs) extracted from the DIS data. Another example is the 
Drell-Yan process \cite{Dre 70}. The Drell-Yan process, massive lepton-pair 
production via electroweak quark-antiquark annihilation into vector bosons 
(photon, W$^{\pm}$, or Z) and then decay, is one of the few processes which have been 
calculated up to next-to-next-to-leading order (NNLO) in perturbative QCD. A 
recent calculation \cite{Rij 95} including some of the NNLO terms has shown a good
agreement with the data. Another example is the inclusive heavy quark production.
Fixed-target studies of heavy-flavor production have provided a wealth of data. 
Total cross sections, single-inclusive distributions, correlations
between the quark and the antiquark have been measured in both hadro- and
photoproduction. All experimental results are in qualitative agreement with
perturbative QCD calculations. A detailed comparison of the fixed-target data
and NLO QCD predictions can be found in \cite{Fri 97}.

Throughout the entire thesis the following symbols are used to describe the 
kinematic variables:

\indent $S$: center-of-mass energy of the beam-target system.\\
\indent $m$: rest mass of the dimuon pair.\\
\indent $p_T$: transverse momentum of the dimuon pair.\\
\indent $x_F$: dimensionless longitudinal momentum of the dimuon pair. It is defined as
the pair longitudinal momentum $P_L$ divided by its maximum kinematically allowed
value $P_{L,max}$ in the beam-target COM frame. It relates to the Bjorken x,
the fraction of the hadron momentum carried by the parton in the hadron boosted to 
the infinite-momentum frame, of the beam parton $x_1$ and of the target parton
$x_2$ by $x_{F}(1 - m^{2}/S) = x_{1} - x_{2}$.\\
\indent $\theta$ and $\phi$: polar and azimuthal angles of the dimuon pair;
described in section 1.2.3. 

\section{Failure of Perturbative QCD in Charmonium Production}
\indent
While QCD has provided successful descriptions of many aspects of the experimental
data, there are still some phenomena which could not be described. The 
production of charmonium at large transverse momentum is one of such processes: 
the observed production cross sections have differed from QCD predictions by 
more than an order of magnitude, even though in \cite{Fri 97} it has been shown 
that the charm production total cross section can be calculated from QCD. This 
discrepancy between experiment and theory has revealed a more complicated picture 
for heavy quarkonium production.

Quarkonium production was conventionally calculated based on the color-singlet
model (CSM) before 1993. However this model has failed to describe 
charmonium-hadroproduction data \cite{Sch 94}. In hadroproduction of charmonium 
at fixed target energy,
$\sqrt{S} < 50$ GeV, the ratio of the number of $J/\Psi$'s produced directly to 
those arising from decays of higher charmonium states is under-predicted by at 
least a factor of five \cite{Van 95}. At Tevatron collider energies, the excess of 
direct $\psi'$ production compared to the CSM prediction is a factor of 
30 \cite{Bra 94,Roy 94}. This excess has been referred to as the 
$\psi'$-anomaly. Figure \ref{Fig:anomaly} compares the CDF $\psi'$ data and some 
theoretical predictions. Figure \ref{Fig:Jpsi} shows the comparison with the 
fixed-target data.

\begin{figure}
{\centering {\resizebox{14.3cm}{!} {\includegraphics*[1.9cm,10cm][19.5cm,24cm]
{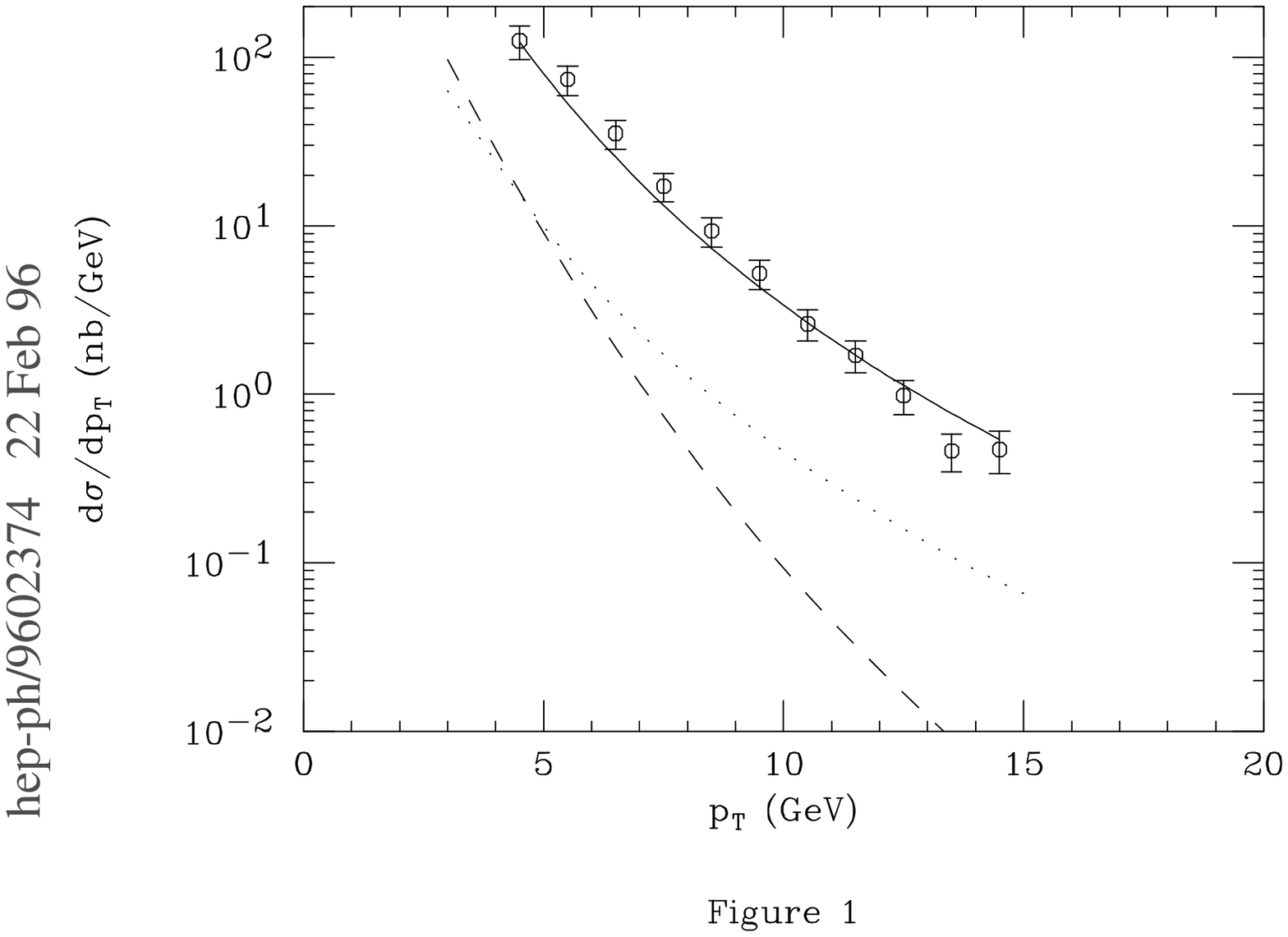}}} \par}
\caption[CDF data on the differential cross section for prompt $\psi'$s]
{CDF data on the differential cross section for prompt $\psi'$s that do 
not come from $\chi_{c}$ decays as a function of $p_T$. The curves are the LO
predictions of the color-singlet model (dashed curve), predictions including
fragmentation in the color-singlet model (dotted curve), and including 
contributions from gluon fragmentation via the color-octet mechanism (solid curve)
with the normalization adjusted to fit the CDF data. (Taken from \cite{Bra 96})} 
\label{Fig:anomaly}
\end{figure}

%-- Fig. 1 from hep-ph/9602374 --

\begin{figure}
{\centering {\resizebox{14.3cm}{!} {\includegraphics*[5cm,14.2cm][16cm,25cm]
{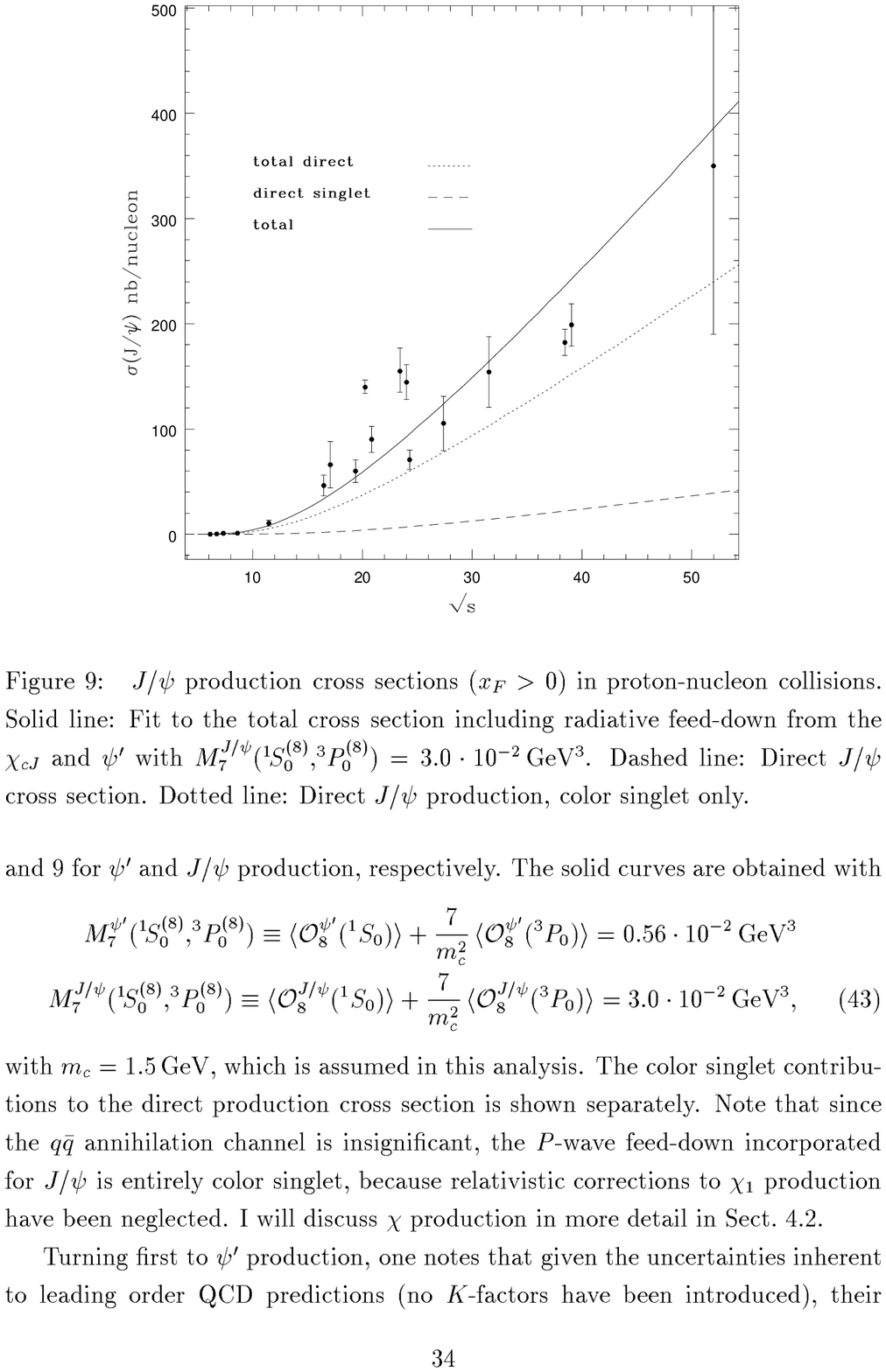}}} \par}
\caption[$J/\psi$ production cross section ($x_F > 0$) in proton-nucleon collisions]
{$J/\psi$ production cross section ($x_F > 0$) in proton-nucleon 
collisions. Solid line: Fit to the total cross section including radiative 
feed-down from the $\chi_{c}$ states and $\psi '$. Dashed line: Direct $J/\psi$ 
cross section. Dotted line: Direct $J/\psi$ production, color-singlet only. 
(Taken from \cite{Bnk 97})} 
\label{Fig:Jpsi}
\end{figure}

%-- Fig.9 from hep-ph/9703429 --
       
\section{Developments of Theoretical Models}
\indent
Data from the Tevatron have revealed that the production rate of $\psi'$ at large
transverse momentum is more than an order of magnitude larger than the early 
theoretical predictions. These results can be understood by taking into account 
two more mechanisms. The first is the realization that fragmentation must 
dominate at large transverse momentum, which implies that most charmonium in the large
$p_T$ region is produced by the hadronization of individual high-$p_T$ partons. 
The second is the development of a factorization formalism for quarkonium 
production based on non-relativistic QCD (NRQCD) that allows the formation of 
charmonium from color-octet $c\overline{c}$ pairs to be treated systematically. In
this section we will summarize these theoretical developments.  A more complete 
review of the development was given by Braaten et al. \cite{Bra 96}.

\subsection{Color Singlet Model}
\indent
A thorough review of the applications of the color-singlet model to 
heavy-quarkonium production was given by Schuler \cite{Sch 94}. To describe the 
color-singlet model, we can think of the production of charmonium as proceeding 
in two steps. The first step is the production of a $c\overline{c}$ pair, and 
the second step is the binding of the $c\overline{c}$ pair into a charmonium 
state. 

We first consider the production of the $c\overline{c}$ pair. The $c\overline{c}$
pair must be produced with relative momentum that is small compared to the mass of
the charm quark in order to have a significant probability to be bound together. 
Assuming that the c and $\overline{c}$ do not exist in the initial state, the 
production of a $c\overline{c}$ pair must involve virtual particles which are 
off-shell by amounts of order $m_c$ or larger. This part of the amplitude is 
called the short-distance part, because the spatial separation of the c and 
$\overline{c}$ is of order 1/$m_c$ or smaller. On the other side, the formation 
of the bound state is considered to be the long-distance part of the amplitude. 
The total amplitude of charmonium production is expected to be dependent on the 
charmonium state H and on the quantum numbers of the $c\overline{c}$ bound pair. 

For any charmonium state, the dominant Fock state is a color-singlet 
$c\overline{c}$ pair in a definite angular-momentum state. We introduce the 
following notation, for example, the dominant Fock state for the $J/\psi$ is 
$|c\overline{c}(\underline{1},^{3}S_1)\rangle$, while for the $\chi_{cJ}$ it is 
$|c\overline{c}(\underline{1},^{3}P_J)\rangle$. The color states are denoted by 
\underline{1} for color-singlet and \underline{8} for color-octet, and the 
angular momentum states are denoted using the standard spectroscopic notation 
$^{2S+1}L_J$. The color-singlet model requires that, only the $c\overline{c}$ 
pair in a color-singlet $^{2S+1}L_J$ state can bind to form the charmonium with 
$|c\overline{c}(1_,^{2S+1}L_J)\rangle$ as the final Fock state.

The color-singlet model has enormous predictive power. The cross section for
producing a quarkonium state is predicted in terms of a single nonperturbative
parameter for each orbital angular momentum multiplet. The amplitude for
producing a color-singlet $c\overline{c}$ pair with small relative momentum (the 
short distance part) can be calculated using perturbative QCD, while the 
long-distance part parameters can be determined from experiments and are 
expected to be process-independent. Thus the long-distance parameters determined 
from decays of the charmonium states can be used to predict the normalized 
production rate of charmonium states.

We should keep in mind that the color-singlet model is only a model. The most
basic assumption, the factorization picture, has never been proven to be
correct, and the relativistic corrections which account for the relative velocity
of the quark and antiquark are neglected. The color-singlet model also assumes
that a $c\overline{c}$ pair produced in a color-octet state will never form 
the final charmonium. However it might be possible that a color-octet 
$c\overline{c}$ pair can transit to a color-singlet state by radiating soft 
gluons. We will include the color-octet mechanism in the coming section.

\subsection{Gluon Fragmentation}
\indent
The first major conceptual advance in recent theoretical developments of quarkonium
production was the idea of ``fragmentation.''  Fragmentation is the formation
of a hadron within a jet produced by a parton with large transverse momentum.
But here this term is used to include general hadronization processes.

The real revolution about the fragmentation mechanism is the realization that
a colored parton, generally a gluon, can result in a color-singlet final state via 
soft-gluon emissions.  This possibility was not considered in the conventional 
wisdom. Once it was accepted, the color-octet $c\overline{c}$ state could 
also result in the color-singlet final quarkonium by the same argument, and thus
the contributions from color-octet components become possible, as opposed to the
color-singlet model. 

When the CSM includes the contributions from gluon-fragmentation,
its prediction qualitatively agrees with the shape of the CDF $\psi'$ data, but is
still off in normalization
by an order of magnitude, in the $\psi'$ $p_T$ differential cross section. The 
prediction from the CSM failed completely in the high $p_T$ region without 
including 
the gluon-fragmentation mechanism \cite{Bra 95}. Figure \ref{Fig:anomaly} shows 
the predictions and data.

\subsection{Color-Octet Mechanism}
\indent
The second major conceptual advance is to realize that the color-octet
mechanism can be important. Contrary to the basic assumption of the 
color-singlet model, a $c\overline{c}$ pair that is produced in a color-octet 
state can bind to form the charmonium final state.

\begin{figure}
{\centering {\resizebox{14.3cm}{19cm} {\includegraphics*[4cm,3.5cm][18.3cm,25cm]
{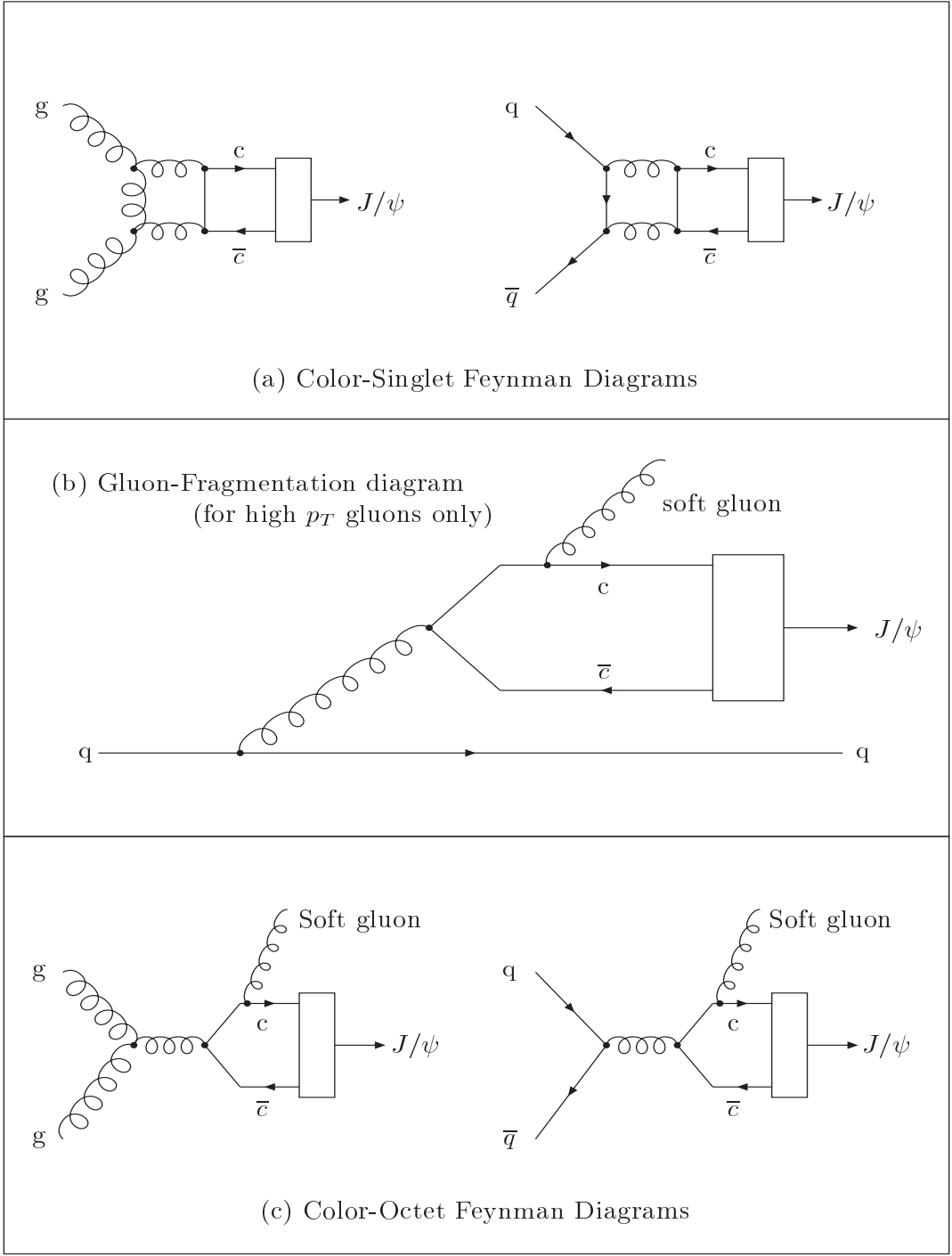}}} \par}
\caption[Feynman diagrams of $J/\psi$ production mechanisms]
{Feynman diagrams of (a) Color-Singlet mechanism, (b) Gluon fragmentation,
(c) Color-Octet mechanism.} 
\label{Fig:Feynman}
\end{figure}

By including the contribution from the color-octet object in the matrix element,
one can make the prediction agree well with the experimental data by leaving
the relative size of the color-octet contribution as an adjustable parameter. Its 
verification now
requires considering quarkonium production in other processes in order to
demonstrate process-independence of the long-distance part of the color-octet 
matrix element. Now the data available from different processes are CDF data, 
fixed-target data, and photo-production data. The size of color-octet 
contributions from these data are not obviously in agreement with each other and 
more sophisticated explanations are needed.

Other problems associated with the color-octet mechanism are the discrepancies 
with the $\chi_{c1}/\chi_{c2}$ production ratio and the $J/\psi$($\psi'$) 
polarization: 
the $\chi_{c1}/\chi_{c2}$ production ratio remains almost an order of magnitude 
too low, and the predicted transverse polarization of the $J/\psi$ and $\psi'$ is 
too large compared to the existing pion data in fixed-target experiments \cite{Ben 96}.
All of these suggest that higher-twist effects may be substantial even after 
including the octet mechanism.
     
A polarization measurement is a crucial test for the color-octet mechanism.
Since the octet production matrix elements of NRQCD lead to a polarization
pattern different from the CSM, a polarization measurement can provide us with
significant information on quarkonium production. For example, $J/\psi$'s 
produced at the Tevatron at large $p_T$ are predicted to be almost fully 
transversely polarized, i.e. $\lambda(J/\psi)$ $\sim$ 1 \cite{Cho 95}, as a 
result of production via gluon fragmentation. At smaller $p_T$, the $J/\psi$'s are
predicted to be produced essentially unpolarized around 
$p_T$ $\sim$ 5 Gev \cite{Ben 97}. The observation of this polarization pattern 
would test the underlying theory (the Factorization Approach). To limit the 
introduction, we will concentrate our attention on fixed-target experiments from 
now on since FNAL E866 is a fixed-target experiment.

The polarization of the quarkonium, measured by analyzing the angular distribution
of the quarkonium decay products in its rest frame, is of the form

\begin{equation}
d \sigma /d \cos \theta \sim 1 + \lambda \cos^2 \theta \label{eq1}
\end{equation}

\noindent where $\theta$ is the polar angle measured in the rest frame of the 
quarkonium. The quarkonium rest frame is well specified except for arbitrary 
three-dimensional rotations. The Collins-Soper frame \cite{Col 77}, in which the 
Z-axis is defined to be parallel to the bisector of the angle between the 
directions of the
interacting hadrons in the quarkonium rest frame, is used in this analysis. In
all other earlier fixed-target experiments the Gottfried-Jackson frame, in which 
the Z-axis is defined to be parallel to the incoming beam axis in the quarkonium 
rest frame, was used. These two frames are equivalent if the quarkonium has zero 
$p_T$ \cite{Fal 86}. For the $p_T$ range of fixed-target experiments, of the order
of 1 GeV, compared to hundreds of GeV of longitudinal momentum, the two frames are
approximately the same \cite{Gee 98}.

\section{Fixed-Target Polarization Experiments and Predictions}
\indent
Polarization measurements have been performed for $J/\psi$ and $\psi'$ production 
in pion and proton scattering fixed-target experiments. From a theoretical point 
of view, the $\psi'$ decay has been more extensively studied because all the 
$\psi'$ data samples are direct $\psi'$s. The observed value of $\lambda$ for 
$\psi'$ is $0.02\pm0.14$, measured at $\sqrt{s}$ = 21.8 GeV in the region 
$x_F > 0.25$ by Heinrich et al. \cite{Hei 91}. When studying the 
polarization of the
$J/\psi$ decay one has to take the polarization inherited from decays of 
the higher charmonium states $\chi_{cJ}$ and $\psi'$ into account and this leaves 
some ambiguity in the interpretation of the results. In the following sections we 
will only review $J/\psi$ polarization experiments to compare with the E866 
results.

\subsection{Model Predictions of Polarization at Fixed-Target Energies}

\subsubsection{Color-Singlet Model}
\indent
The polarization of $J/\psi$ has been calculated from perturbative QCD by 
Vanttinen et al. \cite{Van 95}. The parameter $\lambda$ in Equation 1.1 was 
calculated from the $c\overline{c}$ production amplitude and the electric dipole 
approximation of radiative $\chi$ decays. 

\begin{figure}
{\centering {\resizebox{14.3cm}{!}{\includegraphics*{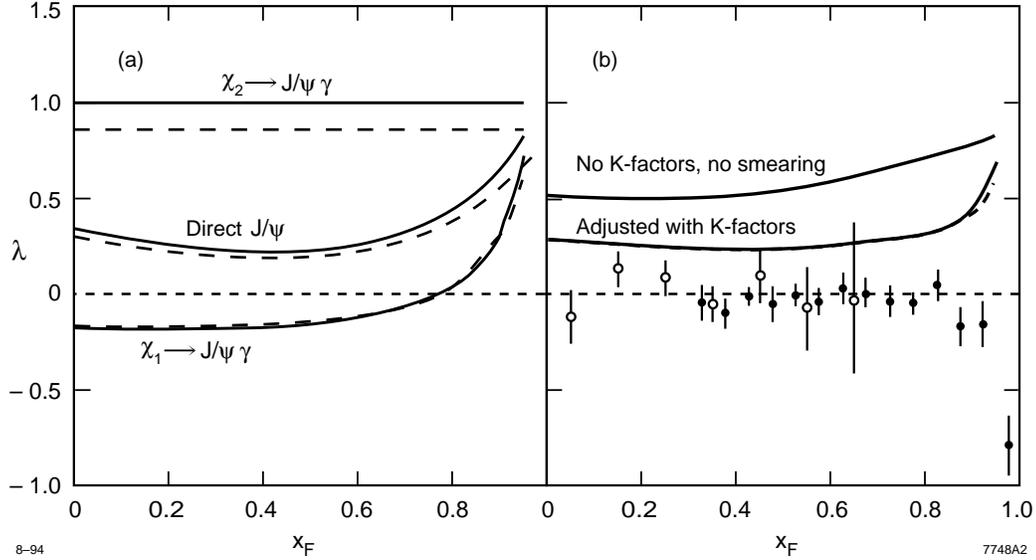}}} \par}
\caption[Leading-twist predictions of $\lambda(x_{F})$ in $\pi N$ collisions]
{Leading-twist predictions of the parameter $\lambda$ in the decay
angular distribution of $J/\psi$'s produced in pion-nucleon collisions at beam
energy
$E_{lab}$ = 300 GeV, plotted as a function of $x_F$. (a)The three solid curves
show the decay distributions of $J/\psi$'s produced via radiative decays of the
$\chi_2$ and $\chi_1$ states and ``directly'' in gluon fusion. The dashed curves
show the effect of smearing the transverse momentum distribution of the beam
parton by a Gaussian function exp[-($k_{\perp}$/500 MeV)$^2$]. (b)The combined
decay distribution of all $J/\psi$'s, including contributions from $\chi_{1,2}$
decays and direct production, is shown here. The lower curve shows the effect
of adjusting the relative normalization of the different contributions to their
measured values by appropriate K-factors. The dashed curve shows the effect of
transverse-momentum smearing and K-factors adjustments. The data are from the
Chicago-Iowa-Princeton (full circles) and E537 (open circles) experiments. (Taken
from \cite{Van 95}).}
\label{Fig:CSM}
\end{figure}

%-- Fig. 1 in hep-ph/9410237 --

Figure \ref{Fig:CSM}a shows the predicted values of the parameter $\lambda$ in 
Equation 1.1 in the
Gottfried-Jackson frame as a function of $x_F$, for the direct $J/\psi$ and the
$\chi_{1,2} \rightarrow J/\psi + \gamma$ processes separately. The dashed lines 
indicate the effect of a Gaussian smearing in the transverse momentum of the beam 
partons. The overall $\lambda(x_F)$ including direct and indirect $J/\psi$ 
processes is shown in Figure \ref{Fig:CSM}b and compared with the 
Chicago-Iowa-Princeton \cite{Bii 87} and E537 data \cite{Ake 93}. The QCD 
calculation gives 
$\lambda \sim 0.5$ for $x_F <0.6$, significantly larger than the measured 
value. The lower curve in Figure \ref{Fig:CSM}b shows the effect of multiplying 
the partial $J/\psi$ cross section with the K-factors obtained from experiments. 
The discrepancies between the calculated and measured values of $\lambda$ suggest 
that the leading-twist processes considered in the calculation are not adequate 
for explaining charmonium production.

\subsubsection{Color-Evaporation Model}
\indent
The color-evaporation model \cite{Fri 77,Hal 77} assumes that the $c\overline{c}$ 
pair in $^{3}S_1$ state
can transit to $^{1}S_0$ state via soft gluon emission, so $J/\psi$ is always 
produced unpolarized. In this model the color and spin quantum numbers of the 
$c\overline{c}$ pair are irrelevant. The fraction of the $c\overline{c}$ pairs 
bound into $J/\psi$ is described by a phenomenological parameter $f_{J/\psi}$.

The color-evaporation model is considered to be an over-simplified model, because
it is not concerned with the details of the particles which initiate the 
reaction.
The evident failure is the prediction of the fraction of $J/\psi$ coming from
$\chi_c$ decays. According to the color-evaporation model, the fraction of $J/\psi$
coming from $\chi_c$ decays should be process-independent. But the experimental 
data both in fixed-target experiments in $p$N and $\pi$N collisions and also in
$p\overline{p}$ collisions at the Tevatron gather around a central value of 
0.3-0.4, while in $\gamma$-p collisions an upper limit of 0.08 was 
obtained \cite{Bar 87}. 

Since this model gives trivial prediction on $J/\psi$ polarization and fails in
predicting ratios of quarkonium production, we will not discuss this model in
later discussions. 

\subsubsection{Non-Relativistic QCD}
\indent
The polarization of $J/\psi$ has been calculated in non-relativistic QCD
by Beneke and Rothstein \cite{Ben 96}. The production cross section 
for a quarkonium state H in the process

\begin{equation}
A + B \longrightarrow H + X  \label{eq2}
\end{equation}

\noindent
can be written as

\begin{eqnarray}
\sigma_{H} = \sum_{i,j} \int_{0}^{1} dx_{1}dx_{2}f_{i/A}(x_{1})f_{j/B}(x_{2})
                \hat{\sigma}(ij \rightarrow H) \\ \label{eq3}
\hat{\sigma}(ij \rightarrow H) = \sum_{n} C^{ij}_{\overline{Q}Q[n]} \langle O^H_n \rangle \label{eq4}
\end{eqnarray}

In Equation 1.3 the summation sums up the contributions by all partons in the colliding
hadrons, and the $f_{i/A}$ and $f_{j/B}$ are the corresponding parton distribution
functions (PDF). The coefficients $C^{ij}_{\overline{Q}Q[n]}$ in Equation 1.4
describe the production of
a quark-antiquark pair in a state n and have expansions in $\alpha_{s}(2m_Q)$. 
The parameters $\langle O^H_n \rangle$ describe the subsequent hadronization of 
the $Q\overline{Q}$ pair into
the quarkonium state H. It is important to test the universality of the
production matrix elements $\langle O^H_n \rangle$ because this is an essential 
prediction of the factorization formula (1.4).

In the calculation of Beneke and Rothstein, the following intermediate $c\overline{c}$
states are considered: ($\underline{1}$,$^{3}S_1$), ($\underline{8}$,$^{1}S_0$), 
($\underline{8}$,$^{3}P_J$), and ($\underline{8}$,$^{3}S_1$). For
each intermediate state the ratios of longitudinal to transverse polarized
quarkonia were computed. To obtain the total polarization, the various
subprocesses have to be weighted by their partial cross sections. Weighting all
subprocesses by their partial cross sections and neglecting the small $\psi'$
feed-down, a sizable polarization is obtained:

\begin{center}
           $0.31<\lambda<0.63$
\end{center}

However the existing data show no sign of polarization. Thus NRQCD including the
color-octet contributions also gives a wrong prediction on the $J/\psi$ 
polarization problem, and one has to seek for explanations from higher-twist processes. 
           
\subsection{Fixed-Target $J/\psi$ Polarization Experiments}

\subsubsection{E537}
\indent
Fermilab experiment E537 has measured the differential cross section 
$d \sigma / d \cos \theta$ for $J/\psi$ production in $\pi^{-}$N 
interactions and in $\overline{p}$N interactions at $\sqrt{S}$ = 15.3 GeV in the 
region $x_F>0$ \cite{Ake 93}. Fitting the angular
distribution to the form of Equation 1.1, $\lambda = -0.115\pm0.061$ for 
$\overline{p}$ and $\lambda = 0.028
\pm0.004$ for $\pi^{-}$ were obtained. The data sample used to obtain this result
contained 12530 $J/\psi$ events produced by the $\overline{p}$ beam and 33820 
$J/\psi$ events by the $\pi^{-}$ beam.

\begin{figure}
{\centering \rotatebox{270} {\resizebox{10cm}{!} {\includegraphics*[6cm,2cm][15cm,12cm]
{thesisfig1.10.ps}}} \par}
\caption[cos$\theta$ distribution for E537 $\overline{p}W$ data]
{cos$\theta$ distribution for E537 $\overline{p}W$ data. 
(Taken from \cite{Ake 93})}
\label{E537a}
\end{figure}

\begin{figure}[h]
{\centering \rotatebox{270} {\resizebox{11cm}{!} {\includegraphics*
[6.5cm,16.5cm][15.5cm,25cm]{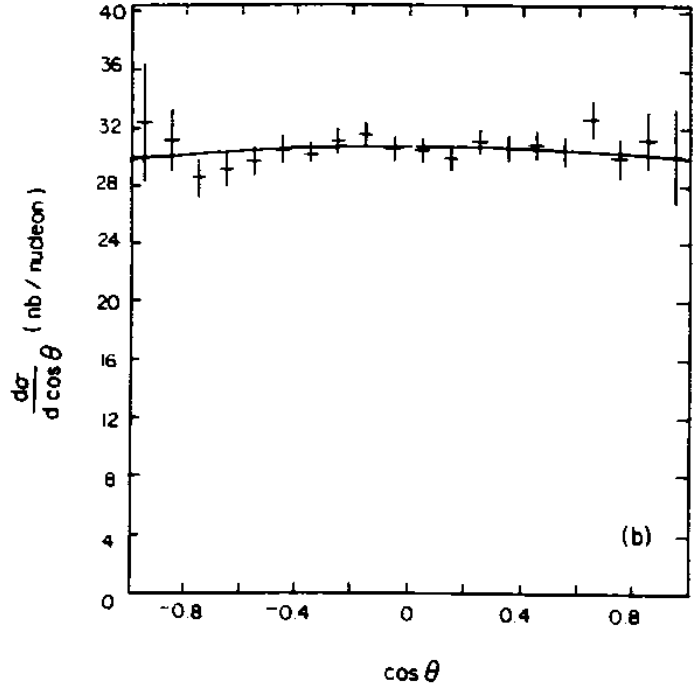}}} \par}
\caption[cos$\theta$ distribution for E537 $\pi^{-}W$ data]
{cos$\theta$ distribution for E537 $\pi^{-}W$ data. 
(Taken from \cite{Ake 93})}
\label{E537b}
\end{figure}

%-- Fig.11 from [33] --

\subsubsection{E672/E706}
\indent
Fermilab experiments E672/E706 have measured the differential cross section
$d \sigma/d \cos \theta$ for $J/\psi$ production in $\pi^{-}$Be 
collisions at $\sqrt{S}$ = 31.5 GeV in the region $0.1<x_F<0.8$ \cite{Gri 96}.
Fitting the angular distribution to the form of Equation 1.1, 
$\lambda = -0.01\pm0.08$ was obtained. The data sample used to obtain
this result contained 9600 $J/\psi$'s.

\begin{figure}
{\centering \includegraphics*[3.0cm,6.5cm][25cm,25cm]{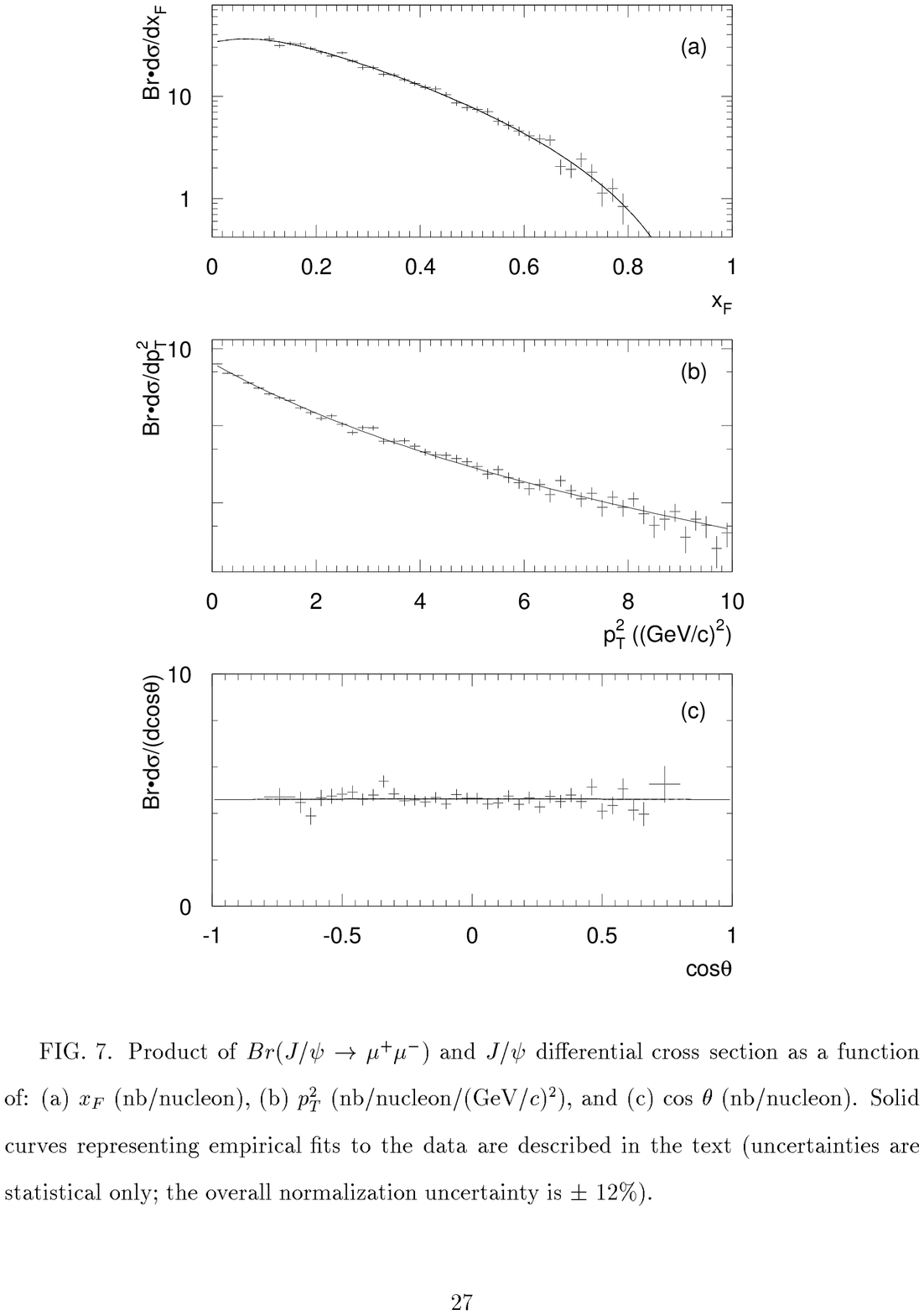} \par}
\caption[$x_F$, $p_T$, and $\cos \theta$ distribution for the E672/706 $\pi^{-}Be$ data]
{$x_F$, $p_T$, and $/cos \theta$ distribution for the E672/706 
$\pi^{-}Be$ data. (Taken from \cite{Gri 96})}
\label{E672/706}
\end{figure}

%-- Fig.7 from [34] --

\subsubsection{E771}
\indent
Fermilab experiment E771 has measured the differential cross section 
$d \sigma/ d \cos \theta$ for $J/\psi$ production in $pSi$ 
collisions at $\sqrt{S}$ = 38.8 GeV in the region $-0.05<x_F<0.25$ \cite{Ale 97}. 
This is the only published polarization measurement for $J/\psi$ produced with a
proton beam. Fitting 
the angular distribution to the form of Equation 1.1,
$\lambda = -0.09 \pm 0.12$ was obtained. The data sample used to obtain this 
result contained 11660 $J/\psi$'s.

\clearpage
\begin{figure}
\vspace*{1.0in}
{\centering {\resizebox{8cm}{4.5cm} {\includegraphics*
[5cm,10.2cm][16cm,18.5cm]{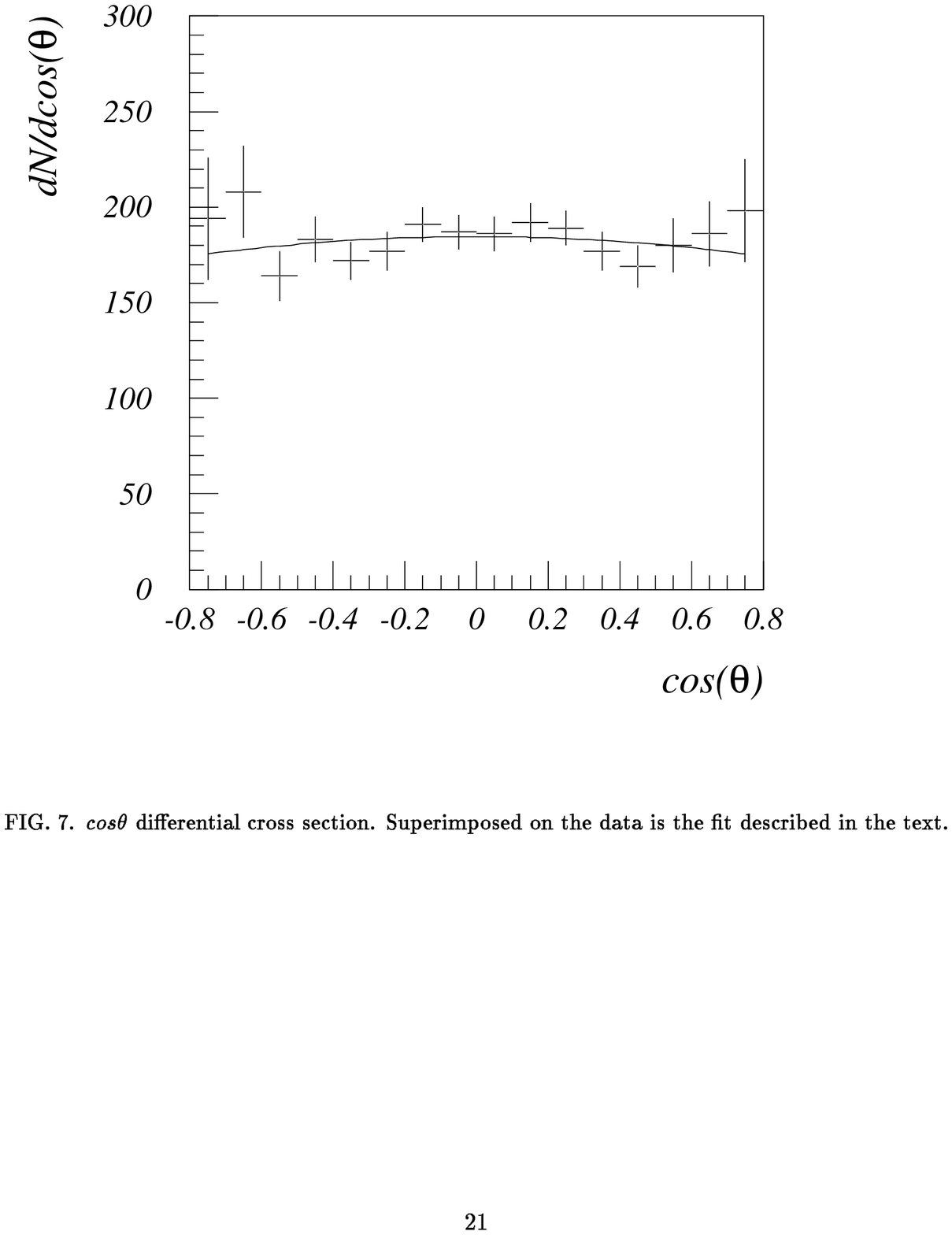}}} \par}
\caption[$\cos \theta$ distribution for E771 $pSi$ data]
{$\cos \theta$ distribution for E771 $pSi$ data. (Taken from \cite{Ale 97})}
\label{E771}
\end{figure}

\subsubsection{Chicago-Iowa-Princeton}
\indent
A dedicated $J/\psi$ decay angular distribution measurement was performed
at Fermilab using a 252 GeV pion beam incident on a tungsten target \cite{Bii 87}. 
The data
sample contains 1600000 $J/\psi$ events from a $\pi^-$ beam and 600000 $J/\psi$ events from
a $\pi^+$ beam. The data are in the kinematic range $x_F>0.25$ and $p_T<5.0$ GeV. To
determine the $J/\psi$ decay angular distribution, the data were divided into
fifteen regions of $x_F$, five regions of $\cos \theta$, and five regions of $\phi$ in
the kinematic range $x_F>0.25$, $-1<\cos \theta<1$, and $-\pi<\phi<\pi$. For each bin of
$x_F$, $\cos \theta$, and $\phi$ the raw $\mu^{+}\mu^{-}$ mass distribution was 
fitted by a seven-parameter function involving a Gaussian distribution for the 
$J/\psi$ and $\psi'$ and a
quadratic polynomial plus an exponential of a first-order polynomial for the
continuum background. The number of $J/\psi$'s in the 375 bins of $x_F$, $\cos \theta$,
and $\phi$ were then corrected for acceptance and for each of the fifteen regions
of $x_F$ the $J/\psi$ angular distribution was fitted by the general form \cite{Lam 78}

\begin{equation}
  d^{2}\sigma/d \cos\theta d \phi \sim 1+ \lambda \cos^{2}\theta + \mu \sin 2\theta
                                  \cos\phi + \frac{1}{2}\nu \sin^{2}\theta \cos 2\phi \label{eq5}
\end{equation}

\begin{figure}
{\centering \includegraphics*[2.5cm,9.5cm][10cm,20.5cm]{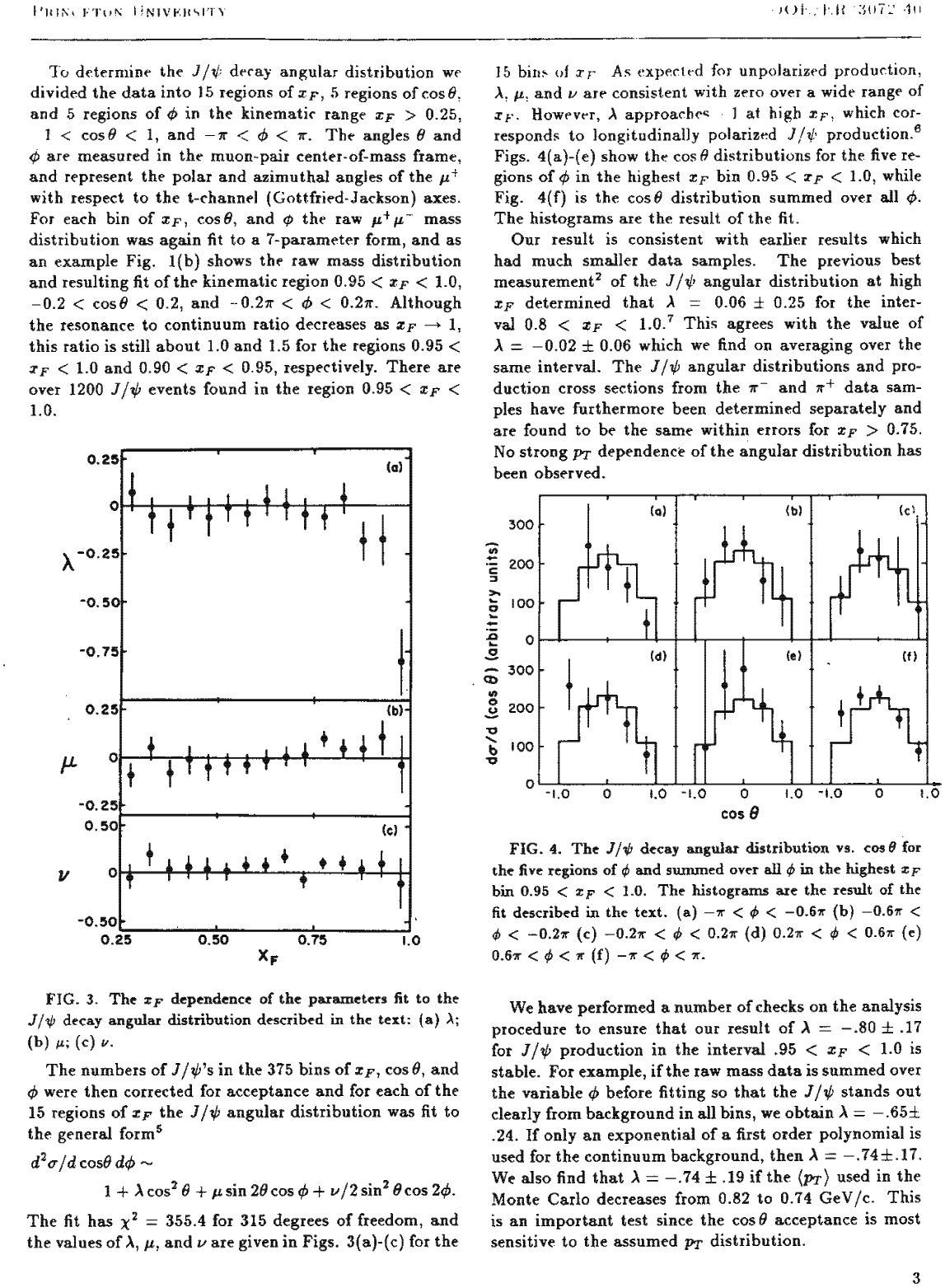} \par}
\caption[$\lambda(x_{F})$ from CIP $\pi N$ experiment]
{The $x_F$ dependence of the parameters fit to the $J/\psi$ decay angular
distribution in Equation 1.5. (Taken from \cite{Bii 87})}
\label{CIP}
\end{figure}

%-- Fig. 3 from [36]
\noindent
The $\lambda$, $\mu$, and $\nu$ are consistent with zero over a wide range of 
$x_F$. Note that
$\lambda$ approaches $-$1 at high $x_F$. This behavior was also observed in a
Drell-Yan
continuum production experiment \cite{Ale 86}. Table \ref{lsummary} summarizes the 
experimental and theoretical results of $J/\psi$ polarization.
\\
\begin{table}[tbp]
\caption{Summary of the experimental and theoretical results.}
\label{lsummary}
\begin{center}
\begin{tabular}{ccccc}
\hline \hline
Experiment & reaction & $\sqrt{S}$ & $x_F$ range & $\lambda$ \\
\hline
E537      & $\overline{p}$ + W & 15.3 GeV & $x_F > 0$ & $-0.115 \pm 0.061$\\
E537      & $\pi^-$ + W        & 15.3 GeV & $x_F > 0$ & $0.028 \pm 0.004$\\
E672/706  & $\pi^-$ + Be       & 31.5 GeV & $0.1<x_F<0.8$ & $-0.01 \pm 0.08$\\
E771      & p + Si             & 38.8 GeV & $-0.05<x_{F}<0.25$ & $-0.09 \pm 0.12$\\
CIP       & $\pi$ + W          & 21.7 GeV & $0.25<x_F<1.0$ & $\sim$0, $\rightarrow$ $-$1 at large $x_F$\\
\hline \hline
\end{tabular}
\end{center}
\end{table}

\begin{table}[tbp]
\begin{flushleft}
\begin{tabular}{ccc}
\hline \hline
Theory & $x_F$ range & $\lambda$\\
\hline
CSM &    $x_F > 0$ &   $\sim$ 0.25\\
CEM &    $x_F > 0$ &   0\\             
NRQCD &  $x_F > 0$ &   $0.31<\lambda<0.63$\\
\hline \hline
\end{tabular}
\end{flushleft}
\end{table}
\section{Fermilab E866 Measurement}
\indent
E866 at Fermilab was designed to measure the $\overline{u}$/$\overline{d}$ 
asymmetry in the nucleon
sea. After the run ended in March, 1997, additional measurements were performed 
in the run extension period. Two major topics in the run extension were angular
distribution of the $J/\psi$ decay and nuclear dependence of $J/\psi$ production. 
The work
presented here is based on the data sample collected during a four week
dedicated beam-dump run, from which the angular distribution of the $J/\psi$ decay
in the
dilepton channel was studied. This angular-distribution measurement is unique
since no high-statistics proton-induced data exists. Also, the $J/\psi$ production 
diagrams are different for pN and $\pi$N interactions. A total of 10 million 
$J/\psi$'s (with $\sim$ 1$\%$ of unseparated $\psi'$'s) were collected, and the 
kinematic coverage of the data extends over $x_F > 0.2$, $p_T< 5GeV$, and 
$-0.95<\cos \theta<0.95$. The quantity of the data sample has allowed us to 
present the $\lambda$ parameter in Equation 1.1 in seven regions of $x_F$ and 
four regions of $p_T$. The results could provide a test of the color-octet 
mechanism, and hopefully will improve our understanding of the higher-twist 
effects. 

\chapter{EXPERIMENTAL APPARATUS}
\indent
The experiment E866 was performed at the Meson-East experimental area of Fermi
National Accelerator Laboratory. The spectrometer, shown in Figure \ref{Fig:spec},
was a modified version of the E605/E772/E789 spectrometer. This spectrometer 
was designed to detect dimuon events with forward $x_F$, though certain
combination of target position and analyzing magnet settings allows finite 
negative-$x_F$ acceptance. The spectrometer primarily consisted of three dipole
magnets, seven hodoscope planes, eighteen drift-chamber planes, and three
proportional-tube planes. The hodoscope planes were used to provide the trigger
information, the drift-chamber planes were used to find the trajectories,
and the proportional-tube planes, which were also part of the trigger system,
were used to identify muons. The SM3 magnet measured the momentum of the muon
pairs while the SM0 and SM12 magnets allowed us to select the desired mass range.
The charged particles produced in the target were split according to the sign of
their charges while going through the set of three magnets.

\begin{figure}[h]
\begin{center}
\mbox{\epsfxsize=6.0in\epsffile{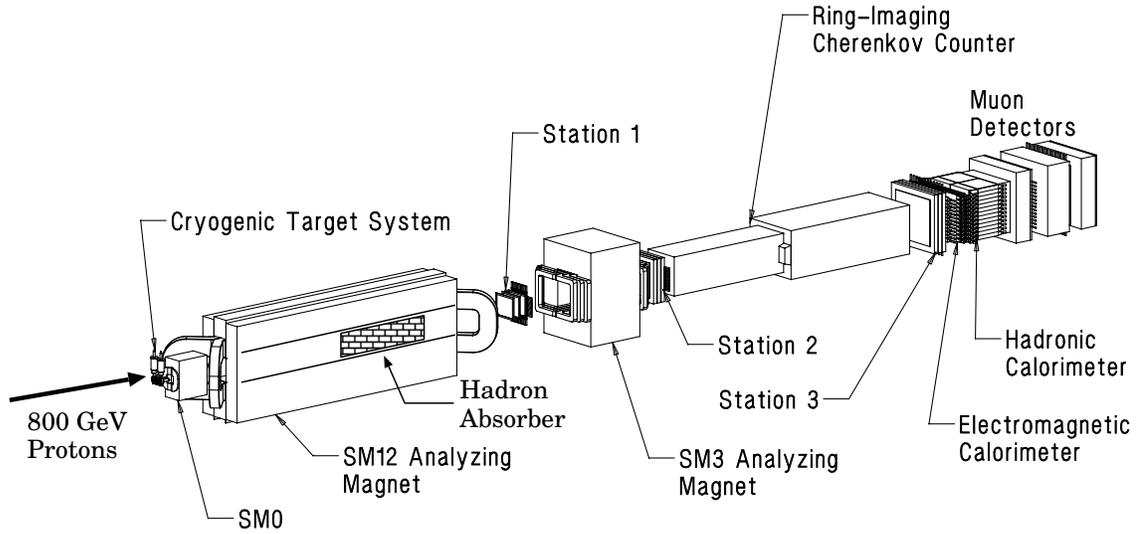}}
\end{center}
\caption{The E866 spectrometer.}
\label{Fig:spec}
\end{figure}

A thick hadron absorber wall was installed in SM12 for the experiment so that
the long-lived hadrons (mainly pions) can be stopped before hitting station 1.
The absorber wall consisted of Cu, C, and (CH$_2$)$_n$ blocks, and gave a hadron
attenuation factor of $e^{-20}$. Not shown in Figure 2.1 is the copper beam dump
sitting in front of the hadron absorber wall. The beam dump was used to stop
the 800 GeV proton beam. For the angular distribution measurement we used the beam
dump as the target. There were also an electromagnetic and a hadronic
calorimeter and a ring-imaging Cherenkov counter (RICH), which, however, were not
operating during this experiment. The RICH counter was filled with helium bags to
reduce multiple scattering. This chapter will discuss only these components that 
were needed for this study. 

Throughout, we will make reference to the spectrometer-fixed coordinate system.
The E866 coordinate system aligns the Z-axis horizontally with the accelerator
proton beamline and the Y-axis with the vertical direction. The X-axis is then 
chosen to form a right-handed Cartesian system. The positive Z direction is 
chosen to be same as the beamline direction, which is also referred to as 
``downstream,'' and the positive Y direction is chosen to be up. The origin is 
located at the center of the upstream face of the SM12 spectrometer magnet.

\section{Accelerator and Beam}
\indent
The high-energy proton beam was produced in the Tevatron, which is a
superconducting proton synchrotron. Protons were first accelerated by a 
pre-accelerator up to about 700 KeV. These protons were then accelerated in a 
linear accelerator to about 400 MeV. Subsequently, a Booster Ring boosted the 
proton energy to 8 GeV. Protons were then injected into the main ring, located in
the same tunnel as the Tevatron but constructed from conventional magnets, in 
which protons could reach an energy of 400 GeV before being transferred into the 
superconducting ring, where the protons were accelerated to 800 GeV. After the
proton beam was accelerated to 800 GeV, it was extracted and split by the 
switchyard for sending three streams of proton beams to the Meson, Neutrino, and
Proton beam lines for the fixed-target experiments. 

Protons in a spill were bunched into RF buckets separated from 
one another by 18.9 ns, with bucket length $\sim$ 1 ns. Each spill contained about
$10^{9}$ buckets. This small scale beam structure was due to the Tevatron 
accelerating radio frequency of 53 MHz. A square wave signal at this frequency,
called the RF clock, was used to synchronize the E866 electronics with the 
Tevatron beam structure. It took $\sim$ 25 seconds to accelerate a fill of protons
up to 800 GeV. These protons were then extracted from the accelerator for 20 
seconds. After that, the superconducting magnets ramped down for ~15 seconds. The 
entire cycle time was approximately one spill per minute. Typical proton 
intensities in the Tevatron were $1-2*10^{13}$ protons per spill.

Within the Meson area, a three-way split divided the proton beam between the
Meson-East line and the rest of the Meson lines. To monitor the beam intensity,
luminosity, position, and beam-spot size, several beamline detectors and 
monitors were used. During the beam-dump running mode the typical beam intensity 
was $6*10^{10}$ protons per spill. The beam intensity was monitored by an ion 
chamber located in the ME3 sector (IC3), a secondary-emission monitor located in 
the ME6 sector (SEM6), and a beam Cherenkov monitor. Both the size and the position of 
the beam were monitored by segmented wire ion chambers (SWICs) and the 
Beam-Position Monitor (BPM). The beam luminosity was monitored by the AMON and 
WMON scintillation counters, which were installed at about 85 degrees from the 
target position.

\section{Beam-Dump Target}
\indent
For this study the beam dump itself was the target. The dump was suspended 
from two of the central magnet inserts inside the SM12 magnet, beginning 
at Z = 68 inches, extending 168 inches downstream, and ending at Z = 236 inches. In 
Figure \ref{Fig:sm12} a picture of the beam dump and absorber wall is shown. The
beam dump was made of pure copper with cooling water tubes running through the 
sides. There was a 12-inch deep rectangular-shaped hole in the center of the 
upstream face of the dump to help contain backscattered particles, so actually the
beam protons did not hit the dump until Z = 80 inches. This still left 156 inches 
thickness of copper, which is equivalent to 26.5 interaction lengths for p + Cu 
collisions, to stop the protons and secondary particles. The probability for a 
primary proton to punch through the entire dump was less than $4*10^{-12}$. Most 
primary protons would interact within the first few interaction lengths. Secondary
particles from the primary interaction would further interact to form showers and 
eventually be stopped in the dump, but the high-energy muons produced would 
penetrate the entire dump with little interaction since muons are not 
strongly-interacting particles. However, these muons would still lose energy and 
suffer multiple scattering on their way through the dump, and thus added 
uncertainties to the reconstructed $x_F$ and $p_T$.

\begin{figure}[h]
\begin{center}
\mbox{\epsfysize=4.3in\epsfxsize=5.0in\epsffile{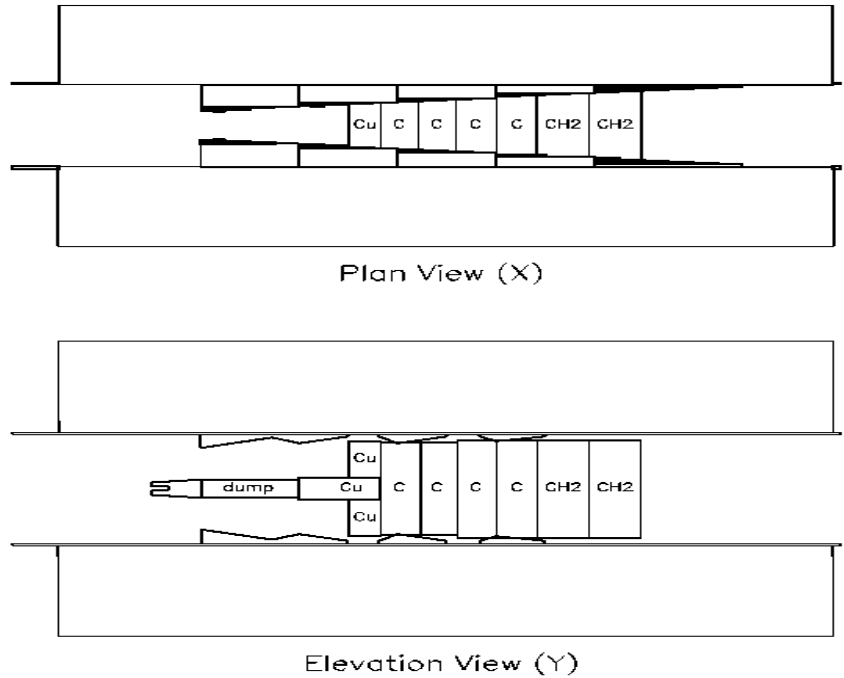}}
\end{center}
\vspace*{-0.25in}
\caption[The SM12 acceptance defining magnet with the absorber wall]
{The SM12 acceptance defining magnet with the absorber wall is
shown. The magnet coils and iron return yoke are only partially shown.
The beam dump is also not shown in the plan view. } 
\label{Fig:sm12}
\end{figure}

%--Fig. 2.2 sm12.eps --

\section{Spectrometer Magnets and Absorber Wall}
\indent
Two dipole magnets, SM12 and SM3 were used in this study (the current of SM0 was 
set
to zero during the beam-dump data taking). The magnetic fields of these magnets
were oriented horizontally. The field strengths of the dipole magnets could be
configured to optimize the mass acceptance for $J/\psi$'s. The bending magnet used
for this study was SM12. The length of the SM12 magnet, made of iron, was 567 
inches. The magnet produced an average horizontal field of up to 1.3 Tesla at a 
maximum current of 4000 Amperes. This corresponds to a 7-GeV transverse momentum 
kick to the charged particles which traveled through its entire length. In this 
study, SM12 was set to 2800 Amperes and 2040 Amperes during two separate 
data-taking periods, delivering a transverse-momentum kick of 4.2 GeV and 3.1 GeV,
respectively.

The momenta of the muon tracks were measured by the analyzing magnet SM3. The
location of the SM3 magnet was between Station 1 and Station 2, as shown in
Figure \ref{Fig:spec}. SM3 delivered a transverse momentum kick of 0.914 GeV to the charged 
particles traveling through when operated at its maximum current of 4260 Amperes.
The field was uniform enough so that the reconstruction of a particle 
trajectory through the field volume can be described by a single bend-plane 
approximation.

The absorber wall was located inside the SM12 magnet directly behind the beam 
dump. It filled the SM12 magnet completely in the x and y direction. The 
absorber wall was constructed of one 24-inch section of copper, three 27-inch
sections of carbon graphite, one 27-inch section of carbon-polyethelene 
compound, and two 36-inch sections of polyethelene.

Both magnets were filled with helium bags to minimize the multiple scatterings 
of the muons.
   
\section{Detector Stations}
\indent
There were four detector stations in the E866 spectrometer, denoted as Station
1 to 4. Station 1-3 each consisted of hodoscopes and drift chambers, while
Station 4 consisted of hodoscopes and proportional tubes. Those stations record
the passage of charged particles in space and time across their active area.
Together with the information provided by the magnet field maps, this allowed the
4-momentum of the individual tracks to be reconstructed. Stations 1-3 were used
for triggering and tracking, while Station 4 was used for muon identification
and triggering.

\subsection{Drift Chambers}
\indent 
Each one of Stations 1-3 consisted of 6 planes of drift chambers. The 6 
planes were arranged in pairs with parallel wire orientation. The second plane of
a pair had its wires offset by half the cell size of the drift chamber. The 
upstream plane of each pair was denoted as the ``unprimed'' plane,
while the downstream plane was denoted as the ``primed'' plane. The Y-Y$^\prime$ 
pair of each station held the wires horizontally to measure the Y-intercept of the
tracks, while the V-V$^\prime$ and U-U$^\prime$ chambers had their wires tilted 
at $-$14 (a slope of $-$0.25) degrees and +14 (a slope of 0.25) degrees from the 
X-axis respectively. These planes determined the X-intercept of the track and also
provided a check on the Y-intercept. The configuration of the drift chambers are 
given in Table \ref{mwpc}.

\begin{table}[tbp]
\caption{Drift chamber parameters. The unit length is one inch.}
\label{mwpc}
\begin{center}
\begin{tabular}{cccccc}
\hline \hline
detector & Z-position & No.of wires & cell size & aperture(X$\times$Y) & operating voltage\\
\hline
V1   &     724.69  &    200  &        0.25   &    48$\times$40      &    +1700\\
V1$^\prime$  &     724.94  &    200  &        0.25   &    48$\times$40      &    +1700\\
Y1   &     740.81  &    160  &        0.25   &    48$\times$40      &    +1700\\
Y1$^\prime$  &     741.06  &    160  &        0.25   &    48$\times$40      &    +1700\\
U1   &     755.48  &    200  &        0.25   &    48$\times$40      &    +1700\\
U1$^\prime$  &     755.73  &    200  &        0.25   &    48$\times$40      &    +1700\\
\hline
V2   &    1083.40  &    160  &        0.388  &    66$\times$51.2    &    $-$2000\\
V2$^\prime$  &    1085.52  &    160  &        0.388  &    66$\times$51.2    &    $-$2000\\
Y2   &    1093.21  &    128  &        0.40   &    66$\times$51.2    &    $-$2000\\
Y2$^\prime$  &    1095.33  &    128  &        0.40   &    66$\times$51.2    &    $-$2000\\
U2   &    1103.25  &    160  &        0.388  &    66$\times$51.2    &    $-$1950\\
U2$^\prime$  &    1105.37  &    160  &        0.388  &    66$\times$51.2    &    $-$1975\\
\hline
V3   &    1790.09  &    144  &        0.796  &    106$\times$95.5   &    $-$2200\\
V3$^\prime$  &    1792.84  &    144  &        0.796  &    106$\times$95.5   &    $-$2150\\
Y3   &    1800.20  &    112  &        0.82   &    106$\times$91.8   &    $-$2200\\
Y3$^\prime$  &    1802.95  &    112  &        0.82   &    106$\times$91.8   &    $-$2200\\
U3   &    1810.24  &    144  &        0.796  &    106$\times$95.5   &    $-$2200\\
U3$^\prime$  &    1812.99  &    144  &        0.796  &    106$\times$95.5   &    $-$2200\\
\hline \hline
\end{tabular}
\end{center}
\end{table}

The drift chambers were all operated with a gas mixture of 49.7$\%$ argon, 49.6$\%$
ethane, and 0.7$\%$ ethanol by volume, which was mixed at a constant 25 $^\circ$F.
The Station-1 anode wires were made of gold-plated tungsten wire, while Stations 
2 and 3 used silver-coated beryllium-copper wires. All the anode wires were 25 
$\mu$m in diameter. The cathode wires for all three stations were silver-coated 
beryllium-copper wire with a diameter of 62.5 $\mu$m. The absolute operating 
voltages were between 1700 and 2200 volts, which gave a typical drift velocity 
about 50 $\mu$m/ns.

The signals of these chambers were read out by a fast amplifier and 
discriminator system. Single-hit TDCs (Time-to-Digital Converters), which only
record the first hit on the wire during an event, were used to measure the drift
time. The combination of good hits together with their associated drift times in 
all three views gave a ``triplet'' hit for a station. The bank of the triplets was
saved to provide information for the track reconstruction.
 
\subsection{Hodoscopes}
\indent
Associated with the drift-chamber planes, there were also hodoscope planes in 
each tracking station. These hodoscopes provided fast tracking signals for use
in triggering. In Stations 1, 3, and 4 there were two hodoscopes planes which 
measured the X and Y intercepts of the tracks, while in Station 2 there was only
one hodoscope plane. Each hodoscope plane was arranged into two half-planes of
parallel scintillator paddles, which were attached to photomultiplier tubes via
plexiglass light guides. During operation, each paddle only gave a single bit of
signal (one or zero).

The hodoscope planes were named according to the tracking station they belonged 
to, preceeded by X or Y depending on the orientation of the paddles. For example,
``Y3 hodoscope'' referred to the Station-3 hodoscope plane in which 2$\times$13
scintillator detectors was positioned horizontally and separated into left and
right side. The parameters of the seven hodoscope planes are given in Table \ref
{hodo}.

\begin{table}[tbp]
\caption{Hodoscope plane layout. Dimensions are in inches.}
\label{hodo}
\begin{center}
\begin{tabular}{ccccc}
\hline \hline
detector & Z-position & No. of counters & cell width & aperture X$\times$Y\\  
\hline
Y1    &     769.78  &   2$\times$16  &           2.5  &       47.50$\times$40.75\\        
X1    &     770.72  &  12$\times$2   &           4.0  &       47.53$\times$40.78\\
Y2    &    1114.94  &   2$\times$16  &           3.0  &       64.625$\times$48.625\\
X3    &    1822.00  &  12$\times$2    &          8.68  &      105.18$\times$92.00\\
Y3    &    1832.00  &   2$\times$13    &         7.5   &      104.00$\times$92.00\\
Y4    &    2035.50  &   2$\times$14    &         8.0     &    116.00$\times$100.00\\
X4    &    2131.12  &  16$\times$2     &         7.125   &    126.00$\times$114.00\\
\hline \hline
\end{tabular}
\end{center}
\end{table}

\subsection{Proportional Tubes}
\indent
Station 4 was also called the muon station. It was located downstream of the
calorimeters and consisted of two hodoscope planes (Y4,X4) and three 
proportional tube planes(PTY1, PTX, PTY2). Each of the three proportional tube 
planes had two layers of 1$\times$1-inch cells. These two layers were offset by a
half-cell spacing to cover the dead region between the adjacent cells. The 
proportional tubes used the same argon/ethane/ethanol gas mixture as the drift
chambers. To minimize the probability of hadron punch-through, an absorber wall
(3 feet of zinc and 4 inches of lead) was placed between the calorimeter and
the muon detector. Furthermore, 3-foot thick concrete walls were placed between
PTY1 and X4, and between PTX and PTY2. This provided a total of 16.6 interaction
lengths upstream of Y4. Thus the only charged particles which could reach 
Station 4 detectors were the muons. Signals from the cell of the proportional
tubes were amplified and shaped by the attached pre-amplifier/discriminator
cards. Signals exceeding the threshold voltage were sent to the Coincidence 
Registers(CRs) to indicate the arrival of muons. The parameters of all 
proportional tube planes are given in Table \ref{proptube}.

\begin{table}[tbp]
\caption{Proportional tube parameters. All dimensions are in inches.}
\label{proptube}
\begin{center}
\begin{tabular}{ccccc}
\hline \hline
detector & Z-position & No.of wires & cell size & aperture X$\times$Y\\
\hline
PTY1  &    2041.75  &   120     &     1.0   &     117$\times$120\\
PTX  &    2135.875  &  135      &    1.0   &     135.4$\times$121.5\\
PTY2  &    2200.75  &   143     &     1.0   &     141.5$\times$143\\
\hline \hline
\end{tabular}
\end{center}
\end{table}

\section{Trigger System}

\subsection{Trigger-System Hardware}
\indent
A new trigger system was implemented for E866 data taking \cite{Haw 98}. A block diagram
of most of the trigger system is shown in Fig. 2.3.

\begin{figure}[hp]
{\centering \rotatebox{90} {
 \begin{minipage}{8.0in}
 \vspace*{+0.75in}
 \mbox{\epsfxsize=7.5in\epsffile{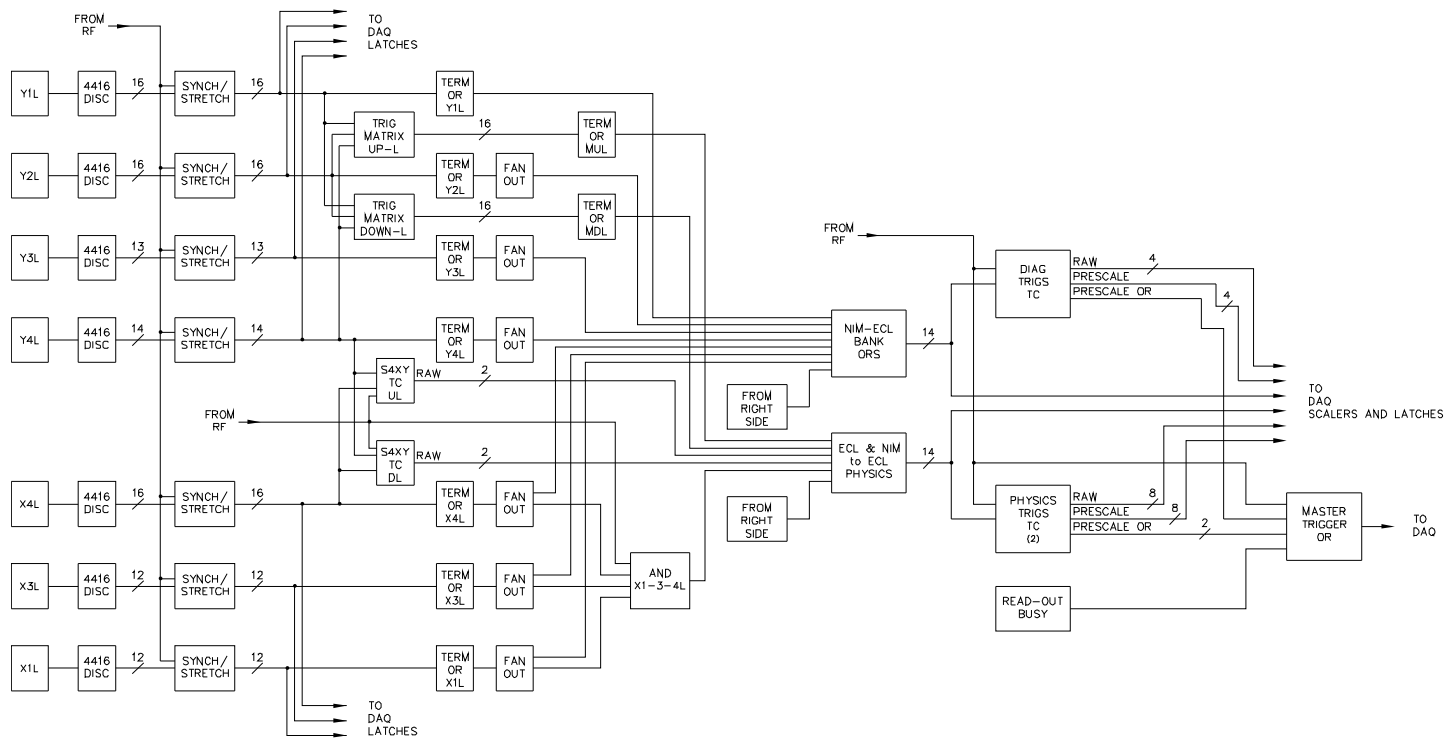}}
 \caption[Block diagram of the E866 trigger system]
 {A block diagram of the E866 trigger system. Note only the
 inputs and associated electronics for the left half of the spectrometer
 are shown.} 
 \label{trig}
 \end{minipage}
} }
\end{figure}

%-- Fig 2.3 diagram of the trigger system --

\subsubsection{Trigger Matrix Module}
\indent
Scintillator counters were used to provide input signals for the trigger 
system. Signals from the photomultiplier tubes attached to the scintillator
counters were brought to LeCroy 4416 16-channel discriminators via coaxial 
cables. Each discriminator output was synchronized to the accelerator RF signal
and shaped to a fixed width of 15 ns by pulse stretchers, and then fanned
out to Coincidence Registers (CRs), Terminator/OR modules, and the Trigger 
Matrix (TM) modules.

The Trigger Matrix modules were the core of the trigger system. The 
pulse-stretcher outputs were grouped as a half-bank (right or left) of Y1, Y2, and
Y4 hodoscope planes. These groups were sent as inputs to the Trigger Matrix to
identify tracks of muons originating in the target. Only Y-view hodo-roads 
were used here, because events with different Z positions and momentum had 
different ``roads'' in the Y view under the deflection of the bending magnets.
This Trigger Matrix was conceptually a lookup table loaded to a set of six 
256$\times$4-bit ECL SRAM chips. All tracks of interest defined a set of valid 
``roads'' going through the hodoscope planes. These roads were identified using a 
Monte Carlo simulation for muon tracks coming from the target, and thus a ``map'' 
of these roads was produced. The ``map'' was then written into a disk file, which 
was loaded into the SRAMs by the Data Acquisition (DAQ) online program during the 
start-run stage. While taking data, the hits on Y2 and Y4 were combined to form 
an ``address'' whose content were the predictions on Y1, which were compared 
to the actual Y1 hit pattern. Any coincidence in the comparison generated the 
Trigger Matrix output.

There were four sets of Trigger Matrix modules called MUL, MUR, MDL, and MDR.
They covered different types of valid muon tracks, namely, up-left, up-right, 
down-left, and down-right, respectively. For finding the target muon pairs, the 
coincident combination of an up and a down track was required. The output 
signals of the Trigger Matrix modules were then sent to the Track Correlator
for further triggering determination.
 
\subsubsection{Track Correlator}
\indent
The Track Correlator (TC) modules were designed by Texas A$\&$M University
\cite{Gag 98}. These programmable modules were used to filter specific 
combinations of Trigger Matrix, Terminator/OR, and S4XY [42] outputs to trigger 
on an event. However, during beam-dump data taking only output signals from the 
TMs were of interest. Four 16-bit patterns, according to the desired trigger 
conditions, were preloaded into a $2^{16}\times 4$ bit SRAM chip inside a TC 
during the start-run stage. Whenever the output combination of the TM modules 
matched one of the preloaded bit patterns, the TC would, prescaled to the desired
frequency and synchronized with the RF clock, send out a signal to the Master 
Trigger OR module to notify the arrival of an interesting event. Each SRAM chip 
could be programmed with up to four independent trigger conditions.

There were three main Track Correlators which were able to trigger on an
event to start the DAQ. The first TC, called Physics TC A, was programmed to 
select two-tracks events, like-sign or unlike-sign. The second TC, named Physics 
TC B, was designated for left-right efficiency studies, which involved the use of X 
hodo planes and single muon events. The third module, which was used to trigger on
cosmic rays to diagnose the trigger and DAQ systems, was called the Diagnostic TC.
It also provided the measurements for scintillator efficiencies.

If the inputs to the Track Correlator fulfilled the triggering criteria, a Trigger
Generate Input (TGI) signal would then be sent to the Master Trigger OR by the TC.
The Master Trigger OR would then synchronize this trigger signal and the drop of 
DAQ System Busy with the RF clock to send out a Trigger Generate Output (TGO). 
Triggers were thus inhibited during event readout; the difference of TGI and TGO 
counts would provide information on readout dead time.

\subsection{Trigger Firing Criteria}
\begin{table}[tbp]
\caption[Correspondence of the SRAM chip bit to various input sources]
{Correspondence of the SRAM chip bit to various input sources of the 
Physics TC A,B module}
\label{sram}
\begin{center}
\begin{tabular}{cc}
\hline \hline
Signal Origin &   TC SRAM bit\\
\hline
S4UL1 &           bit0\\
S4UL2 &           bit1\\
S4DL1 &           bit2\\
S4DL2 &           bit3\\
S4UR1 &           bit4\\
S4UR2 &           bit5\\
S4DR1 &           bit6\\
S4DR2 &           bit7\\
MUL   &           bit8\\
MDL   &           bit9\\
MUR   &           bit10\\
MDR   &           bit11\\
X134L &           bit12\\
X134R &           bit13\\
- unused - &      bit14\\
- unused - &      bit15\\
\hline \hline
\end{tabular}
\end{center}
\end{table}
\begin{table}[tbp]
\caption[Prescale factors and trigger descriptions for Physics TC modules]
{Prescale factors and trigger descriptions for Physics TC A,B
module. The ``*'' represents a logical AND and the ``+'' represents a logical OR.}
\label{tcAB}
\begin{center}
\begin{tabular}{ccc}
\hline \hline
Trigger name &   prescaler factor &  description\\
\hline 
PhysA1     &     1        &          (MUL*MDR) + (MUR*MDL)\\
PhysA2     &     1        &          (MUL*MUR) + (MDL*MDR)\\
PhysA3     &     1        &          (MUL*MDL)\\
PhysA4     &     1        &          (MUR*MDR)\\
\hline
PhysB1     &     10       &          (X134L*X134R)\\
PhysB2     &     1000     &          MUL + MDL + MUR + MDR\\
PhysB3     &     0        &              --\\
PhysB4     &     0        &              --\\
\hline \hline
\end{tabular}
\end{center}
\end{table}

\subsubsection{PhysA Trigger}
\indent
From Table \ref{tcAB} the definitions of PhysA1,2,3,4 triggers are 
self-explanatory. PhysA1 trigger required that two tracks went through two 
diagonally-opposite quarters of the spectrometer, while the PhysA3 and PhysA4 
required that two tracks went through the same side, left or right, of the 
spectrometer with one track going up and the other going down. These tracks were 
identified as unlike-sign muon pairs and were treated as possible candidates of 
target events. The PhysA2 trigger required that both tracks went up or down, and 
thus gave like-sign muon pairs. This information was especially important for 
rate-dependence studies in extracting cross sections.

\subsubsection{PhysB Trigger}
\indent
The Physics TC B was used for recording events for studies. In the PhysB1 trigger,
the symbol ``X134L(R)'' represented a track that went through the left(right) side
of X1, X3, and X4 hodoscope planes. The signals fed into the TC B were outputs
of some Terminator/OR modules, whose outputs represented the logical ORs of the 
signals of the X hodo scintillators. The trigger requirement, X134L*X134R, was 
designed to measure the random muon coincidences. The other trigger PhysB2 only 
required a single hit on any of the four quarters to fire. It had a prescale 
factor of 1000 and was used to measure the rate of single muons.   

\section{Data-Acquisition System}
\indent
The Data-Acquisition System could be divided into three parts by functionality: 
event readout, data archiving, and online analysis. The first 
part was based on a Nevis Transport system, the second was a VME-based 
data-transferring and controlling system, and the third was built on the SGI 
workstations.

\subsection{Readout System}
\indent
The backbone of the E866 readout was a Nevis transport system \cite{Kap 82}. All 
detector subsystems ultimately fed data into the Transport. The subsystems 
included Time-to-Digital Converter (TDC) readouts from drift chambers and 
Coincidence Registers (CRs) from 
hodoscopes and muon proportional tubes signals. Bus arbitration was maintained by 
a hard-wired daisy chain, with the bus mastership determined by the Carry signal.
This scheme not only prevented multiple subsystems from attempting to place data
on the Transport simultaneously, but also guaranteed that events appeared on the
readout bus in a well-defined order. The data bus was 16-bit wide, and the 
system clock was set to 10 MHz. All the data fed into the Transport Bus 
were then transferred to a VME-based archiving system \cite{Car 91}.

Upon receipt of the TGO signal, the first module in the Transport Bus Carry
chain, the Event Generator Source (EGS), would raise the System Busy signal to
inhibit any further triggers and take control of the Carry signal. The EGS then
put a special ``first-word'' into the Transport bus to indicate the beginning of
a new event in the data stream. After a few more words from the EGS, the Carry
signal was passed to the first branch of the readout subsystem to begin 
transferring event data into the Transport.

Upon receipt of the TGO signal, the EGS module also fanned out ``START'' signals
to CRs and TDCs to begin digitizing the pulse signals. For each hit in the hodo or
prop tube the CR would generate one word in the event output, containing the 
scintillator ID or wire number of the muon proportional-tube hit. Each event
also contained a record of which trigger caused it to be readout via the Trigger
Bit Latch (TBL). In addition to the CR's, each TDC would begin incrementing a
Gray-code counter once every 4 ns upon receiving the START signal. The 
incrementing process would be stopped by the amplified signal from the drift
chamber. Each hit in the drift chamber would also produce a one-word output, 
containing the wire number of the hit and the Gray-code value of the TDC timer.
This measured the drift time. All the data were transferred into the Transport
bus in the Carry chain order.  

\subsection{Data Archiving System}
\indent
Events from the Transport bus streamed into the VME through a pair of 
``ping-ponging'' triple-ported VME high-speed memory boards by way of a 
front-panel ECL interface. Interrupt-driven software would initiate DMA transfers 
of packets of events from the high-speed memories across the VME bus into a 
128-megabyte ring buffer. This buffer was continuously being drained across the 
VME bus into a single-board computer by a concurrent task which performed all the
data formatting. From there, formatted packets of events were queued in a small
pool for distribution to the taping subsystem, where up to four Exabyte 8mm tape
drives would record the data.

Unlike in E789, the communications to the readout system and the run-control
capabilities were all built on the VME single-board computer in E866. In 
addition, the scaler data were injected into the data pipeline, from the CAMAC
system, as regular logical records on a spill-by-spill basis. These scalers 
included target, beam, magnet parameters and counts from varieties of trigger 
conditions. A small fraction of event packets were fanned to the UNIX workstations
for online data sampling and analysis. This provided the capability of online
monitoring.

For the beam-dump running, the average data-taking rate was about 20000 events
per spill. The average event size during the beam-dump run was 192 16-bit words.

\subsection{Data Monitoring System}
\indent
The E866 online database system was based upon the ADAMO library distributed
by CERN, with a graphical interface package called ``PinKy.'' The database for E866
recorded various data streams, including the beginning-of-run (BOR), end-of-spill 
(EOS) scalers, and beamline data (EPICURE). The online monitoring tools
included 1) ``runstatus,'' a graphical display of certain critical data 
(magnet settings, beam intensity, luminosity, live-time, and duty-factor 
calculations) updated at each EOS, 2) ``scan,'' a graphical display of scalers
refreshed at each EOS, 3) ``plot,'' a plotting tool for monitoring any entity 
stored in the database, 4) ``review,'' a tool for fetching data for series of
runs for plotting or exporting to the CERN Physics Analysis Workstation software
package (PAW) ntuple file, in which the interested quantities of an event were
stored in an array, and 5) ``dd,'' a tool to receive and distribute data to the 
backend. These advanced monitoring tools provided the capabilities for us to 
reconstruct and monitor a fraction of events online during the data taking. The 
shift taker could, for example, see the mass spectrum, hits and multiplicities on 
the drift chambers and hodoscope planes, and format errors due to transport 
readout problems while the data were being taken. Thus this capability helped us 
to diagnose the hardware problems and improved the quality of the data.

\chapter{MUON-TRACK RECONSTRUCTION}
\indent
The data recorded in 8mm tapes are logical records and have to be decoded before
using. The decoding was accomplished by the analysis code developed for E866.
The most important information stored in the data stream is actually
the space-time marks of the electronic signals traveling through the detector
stations. These marks in reality present the trajectories of the charged 
particles. From the trajectories the kinematic quantities of the particles can
then be determined if the mass of the charged particle and the strength of the
magnetic field is known.  

In this chapter we will first summarize the data taken in April of 1997 for this
study, and then describe in detail the method of track reconstruction in this 
experiment.

\section{Data-Set Summary}
\indent
During beam-dump data taking the typical beam intensity request was 6E10, 
$6*10^{10}$ protons per spill. The average triggering rate was about 20000 
triggers
per spill and the average event size was 192 16-bit words, as mentioned in the 
last chapter. A total of 82 magnetic tapes of raw data were used, with an average 
of 1.8 Gb of raw data written onto one tape. About 400M events in total were
recoded during the beam-dump run. 

One data set was distinguished from another by changing magnet-setting 
configurations or trigger-matrix configurations. There were four data sets in
the beam-dump data sample. The specifications are given in Table \ref{dataset}.

\begin{table}[tbp]
\caption[Magnet currents and trigger-matrix of different data sets]
{Magnet currents and trigger matrix files for different data sets of the
beam-dump data sample. The magnet currents are in Amperes. SM0 was off during the
beam-dump run.}
\label{dataset}
\begin{center}
\begin{tabular}{cccc}
\hline \hline
data sets  &  SM12 current &  SM3 current  &  trigger matrix\\
\hline 
12     &        $-$2800     &     $-$4230   &     trigmat.psidump, trigmat.psi2800\\
13     &        $-$2040     &     $-$4230   &     trigmat.psi2000\\
14     &        +2040     &     +4230   &     trigmat.psi2000\\
15     &        +2800     &     +4230   &     trigmat.psi2800\\
\hline \hline
\end{tabular}
\end{center}
\end{table}

The data were taken under two different SM12 settings and two polarities in order 
to reduce possible systematic errors. It was known that the incident beam was
not perfectly lined-up with the Z-axis of the spectrometer, so the data show
an up-down asymmetry in the event distribution with respect to the Y = 0 plane. The
flipping of the magnet polarity thus provided crucial information on the 
measurement of this asymmetry. The changing of SM12 current also changed the 
acceptance of the spectrometer. The consistency on the results obtained from the 
two magnet settings would provide a test, since the physics should be 
acceptance-independent. Each of the four data sets contained about equal amount of
data.

Data set 12 was further divided into data set 12a, 12b, and 12c, and data set
15 was divided into 15a and 15b, according to the incident beam angle and 
trigger matrix file. However the beam angle was only determined after the data 
were analyzed, so not till the later chapters will such division be used.

\section{Track Reconstruction}
\indent
In this section the methodology of track reconstruction applied in the E866 
analysis is described. For each event, the procedure can be considered as two 
main steps. The first step involved track finding, which was based on the 
drift-chamber hits and muon identification from the proportional-tube signals, and
track fitting, in which the possible candidates of track segments between Stations 
2 and 3 were found. In the second step a trace-back procedure was applied to the 
track candidates, so that the complete trajectories through the SM12 and SM0 to 
the target position were reconstructed.

\subsection{Identifying Drift Chamber Hits}
\indent
Each drift chamber station consisted of six planes: Y, U, V, and their
associated prime planes Y$^\prime$,U$^\prime$,V$^\prime$. When a charged particle 
traveled through the drift-chamber array in one station, correlated signals from 
different planes were produced. The subroutine DCTRIPS searched station 3 and 
station 2 for the correlated hit patterns. Only the patterns that consisted of at 
least 4 crossed hits whose cross-intersections were very close to a space point 
were registered. A triplet pattern was defined as having hits in all three views, 
while a doublet pattern was defined as having hits in two views only. An 
associated hit was defined as a particle that hits both the prime and unprimed planes.

\subsection{Fitting the Tracks}
\indent
Once all the valid hit patterns were registered, the next step was to link the 
registered hits from Station 3 and Station 2. The subroutine DCTRAX looped over
the triplets and doublets in Stations 2 and 3 to construct the track candidates,
called DC track segments. Several constrains were imposed on the track 
candidates: 1) if a doublet in one station was found, it was only allowed to 
connect to a triplet from another station; 2) at least 3 associated hits from 
Stations 2 and 3 were required to construct the track segment; 3) the segment was
extrapolated to Station 4 and was required to fire at least 3 out of 5 planes in
the desired location; 4) the segment was approximately pointing to the target 
location. For this study, very loose cuts were made to confine the segment 
vectors, and muon identification was done in DCTRAX.  

The next step was to link the track segments with the identified hits in Station
1. The subroutine WCTRAX required each of the track segments to be lined up with
a valid hit of Station 1 in the X-Z view (non-bend plane) within a vertical
band. Only hits within this Station-1 window were further considered. A single
bend-plane approximation was used to account for the SM3 momentum kick. Once a
valid hit was identified in station 1, the entire track was refit into two
straight-line segments joint at the bend plane. The fitting routine, FITTIME,
using all 18 planes of drift chambers to fit the track, and routine SM3 
calculated the momentum kick and the Z-coordinate of the bend plane. The result
of this final fit gave the coordinate of the track at the SM3 bend plane,
the Y-slopes before and after the bend plane, and the X-slope at the intersection
point. With the knowledge of the SM3 field map, together with the slope
information, the track momentum at Station 1 was determined.

\subsection{Tracing Back through SM12}

\subsubsection{Energy-Loss Correction}
\indent
From the SM3 bend plane to the target position, the track was reconstructed in a 
routine called PBSWIM. Given the field map of SM12, the coordinates and the 
momentum of the track were reconstructed in the field-map grid step by the 
routine TRACER. During the procedure of tracing back, the effect of energy loss in
the absorber wall and in the beam dump material was taken into account. The lost 
energy, calculated by an empirical formula 

\begin{equation}
E_{loss} = a + b*\log (P_{in}) + c*\log (P_{in})^{2}
\end{equation}

\noindent
for each layer of the absorber materials, was added back to the track after
TRACER had traced through that layer. The coefficients in the formula were
determined from dedicated Monte Carlo studies. The total energy loss in the 
beam dump was estimated in the same way. For each step inside the dump, a 
fraction of the total estimated energy loss, proportional to the step size, was 
added back to the track.

\subsubsection{Multiple Scattering}
\indent
Due to multiple scattering, it was impossible to trace back to the exact 
event-producing location. So it was assumed that all the events came from a point
located at one
interaction length into the dump. In this case Z$_{target}$ was set to 86 inches in the 
E866 coordinate frame. To correct for the effects of multiple scattering, a 
scattering bend-plane approximation was used. After the initial traceback, the 
intercepts of the track at Z$_{target}$ were compared with the beam centroid,

\begin{eqnarray}
dX = X_{target} - X_{centroid},\\
dY = Y_{target} - Y_{centroid}.
\end{eqnarray}
\noindent
Based on these differences, an angular correction to the track direction at the
scattering bend plane (located at Z$_{scatter}$) was calculated:

\begin{eqnarray}
d\theta_{x} = dX/(Z_{target} - Z_{scatter}),\\
d\theta_{y} = dY/(Z_{target} - Z_{scatter}).
\end{eqnarray}
\noindent
After the angular correction was applied, the track was traced again to 
$Z_{target}$ starting from $Z_{scatter}$. The iteration procedure was repeated 
until the intercept errors became negligible. The value of $Z_{scatter}$ was 
determined by optimizing the angular resolution at the target point.

\subsubsection{Additional Angle Corrections}
\indent
Further Monte Carlo study had shown that the single scattering bend-plane 
approximation actually over-calculated the reconstructed angle, as shown
in Figure \ref{Fig:angcorr}. 

\begin{figure}
\begin{center}
\mbox{\epsfxsize=5.7in\epsfysize=6.5in\epsffile{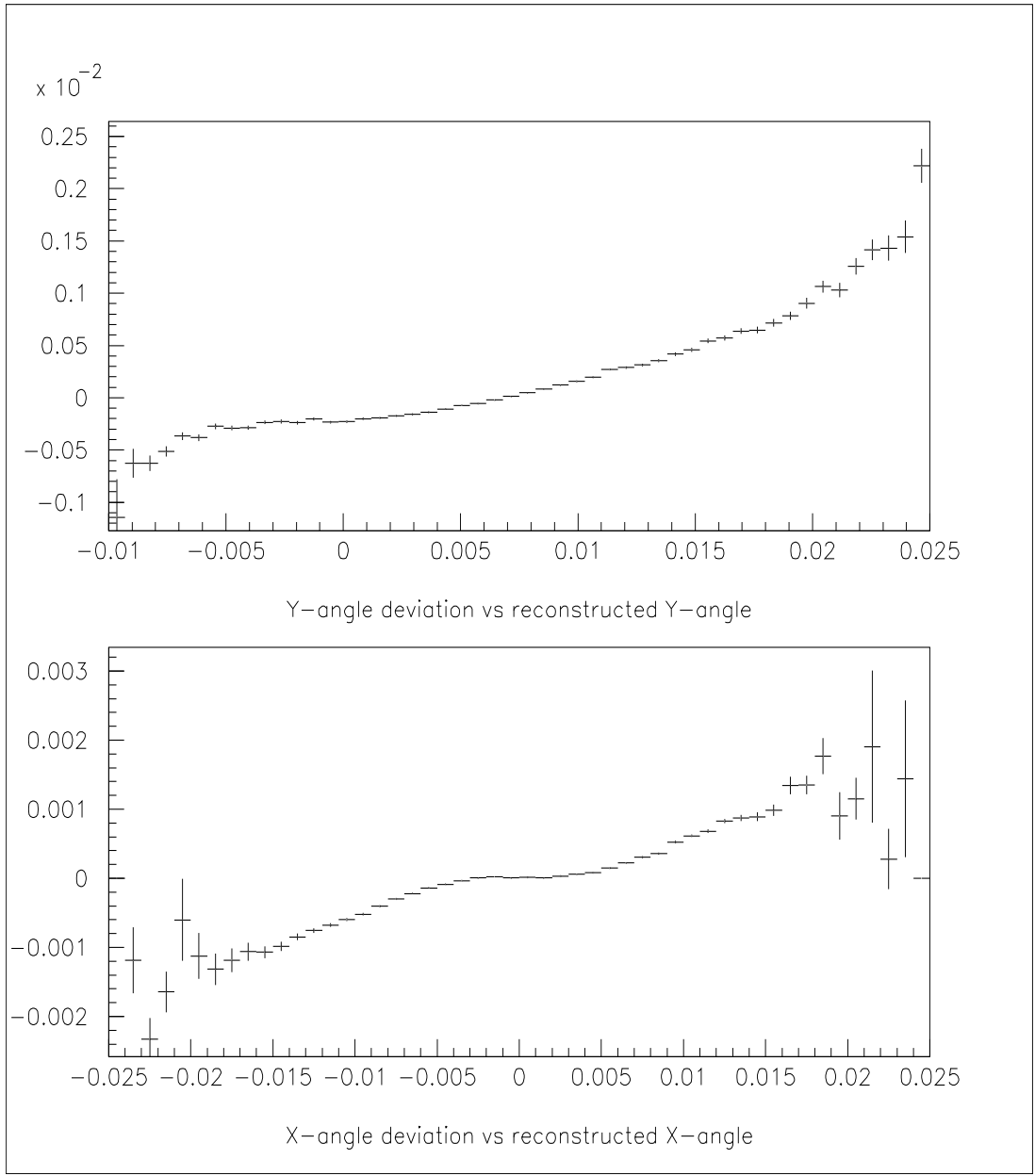}}
\end{center}
\vspace*{+0.25in}
\caption[Angle corrections from single scattering-plane approximation]
{The single scattering-plane approximation fails at large angles. The 
reconstructed angles are greater than the thrown angles by about 10$\%$ at 
large angles. The figures shown here are taken from the up-going muon tracks.} 
\label{Fig:angcorr}
\end{figure}

%-- Fig 3.1  A angle correlation plot, x and y angle before correction --

In order to reconstruct the opening angle correctly, an empirical formula, the
angle deviation expressed as a polynomial function of the reconstructed angle, was
used to adjust the angle that came out of the initial scattering-plane 
approximation.
The formula was purely empirical and relied completely on Monte Carlo 
studies, so it was important to test whether these corrections gave back the 
thrown angular distribution for Monte Carlo events after the events were analyzed,
even though this
self-consistency check is only necessary but not sufficient. 
Figure \ref{Fig:reconth} shows the reconstructed $\cos \theta$ distributions 
for both magnet settings. Those plots were obtained from the 
thrown-reconstructed events divided by the unsmeared acceptances. The function
$p1 \times (1 + p2*\cos ^{2}\theta )$ was used to fit the plots to test whether 
there is any systematic bias while reconstructing the $\cos \theta$ distributions.
As a result, a nearly flat distribution of $\cos \theta$ 
($\lambda = -0.02 \pm 0.017$) is recovered for the 2040Amp data set, and 
$\lambda = 0.02 \pm 0.018$ for the 2800Amp data set.

\begin{figure}
\begin{center}
\mbox{\epsfxsize=5.7in\epsfysize=6.5in\epsffile{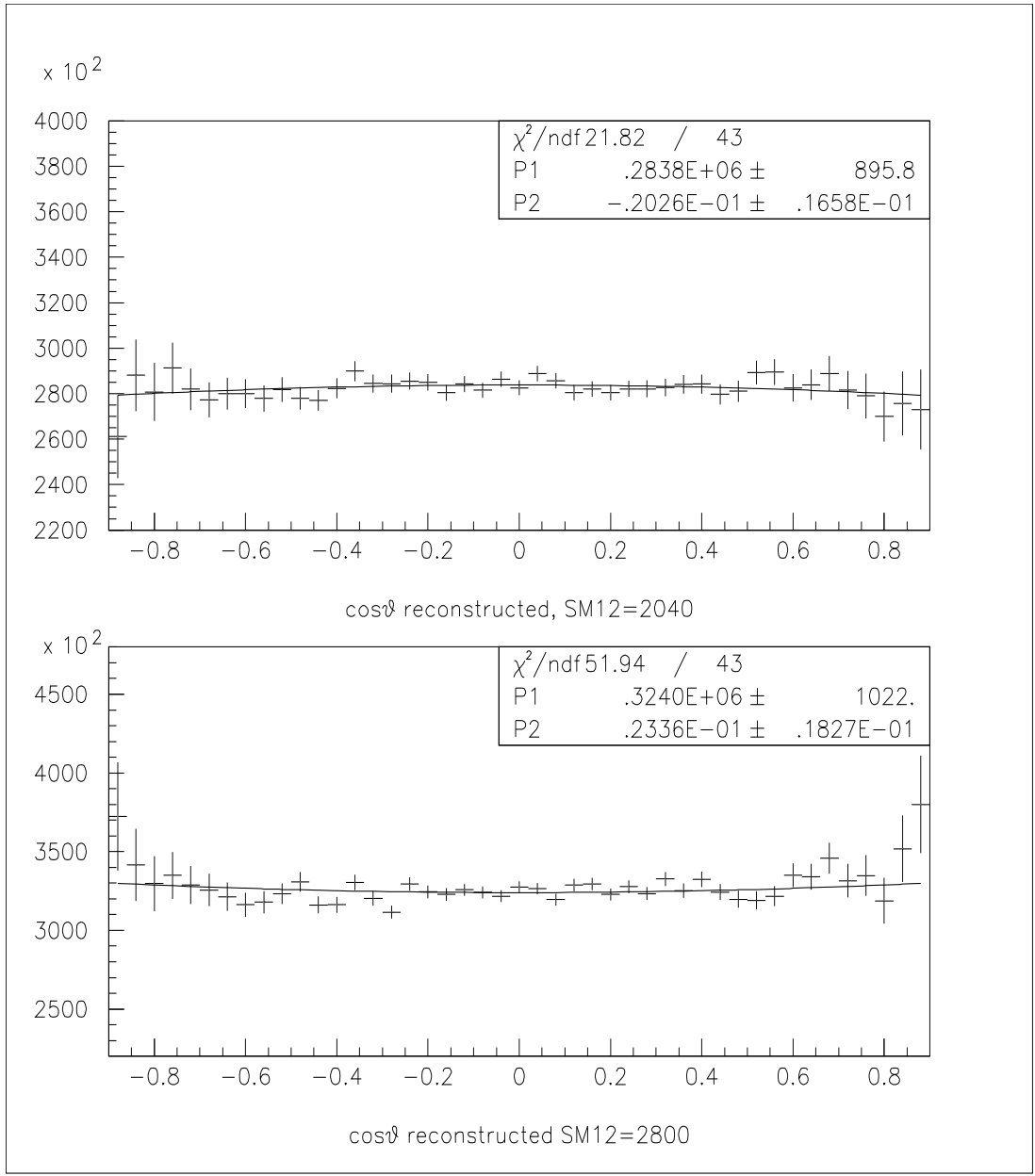}}
\end{center}
\vspace*{+0.25in}
\caption[Reconstructed $\cos \theta$ distributions for both magnet settings]
{Reconstructed $\cos \theta$ distributions for both magnet settings.
The reconstructed distributions recover the thrown (flat) distributions.} 
\label{Fig:reconth}
\end{figure}

It was also important to test with Drell-Yan data, where we believe we 
know the angular distribution, to search for additional systematic problems not
revealed from Monte Carlo studies alone. The angular distribution study of 
Drell-Yan events as a confidence check will be presented in Chapter 5.

\chapter{DATA ANALYSIS}

From the raw-data tapes to the final physics results, several types of 
computing hardware and software were used. The whole procedure can be 
divided into three passes in the analysis, which will be described in this 
chapter.

\section{Pass-1: Fermilab IBM Farm}

The first pass of data analysis was performed on the Fermilab parallel computing
farms. The computing farm system is a cluster of IBM workstations which can
distribute the raw data into all the computing nodes and analyze the data 
simultaneously. The events passing the first-pass analysis were then written on 
to Data Summary Tapes (DSTs) for the second-pass analysis.

The tasks of the first-pass analysis were mainly to find dimuon events 
originating in the beam dump within the desired mass range. An 18-inch grid SM12 
magnet map was used to trace the tracks in this pass, and very loose aperture 
and target cuts were applied. However it required the events to have 
two muon tracks and the mass of the muon pair has to be greater than 2 GeV. As
a consequence only about 5$\%$ of the events passed the cuts and were written onto
DSTs.
 
\section{Pass-2: Hewlett-Packard Workstation}

The second pass of the data analysis was performed on the Hewlett-Packard
Workstation located in New Mexico State University. The inputs of this phase
were DSTs, and the outputs were the ntuple files. The main task of this phase of
analysis was to reconstruct the kinematics of the DST events as accurately as
possible. A 2-inch grid SM12 magnet map as well as a Y-field map were used
to reconstruct the events instead of the 18-inch map. No other tighter cuts 
were applied, but many fine-tuning tasks were done in this pass of analysis.

\subsection{Determining Tweeks}

The ``tweek'' is an overall correction factor for the magnetic field strength
provided by the field map. The field maps provided by the ANL (Argonne National
Lab) group assumed that the magnets were operated at the preset currents,
SM3 at 4260 Amperes and SM12 at 2800 Amperes, for example. But in reality the
operating currents were not precisely equal to the desired currents, and there
were uncertainties in the mapping, 
therefore it was necessary to apply the corrections to the magnet maps for
analyzing the events or to generate Monte Carlo events. 

Since the actual currents were not known, this whole subject relied on careful
Monte Carlo studies. There were two unknown quantities to be determined: the tweek
of SM12 and the tweek of SM3. The two conditions used to determine these two 
quantities were 1) reconstructed $J/\psi$ mass and 2) the uniterated Z-vertex
(ZUNIN). By adjusting the magnet map in the MC event-generating phase, it was 
required that the reconstructed experimental data have the same mass 
and ZUNIN location as the MC reconstructed events; the tweeks used in the 
data event reconstruction were the same as in the MC event generation and 
reconstruction. Figures \ref{Fig:zunincomp} and \ref{Fig:masscomp} show the 
ZUNIN and mass peaks of the real events and MC events from both magnet 
settings. The shape of the ZUNIN peaks is not symmetric around its central value
because of energy loss and multiple scattering of the muon tracks. The peaks
were fitted to a Gaussian using asymmetric boundaries, $-$10 inches to 33 inches,
in order to locate the peak centroids without being affected by the non-Gaussian
tails. The mass peaks of the data distributions were fitted
to a second-order polynomial plus a Gaussian function, while the mass 
distributions from Monte Carlo were fitted to Gaussians since the Monte Carlo did
not include any background events. As one can see from Figures
\ref{Fig:zunincomp} and \ref{Fig:masscomp}, the
agreement between Monte Carlo and the experimental data is satisfactory. 
The tweek values are given in Table \ref{tweeks}.

\begin{table}[tbp]
\caption{The tweek values of SM12 and SM3 for all the data sets.}
\label{tweeks}
\begin{center}
\begin{tabular}{ccccc}
\hline \hline  

Data Set &  SM12(Amp) &  SM3(Amp)  & TWEEK12  & TWEEK3\\
\hline
12   &      $-$2800    &    $-$4260  &    1.006  &  0.986\\
13   &      $-$2040    &    $-$4260  &    1.019  &  1.002\\
14   &      +2040    &    +4260  &    1.019  &  1.002\\
15   &      +2800    &    +4260  &    1.006  &  0.986\\
\hline \hline
\end{tabular}
\end{center}
\end{table}

\begin{figure}
\begin{center}
\mbox{\epsfxsize=5.7in\epsfysize=6.5in\epsffile{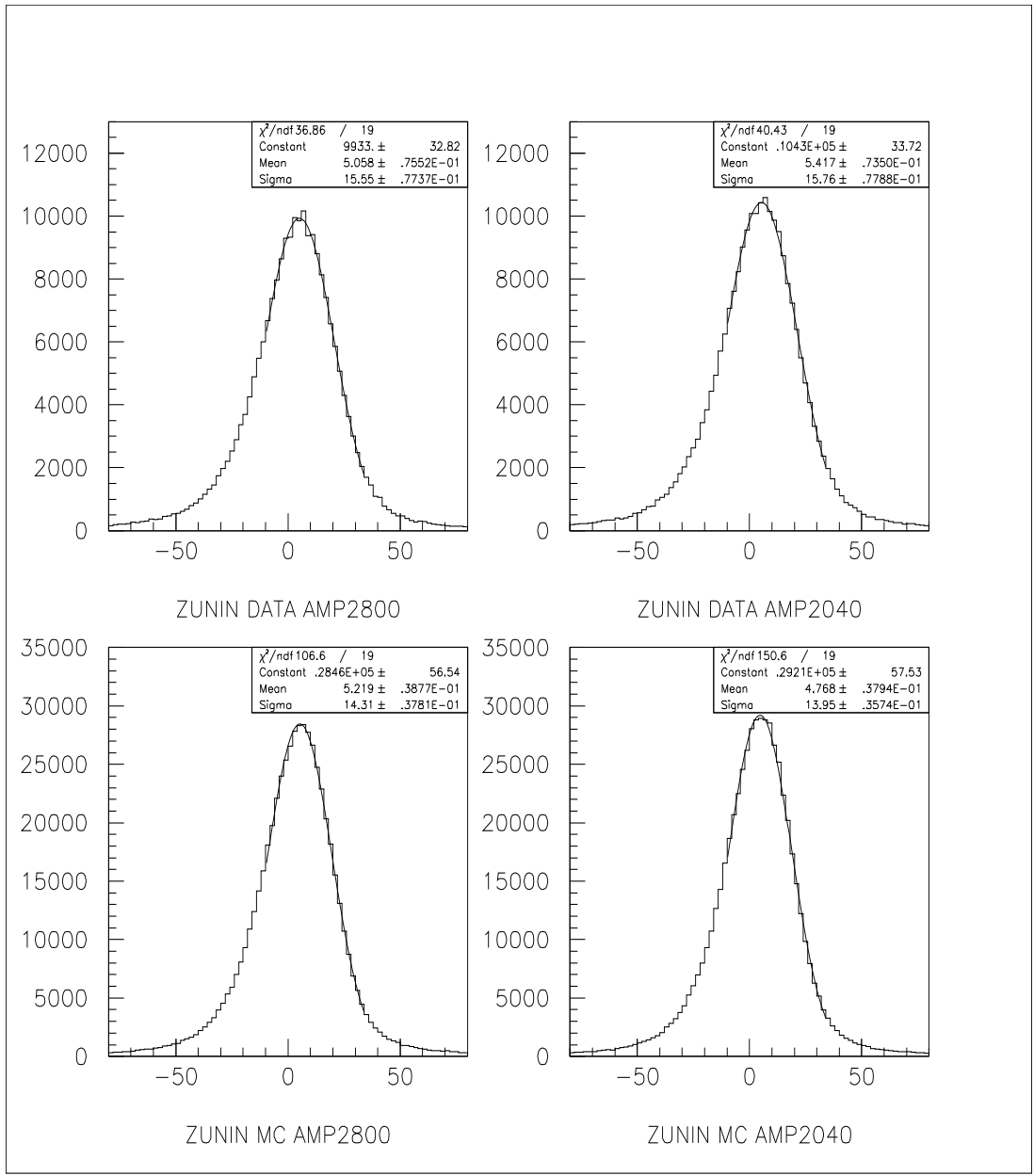}}
\end{center}
\vspace*{+0.25in}
\caption{ZUNIN peaks of Data and MC from both magnet settings.} 
\label{Fig:zunincomp}
\end{figure}
\begin{figure}
\begin{center}
\mbox{\epsfxsize=5.7in\epsfysize=6.5in\epsffile{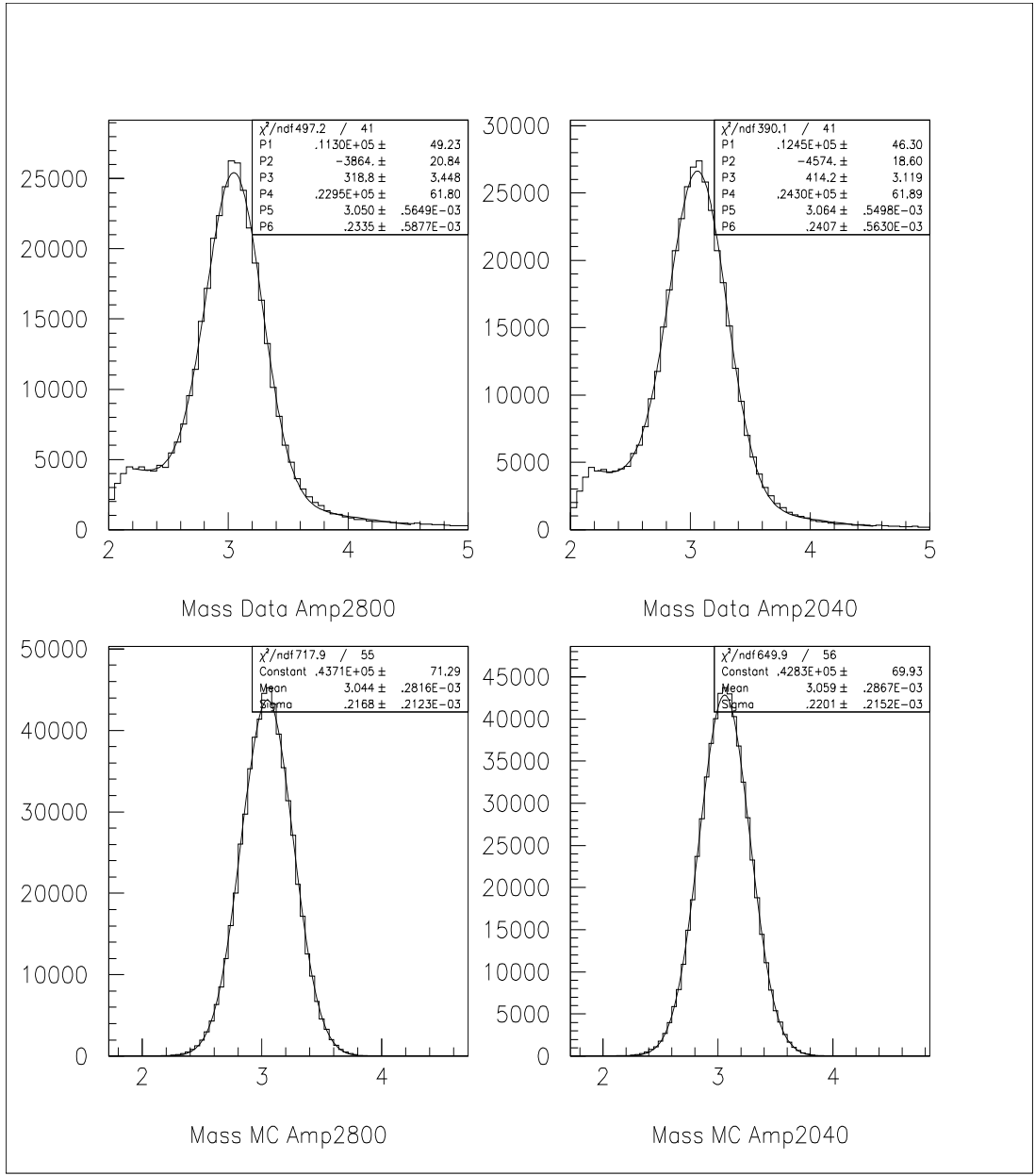}}
\end{center}
\vspace*{+0.25in}
\caption{Mass peaks of Data and MC from both magnet settings.} 
\label{Fig:masscomp}
\end{figure}

%-- Fig 4.1 and 4.2 Zunin and Mass peak of Data and MC from both magnet 
%   setting --

\subsection{Determining Beam Positions}

During the course of analyzing the raw data, it was found that the beam 
centroid was not steadily fixed at a single location over the entire period of 
running. The moving range of the centroid is greater than three sigma
of the beam profile, which is 0.14 inches in X and 0.07 inches in Y at the target
position, so this was due to the beam-line magnet-current fluctuations. A 
typical reconstructed beam centroid distribution during one run is shown in 
Figure \ref{Fig:beamcenter}. 

\begin{figure}
\begin{center}
\mbox{\epsfxsize=5.7in\epsfysize=6.5in\epsffile{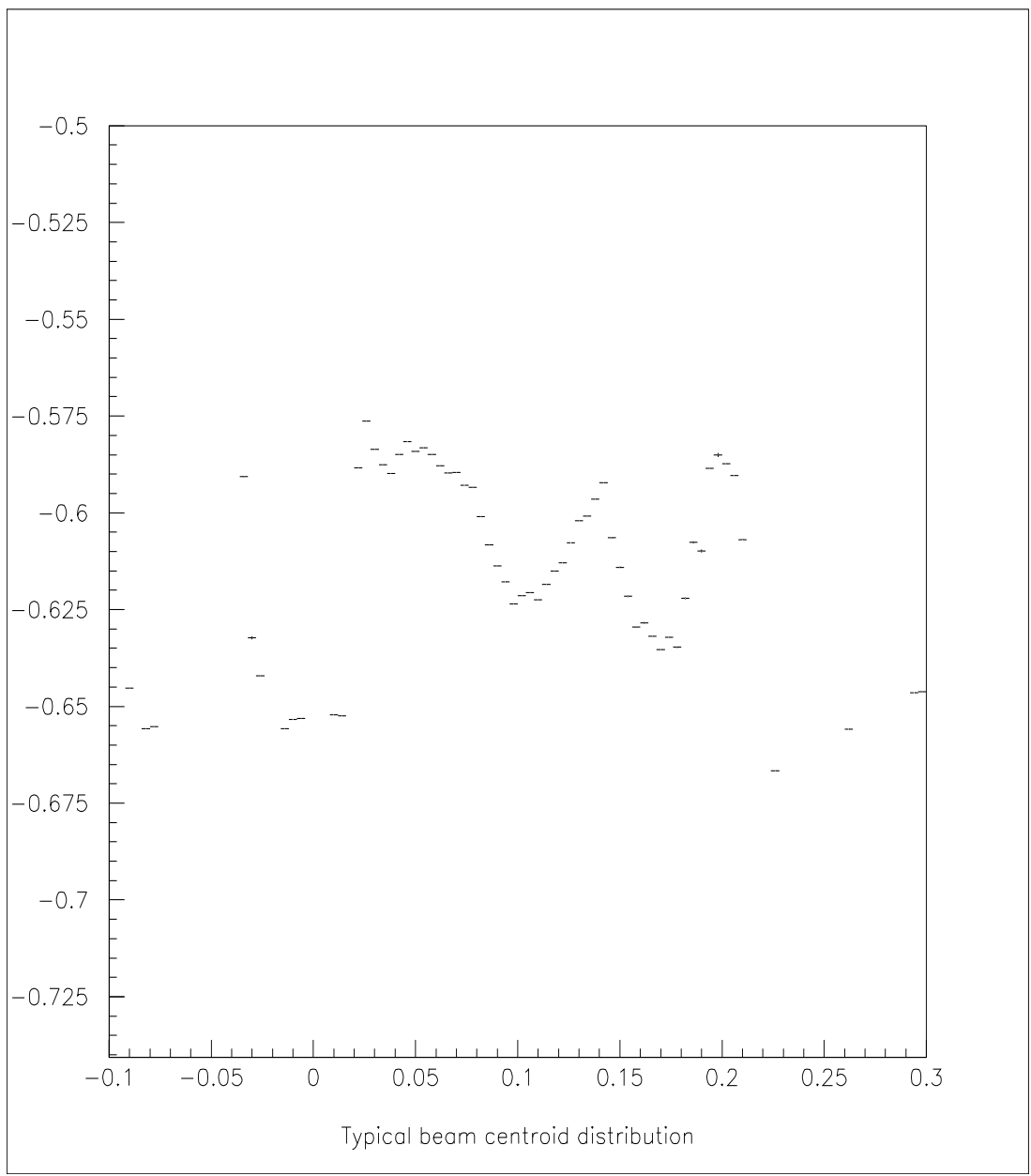}}
\end{center}
\vspace*{+0.25in}
\caption[Beam-centroids positions during data taking] 
{Beam centroids were moving during the data taking. The vertical scale
gives the Y-position and the horizontal scale gives the X-position in inches
in the lab frame.} 
\label{Fig:beamcenter}
\end{figure}

%-- Fig 4.2  a plot of beam centroid distribution --

Since the angular distribution was affected by the beam angle
(a slight offset of the beam angle will generate a linear term of 
$\cos \theta$ in the final $\lambda$ extraction) and a single fixed target point 
in the analysis was not able to account for the angle fluctuations due to the 
beam-line movement, the beam centroids at the target were determined in a 
spill-by-spill basis by fitting the beam profile of the raw data at the target 
position. In the first second-pass analysis, the beam centroid of each 
event in each spill was determined by averaging the X and Y coordinates of the 
two muon tracks at the Z = 86 inches plane and then saved the centroid distributions
(one for X and one for Y) in a temporary histogram. Then these distributions were 
fitted using Gaussian functions to determine the central values. These central 
values were tagged with the run number and the spill number and were saved in a 
2-D lookup table. This procedure was applied to every spill of the raw data. If 
there were not enough events (the threshold number of events to perform the 
fitting was set to 50) in a spill to perform the fit, the centroid value was taken
from the previous spill. If the first spill in a run did not have enough 
statistics, the centroid value was set to some default value depending on the data
set. Then a second second-pass analysis was performed, using the centroid values 
stored in the lookup table, for each spill of data. The beam angles 
reconstructed in this procedure, on a run-by-run basis, were approximately constant
within a data set. Some runs within a data set had very different angles and had 
to be treated separately. Data set 12 was broken into 12a, 12b, and 12c, and data 
set 15 was broken into 15a and 15b. The reconstructed beam angle for each data 
set is shown in Table \ref{beamang}.

\begin{table}[tbp]
\caption{Beam angles of each data set.}
\label{beamang}
\begin{center}
\begin{tabular}{ccc}
\hline \hline

Data set   &   X angle(1E-3)  &   Y angle(1E-3)\\
\hline
12a    &       0.12      &       $-$1.09\\
12b    &       0.11      &       $-$1.05\\
12c    &       0.11      &       $-$0.94\\
13     &       0.00      &       $-$0.76\\
14     &      $-$0.02      &        0.11\\
15a    &       0.10      &        0.12\\
15b    &       0.05      &        0.31\\
\hline \hline
\end{tabular}
\end{center}
\end{table}

\subsection{Determining Beam Angle}

The determination of the beam angle for each data set involved two phases. In
the first step we obtained the initial value of the angle recovered by plotting
the momentum vector from the ntuple. Note that the beam-centroid-fitting procedure 
described in the
previous section had to be applied first. Then the second-pass analysis was
done again using the initial angle. It was found that after one iteration the
reconstructed angles converged within 0.03 mrad. Thus a second
iteration was not necessary. Figure \ref{Fig:anglerecon} shows the beam angle
reconstructed from the experimental data set 15b as an example.

\begin{figure}
\begin{center}
\mbox{\epsfxsize=5.7in\epsfysize=6.5in\epsffile{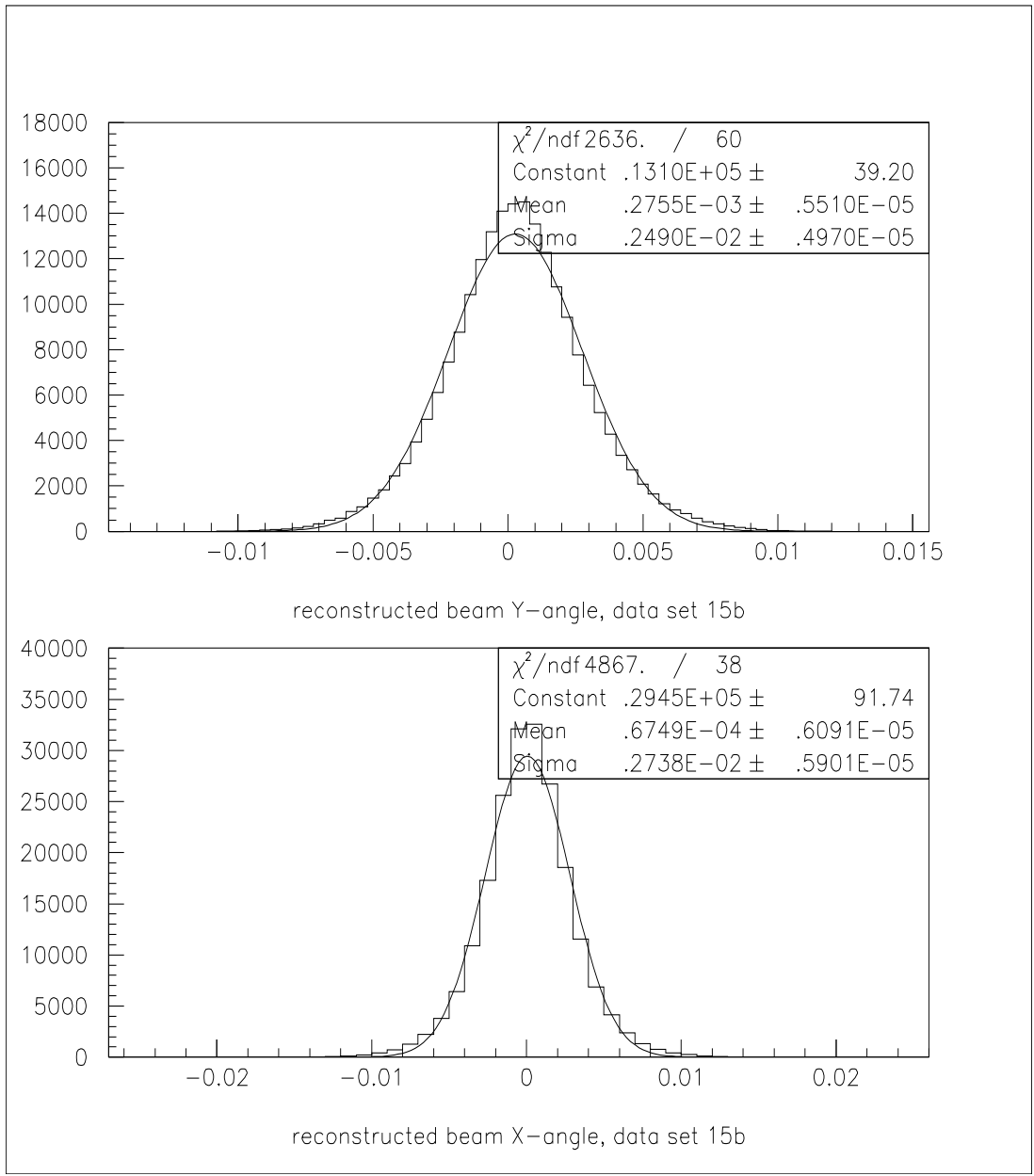}}
\end{center}
\vspace*{+0.25in}
\caption[Beam angle reconstruction plots]
{Angle reconstruction plots for data set 15b as an example. The peaks
are fitted to a Gaussian function to determine the central values.} 
\label{Fig:anglerecon}
\end{figure}

%-- Fig. 4.3  Data angle reconstruction plot for all data sets ---

The second step in determining the beam angle was to look at the production-$\phi$
distribution. The production-$\phi$ (PPHI) distribution was very sensitive to 
the input beam angle. Monte Carlo events were generated to compare with the PPHI
plots of the real data. The $\phi$ distribution of the real data is expected to be
isotropical because both the beam and target were not polarized, and in the Monte 
Carlo the PPHI distribution was thrown isotropically. It was required that the 
ratio of the PPHI plots be flat, so there was no $\phi$-term contribution in 
extracting the angular distribution.  Usually small changes in the beam angle had 
to be added to the Monte Carlo in the generation phase to obtain good PPHI 
agreement. However those changes were small compared to the beam angle variation of
0.3 mrad, so it was not necessary to repeat the second-pass analysis. The ratio of
real-data PPHI over MC data PPHI is given in Figure \ref{Fig:pphicomp}. The plot 
was fitted to $p1 \times \sin (p2 + \theta ) + p3$. The amplitude shows that the 
uncertainty of the beam angle values used in Monte Carlo is about 2$\%$.

\begin{figure}
\begin{center}
\mbox{\epsfxsize=5.7in\epsfysize=6.5in\epsffile{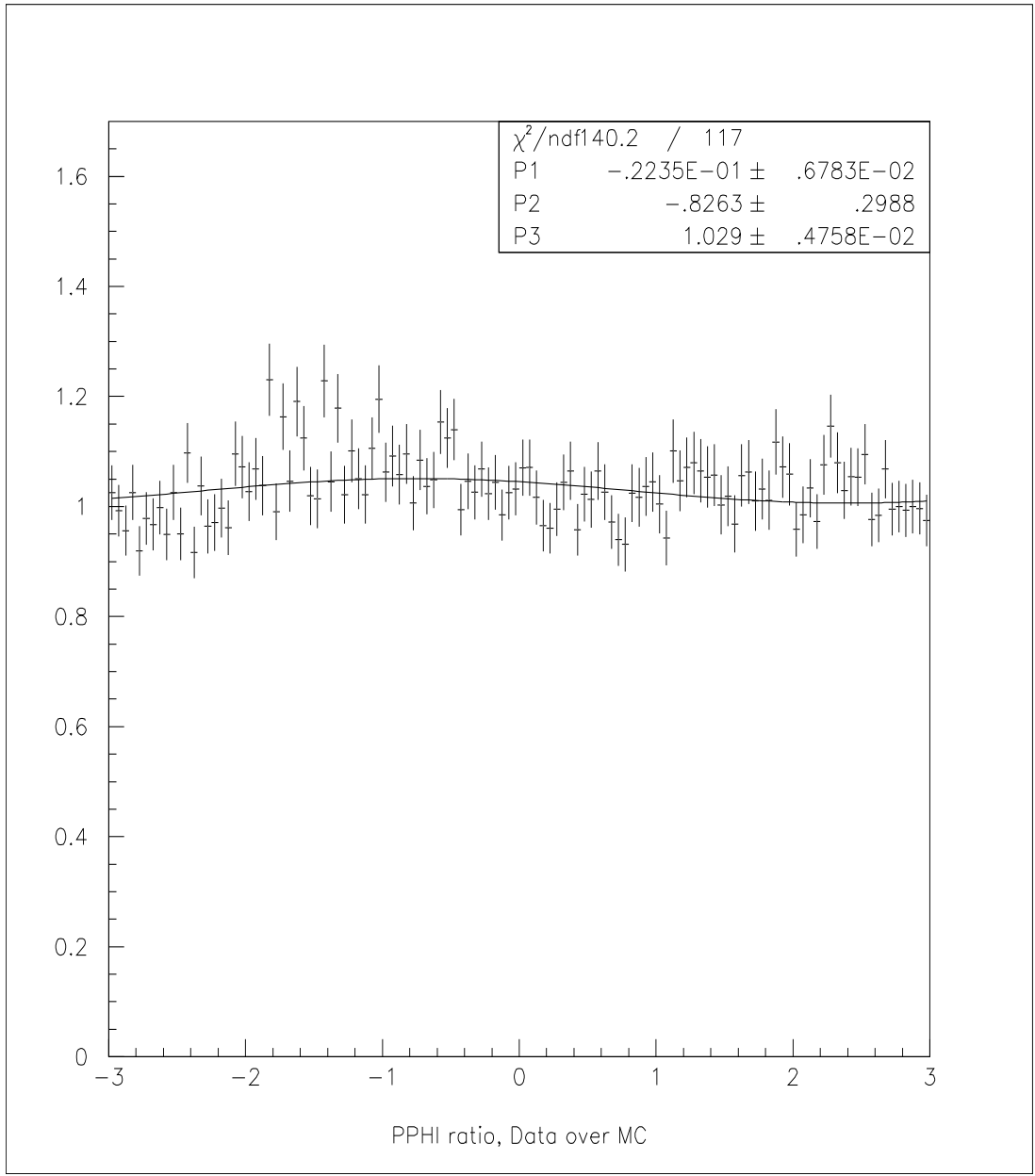}}
\end{center}
\vspace*{+0.25in}
\caption[Production-$\phi$ distribution ratio]
{Production-$\phi$ distribution: experimental data divided by Monte
Carlo. The plot is fitted to a sine function. The result shows a 2$\%$ uncertainty
in the incident angle.} 
\label{Fig:pphicomp}
\end{figure}

%-- Fig 4.4  PPHI plots: Data/MC ---

\section{Pass-3: PAW}

The third and final pass of the analysis was performed with the Physics Analysis
Workstation (PAW) program. This program provided a way to present a graphical 
output of the analysis. The input of this program can either be an existing 
histogram file, which is simply a vector, or an ntuple file, which consists of 
columns
of vectors. When filtering the final candidates, some additional 
cuts were applied to the events. Target vertex cuts and trigger-bit cuts were
imposed. Events originating outside a 
2$^{\prime\prime}$(X)$\times$2$^{\prime\prime}$(Y)$\times$200$^{\prime\prime}$(Z) range 
centered at the nominal target position were discared. The trigger-bit 
information TBRAW$\_$CK of each event was checked to assure that it satisfied 
the
physics-trigger requirements (satisfies Physics Trigger A 1, 2 or 4, see section
2.5.2). Like-sign pairs were discarded by the trigger cut at this point. As a 
result, about 94$\%$ of events survived the vertex cut and 91$\%$ of events 
survived both vertex and trigger cuts.      
 
Those events left were then distributed according to their $p_T$ and $x_F$
values. The $x_F$ range in this study was from 0.25 to 1.0 binned in intervals of 
0.1; $p_T$ was binned in intervals of 1 GeV. The last bin of $x_F$ combined all 
the data
above 0.85, and the last bin of $p_T$ included the data having $p_T > 3.0$ GeV.
The mass spectrum, from 2.0 GeV to 7.0 GeV plotted in 50-MeV bins, for each $x_F$
and $p_T$ bin was then fitted to a Gaussian plus some background function. 
The $J/\psi$ peak was described by the Gaussian function, but the 
background shape varied as the kinematic range changed. Listed below are the 
functional forms used to do the background fitting in this study:

\begin{eqnarray}
f(x) &=& exp(p1 + p2*x),\\
f(x) &=& p1 + p2*x + p3*x^{2},\\
f(x) &=& p1/(1 + (x/p2)^{p3}).
\end{eqnarray}

The uncertainties caused by the background function forms are discussed in 
section 4.6. The counts of $J/\psi$'s in each bin of $x_F$ and $p_T$ were then
determined by the formula

\begin{equation}
COUNTS = N*(bin ~ width/\sqrt{2\pi}\sigma)*exp(-(x-centroid)^{2}/2\sigma^{2})
\end{equation}
\noindent
where N, centroid, and $\sigma$ are free parameters to fit. The value 
$N + \Delta N$
returned from the PAW fitting program provided the population of $J/\psi$'s and
the statistical uncertainty in that bin. Tables \ref{NJ12} to \ref{NJ15} give 
the approximate number of $J/\psi$s in each bin for each data set.

\begin{table}[tbp]
\caption{Number of $J/\psi$'s in each bin of data set 12.}
\label{NJ12}
\begin{center}
\begin{tabular}{ccccc}
\hline \hline 
$x_F$ / $p_T$(GeV) &  0 - 1 &  1 - 2  & 2 - 3  &  $> 3$\\
\hline 
0.3      &      322000 & 292000 &  78800 &  16600\\
0.4      &      531000 & 497000 & 137000 &  28000\\
0.5      &      323000 & 310000 &  84800 &  16900\\
0.6      &      142000 & 139000 &  36500 &   6900\\
0.7      &       43100 &  42000 &  10700 &   1900\\
0.8      &        8800 &   7800 &   2200 &    300\\
$>0.85$  &         900 &    900 &     -  &     -\\
\hline \hline
\end{tabular}
\end{center}
\vspace*{+1.0in}
\caption{Number of $J/\psi$'s in each bin of data set 13.}
\label{NJ13}
\begin{center}
\begin{tabular}{ccccc}
\hline \hline 
$x_F$ / $p_T$(GeV) &  0 - 1 &  1 - 2  & 2 - 3  &  $> 3$\\
\hline 
0.3      &      363000 & 336000 &  91800 &  17800\\
0.4      &      244000 & 232000 &  66400 &  14000\\
0.5      &      119000 & 116000 &  32400 &   7100\\
0.6      &       47500 &  45900 &  13100 &   2600\\
0.7      &       13400 &  13100 &   3600 &    700\\
0.8      &        2500 &   2800 &    700 &    100\\
$>0.85$  &         300 &    300 &     -  &     -\\
\hline \hline
\end{tabular}
\end{center}
\end{table}
\begin{table}[tbp]
\caption{Number of $J/\psi$'s in each bin of data set 14.}
\label{NJ14}
\begin{center}
\begin{tabular}{ccccc}
\hline \hline 
$x_F$ / $p_T$(GeV) &  0 - 1 &  1 - 2  & 2 - 3  &  $> 3$\\
\hline 
0.3      &      372000 & 355000 &  99300 &  19100\\
0.4      &      244000 & 243000 &  71000 &  15200\\
0.5      &      120000 & 120000 &  34200 &   7300\\
0.6      &       47600 &  47300 &  13200 &   2800\\
0.7      &       13300 &  13600 &   3800 &    700\\
0.8      &        2600 &   2700 &    800 &    100\\
$>0.85$  &         400 &    300 &     -  &     -\\
\hline \hline
\end{tabular}
\end{center}
\vspace*{1.0in}
\caption{Number of $J/\psi$'s in each bin of data set 15.}
\label{NJ15}
\begin{center}
\begin{tabular}{ccccc}
\hline \hline 
$x_F$ / $p_T$(GeV) &  0 - 1 &  1 - 2  & 2 - 3  &  $> 3$\\
\hline 
0.3      &      260000 & 235000 &  62300 &  13200\\
0.4      &      413000 & 387000 & 107000 &  22800\\
0.5      &      250000 & 240000 &  65200 &  13400\\
0.6      &      109000 & 106000 &  27700 &   5400\\
0.7      &       33100 &  32100 &   8400 &   1500\\
0.8      &        6400 &   6300 &   1500 &    300\\
$>0.85$  &         700 &    600 &     -  &     -\\
\hline \hline
\end{tabular}
\end{center}
\end{table}

Figure \ref{Fig:rawsptrm} shows the reconstructed spectra of some kinematic
variables. The mass spectrum, on the upper-left, presents all the dimuon pairs
recorded during the beam-dump run with masses up to 7.0 GeV. The other three 
variables,
namely $x_F$, $p_T$, and $\cos \theta$, are plotted for the events that
satisfied the vertex cuts and the trigger cuts and have a mass in the range 
between
2.5 GeV and 4.0 GeV. The purpose of the mass cut was to reduce the 
contributions from non-$J/\psi$ events in these variables. One can see from
the figure that this data sample contains a large collection of $J/\psi$'s that
extend over a wide kinematic range.

\begin{figure}
\begin{center}
\mbox{\epsfxsize=5.7in\epsfysize=6.5in\epsffile{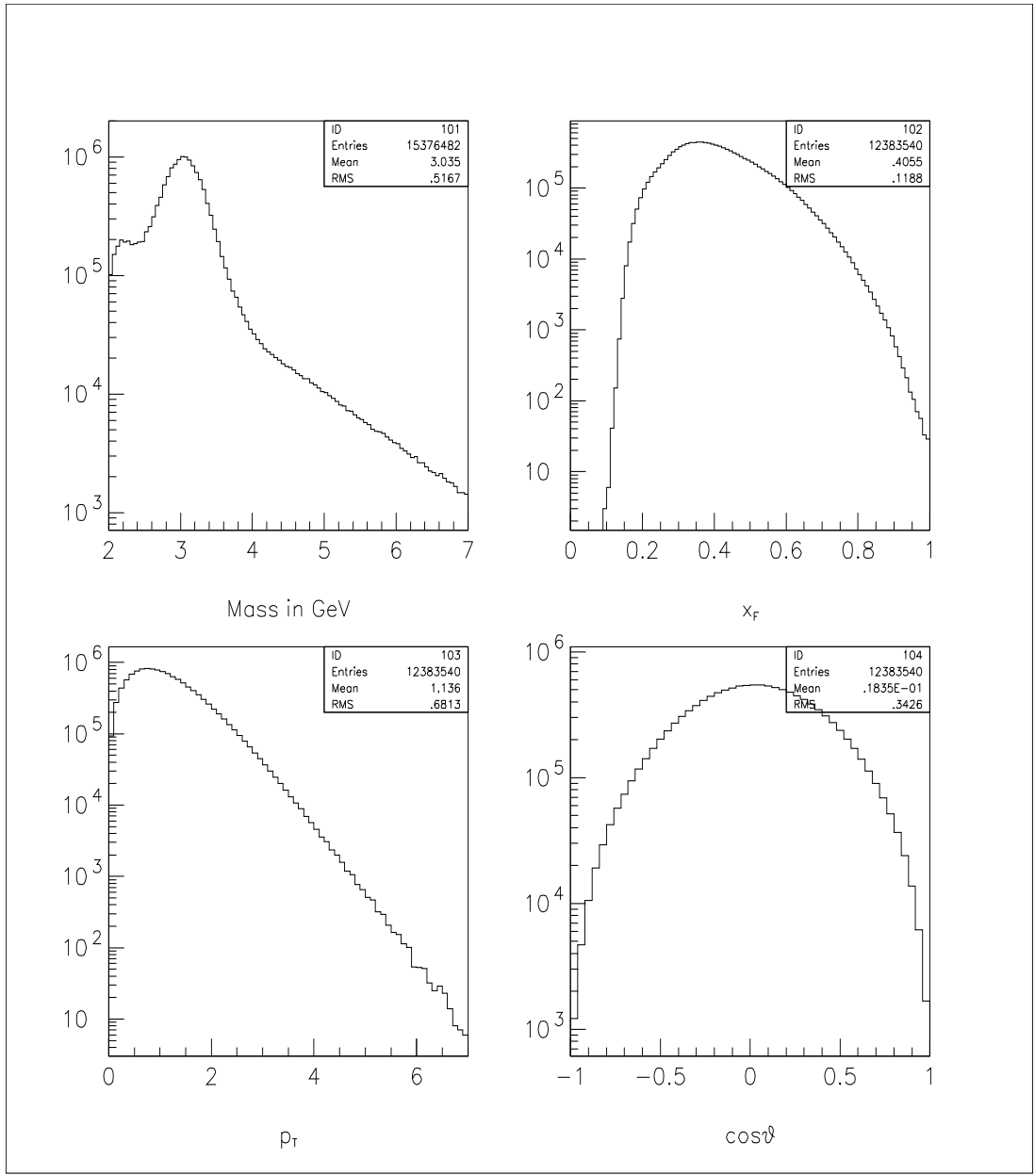}}
\end{center}
\vspace*{+0.25in}
\caption[Reconstructed spectra of some kinematic variables]
{Reconstructed spectra of some kinematic variables. Mass (2.5-4.0
GeV), vertex, and trigger cuts have been applied to the $x_F$, $p_T$ and
$\cos \theta$ plots.} 
\label{Fig:rawsptrm}
\end{figure}

%-- Figure 4.5 mass, xf, pt, cos(theta) plots ---

\section{Monte Carlo}

To extract $J/\psi$ angular distributions correctly we rely on good knowledge of
the angular acceptance. Monte Carlo simulation is the standard technique to
obtain the acceptance. In the Monte Carlo study the experimental apparatus 
setup was programmed as close as possible to the real experiment, and all the 
apparatus input parameters, some of which were physically measurable, were
tuned according to the best of our knowledge. However the physics part of the Monte
Carlo generation, which is the part of real interest, was unknown and relied 
completely on theoretical model calculations. The output of the Monte Carlo was 
then compared with the experimental data, assuming that the simulation of the 
apparatus part was reliable and trustworthy. The difference between the Monte 
Carlo results and the experimental measurements was then used to improve the 
various thrown physics kinematic quantities. The whole procedure was an iterative 
process since the acceptance depended on the thrown distributions, and the thrown 
distributions, usually taken from the experimental data distributions, relied on 
the knowledge of the acceptance. The accepted Monte Carlo events were stored in 
the same format as the raw data, and then were analyzed as the experimental data.
The final output was stored in the form of ntuple files, like the real data.
The acceptance correction of a physical variable to be applied to the experimental
data was given by the reconstructed MC distribution divided by the thrown 
distribution. 

In this section we will describe the thrown functions of the physical variables 
and compare them with the experimental data.

\subsection{$x_F$ Distribution}

$x_F$ can be understood as the $J/\psi$ longitudinal momentum $P_L$ divided by its
maximum kinematically allowed value $P_{L,Max}$, which is approximately equal to
half of the square root of the center-of-mass energy S, in the beam-target 
center-of-mass frame. Theoretically the $x_F$
differential cross section is of interest because it can be calculated based on
the knowledge of the parton distributions and some phenomenological models. In
this study we used an empirical formula for the thrown $x_F$ distribution
for $J/\psi$ within the range $0.25< x_F <1.0$:

\begin{eqnarray}
&x_{F}& \equiv ~ \frac{P_{L}}{P_{L,Max}} \approx \frac{P_{L}}{\sqrt{S}/2}~,\\
&d\sigma/dx_{F}&=~ P3(1 - 0.82x_{F})^{8.7}~,\\
&P3&=~ 2.784 - 10.14x_{F} + 17.81x_{F}^{2} - 9.585x_{F}^{3}~.
\end{eqnarray}
\noindent
The third-order polynomial was used to describe better the high-$x_F$
part. The comparison of the Monte Carlo and the real data is shown in 
Figure \ref{Fig:xfratio}. At each bin of $x_F$ the counts of $J/\psi$'s for the
real data were obtained by fitting the mass spectrum to a Gaussian plus a
exponential background. The agreement is good to about 3$\%$ for 
$x_{F}< 0.7$, which includes 98.2$\%$ of all the data. The largest
discrepancy, however, about 20$\% $, for $x_{F} > 0.7$ comes from the thrown 
shape in the Monte Carlo and the background uncertainty in the data. This has
no effect on our results of $d\sigma /d\cos\theta dx_{F}$ and
$d\sigma /d\cos \theta dx_{F}dp_{T}$ since these are differential quantities.
The effects on $d\sigma /d\cos \theta$ are also expected to be very small
because only a few events are at that high $x_F$.

\begin{figure}
\begin{center}
\mbox{\epsfxsize=5.7in\epsfysize=6.5in\epsffile{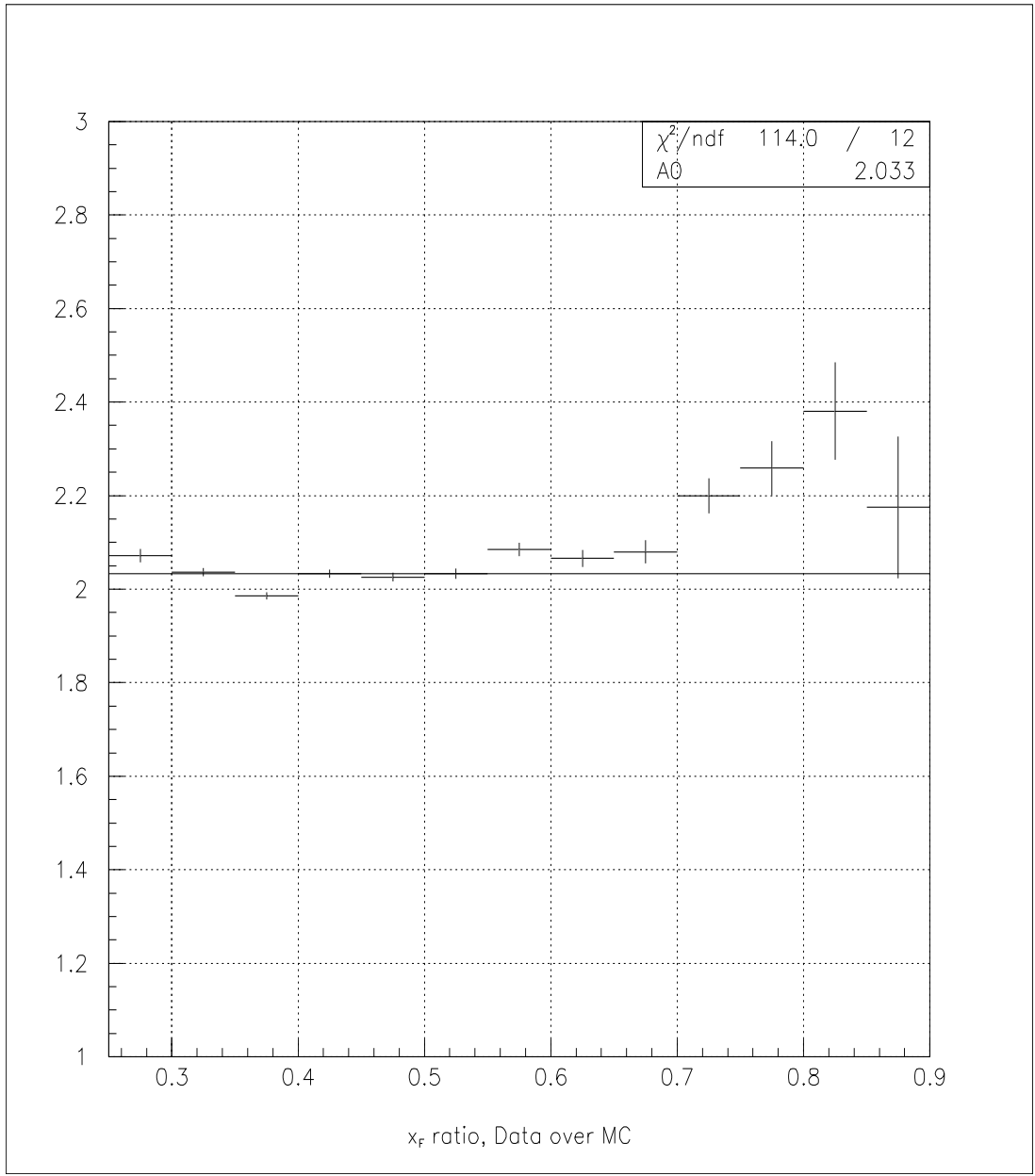}}
\end{center}
\vspace*{+0.25in}
\caption[$d\sigma /dx_{F}$ of data over $d\sigma /dx_{F}$ of Monte Carlo]
{$d\sigma /dx_{F}$ of data over $d\sigma /dx_{F}$ of Monte Carlo. The
agreement is good within 3$\%$ for $x_{F}< 0.7$, in which contains 98.2$\%$ of
the data.} 
\label{Fig:xfratio}
\end{figure}

%- Figure 4.5  $d\sigma /dx_{F}$ of data over $d\sigma /dx_{F}$ of Monte Carlo ---

\subsection{$p_T$ Distribution}

$p_T$ is the transverse momentum of the dimuon pair. The origin of $p_T$ is
understood as a combination of intrinsic transverse motions of the partons
inside the hadrons and higher-order QCD processes. Naively $p_T$ was expected to 
be an independent variable from
$x_F$, which accounts for the longitudinal part of the dimuon pair momentum, 
except at some extreme kinematic ranges where the maximum available energy 
becomes a constraint. However experimentally it was found that $<p_T>$ was 
correlated with $x_F$ beyond pure acceptance effects. In our Monte Carlo code
we have the following form for the $p_T$ thrown function:

\begin{eqnarray}
&d\sigma/dp_{T} = p_{T}/(1 + (p_{T}/p0(x_{F}))^{2})^{6},\\
&p0 = 1.43 + 8.28x_{F} - 15.3x_{F}^{2} + 2.66x_{F}^{3} + 13.9x_{F}^{4} - 9.23x_{F}^{5}.
\end{eqnarray}
\noindent
p0 was expressed in a polynomial form of $x_F$. This form attemped to fold
in the real physical correlations between $x_F$ and $<p_T>$ as well as possible
residual acceptance effects. Figure \ref{Fig:ptbarxf} shows $<p_T>$ as a function 
of $x_F$ for the experimental data and Monte Carlo data. The overall integrated 
$p_T$ distribution from MC was then compared to the integrated $p_T$ distribution
of the data; this is shown in Figure \ref{Fig:ptratio}. For each bin of $p_T$, the
mass spectrum of the data was fitted to a Gaussian plus a second-order 
polynomial as the background. The number of $J/\psi$'s was then calculated
from the Gaussian parameters. The Monte Carlo $\cos \theta$ distribution was 
weighted according to the normalized ratio of Figure \ref{Fig:ptratio}.  

\begin{figure}
\begin{center}
\mbox{\epsfxsize=5.7in\epsfysize=6.5in\epsffile{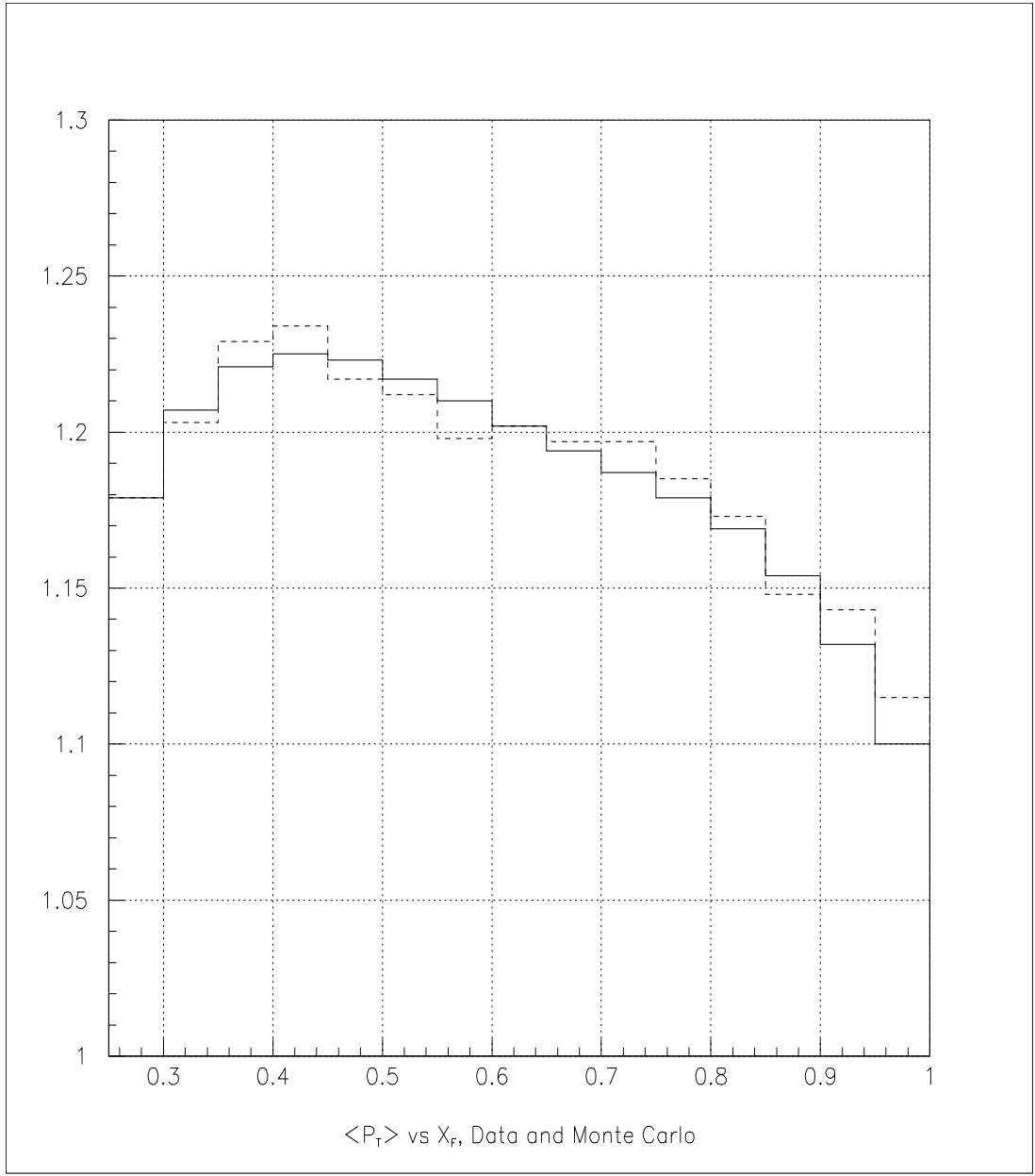}}
\end{center}
\vspace*{+0.25in}
\caption[$<p_{T}>$ vs $x_F$]
{$<p_{T}>$ vs $x_F$. Solid line: Data. Dashed line: Monte Carlo.} 
\label{Fig:ptbarxf}
\end{figure}

\begin{figure}
\begin{center}
\mbox{\epsfxsize=5.7in\epsfysize=6.5in\epsffile{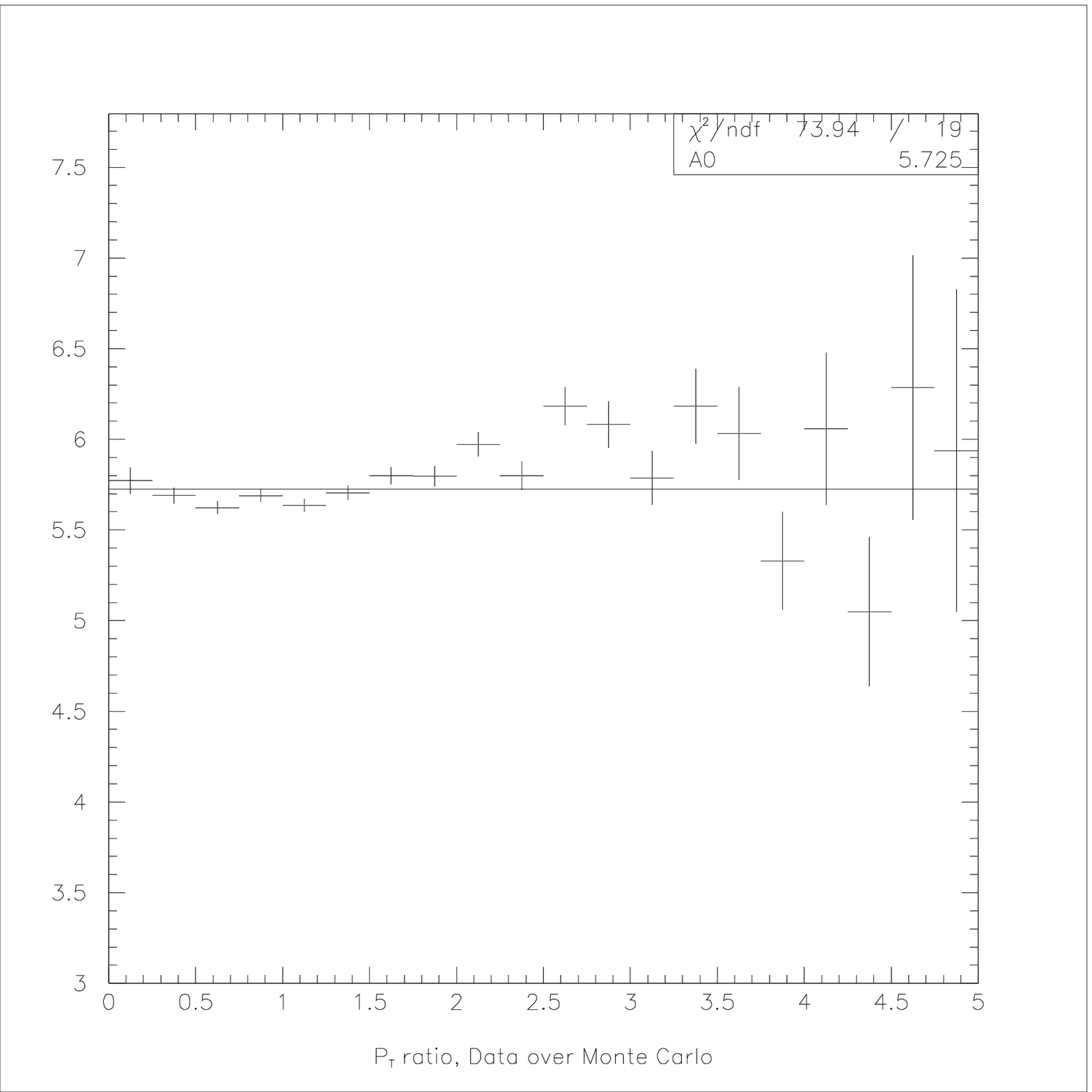}}
\end{center}
\vspace*{+0.25in}
\caption[$d\sigma /dp_{T}$ of data over $d\sigma /dp_{T}$ of  Monte Carlo]
{$d\sigma /dp_{T}$ of data over $d\sigma /dp_{T}$ of  Monte Carlo. The
 plot is fitted to a zeroth-order polynomial. The result of the fit is the 
 normalization constant used in the $\cos \theta$ weighting.}
\label{Fig:ptratio}
\end{figure}
  
%-- Figure 4.6  $<p_T>$ as function of xf, Data and MC--

%-- Figure 4.7  $d\sigma /dp_{T}$ of data over $d\sigma /dp_{T}$ of  Monte Carlo -

\subsection{Angular Thrown Distributions}

There are three independent angles used to specify the dilepton pair 
production: decay-$\theta$ (DTHETA), decay-$\phi$ (DPHI), and production-$\phi$
(PPHI). The DTHETA variable was defined as the polar angle in the Collin-Soper 
frame, and the DPHI variable was defined by the azimuthal angle with DPHI = 0 
pointing up with respect to the Z axis in the C-S frame. The PPHI gave the 
azimuthal angle of the virtual photon around the lab Z-axis. PPHI = 0 was chosen to 
be the positive X direction of the lab coordinate system
and it is a lab-frame variable. All three angles in this MC code were thrown
as flat distributions. The angular distributions obtained from the final ntuple
thus gave the angular-acceptance shape directly.

\section{Extracting Angular Distributions}

In this section the technique for obtaining the polarization parameter $\lambda$
is described.

\subsection{General Procedure}

$d\sigma/d \cos \theta dp_{T}dx_{F}$ of $J/\psi$ was determined by taking the 
accepted events divided by the acceptance, which was obtained by taking the Monte 
Carlo $\cos \theta$ distribution divided by the flat thrown distribution. The 
statistical uncertainty of the acceptance was small compared to the statistical 
uncertainty of the data. Before being applied to the data, the acceptance curves 
were weighted by the $p_T$ distribution of the real data. The uncertainties 
associated with the $p_T$ thrown function were thus expected to be minimized. In 
addition, since the $\cos \theta$ acceptances in this study were calculated in 
small $x_F$ and $p_T$ bins, the uncertainties due to the shape of the thrown $x_F$ and
$p_T$ distributions should be reduced. The acceptance of $J/\psi$ as a function of
$\cos \theta$ for each bin of $x_F$ and $p_T$ of the Monte Carlo simulation is 
given in Appendix A.

To extract the counts of $J/\psi$, the fitting procedure described in section
4.3 was applied to each bin of $x_F$, $p_T$, and $\cos \theta$, with bin width 
being
0.1, 1.0 GeV, and 0.1, respectively. At large $x_F$ and $p_T$ bins, the bin width 
of
$\cos \theta$ was increased to 0.2 to give better statistics. The fitting of each
histogram is shown in Appendix B. The number of counts, calculated according to 
Equation
4.4, in each bin of $\cos \theta$ was plotted to obtain the accepted 
$\cos \theta$ distribution. Each accepted $\cos \theta$ distribution was then
divided by the corresponding acceptance, and the shape of 
$d\sigma/d \cos \theta dp_{T}dx_{F}$ of $J/\psi$ was thus obtained.
The $\cos \theta$ distribution for each bin of $x_F$ and $p_T$ was then fitted
according to equation (1.1):

\begin{center}
$d\sigma/d \cos \theta \approx 1 + \lambda \cos^{2} \theta $.\\
\end{center}
\noindent
The fitting plots of $d\sigma/d \cos \theta dp_{T}dx_{F}$ are shown in Appendix C.

\subsection{Combined Data Set}

As mentioned in section 4.2, there were seven data sets differing by the incident
beam angle, the trigger matrix, or the magnet settings. To obtain the final 
results, those sets were combined according to the SM12 currents. Data 
sets 12a, 12b, 12c, 15a, and 15b together formed the ``SM12 = 2800'' set and data 
sets
13 and 14 formed the ``SM12 = 2040'' set. The events from different 
experimental data sets were directly added together, and the average $\cos \theta$
acceptance was calculated by

\begin{equation}
<Acceptance> = \sum _{i} f_{i} \times a_{i}(\cos \theta)/N_{i}(\cos \theta)
\end{equation}
\noindent
with

\begin{equation}
\sum _{i} f_{i} = 1,
\end{equation}
\noindent
where $f_i$ is the fraction of the accepted events of data set i among all 
accepted
events, and $a_{i}(\cos \theta)/N_{i}(\cos \theta)$ is the $\cos \theta$ 
acceptance 
distribution of data set i. The combined results were obtained by dividing the
sum of the accepted events by the average acceptance. The results for the two 
different magnet settings were obtained separately. 

In the next chapter and throughout Appendix A to C, the results are presented for 
the two magnet settings. The agreement between the two sets of results provides an 
important check that our results are not affected by apparatus effects. 
Combined results from the two magnet settings are derived to
compare with results from other fixed-target experiments.

\section{Uncertainties}

In this section two types of errors are discussed: statistical uncertainties and
systematic uncertainties. These different sources of errors are summarized in
tables.

\subsection{Statistical Uncertainty}

Statistical errors in determining $\lambda$ are the direct results of statistical
uncertainties in the counts of $J/\psi$. The statistical errors of the 
acceptance are supressed by outnumbering Monte Carlo events over the 
experimental data. The statistical errors of $\lambda$ were obtained by
including only the statistical uncertainties of the $J/\psi$ counting in the 
$\lambda$ fitting. The $\lambda$'s and the statistical errors are shown 
in Table \ref{tblxfpt2040} and Table \ref{tblxfpt2800}.

\subsection{Systematical Errors from Analysis and MC Inputs}

To estimate the errors caused by the uncertainties of the magnetic fields, the 
tweek values used in the analysis were varied by 1$\%$ in the analysis and Monte 
Carlo.  The 1$\%$ uncertainty is a reasonable upper limit for the magnetic-field 
strength because the tweek values were tightly constrained by the reconstructed 
$J/\psi$ mass and the uniterated Z-vertex position. It was found that the 
$\lambda$ values changed by $\pm 0.01$ as the field strength of SM12 was 
adjusted, and varied by $\pm 0.01$ as SM3 was changed. 

The incident beam angle has a strong effect on the $\cos \theta$ distribution.
Although the value of the beam angle was tuned to remove any asymmetry in the
ratio plots of decay $\theta$ and production $\phi$ between experimental data
and Monte Carlo, the precise incident angle is actually unknown. To study the
effect of this uncertainty, the beam angle used in the analysis and Monte 
Carlo was varied by $\pm 0.0002$, which is twice the beam angle-spread sigma. The
results showed that the $\lambda$ values were changed by $\pm 0.02$ for large 
$x_F$ and by $\pm 0.04$ for small $x_F$.

The target position in the X-Y plane used in the Monte Carlo was an average value 
determined from the data. In data reconstruction the beam center was calculated
for each spill. To study this uncertainty, the beam centroid of Monte Carlo
events was moved by $\pm 0.1$ inch in both the X and Y direction. The circle of 
the 0.1 inch confinement was determined by the data distribution. The net effect 
on the $\lambda$ values is $\pm 0.02$.

The $p_T$ dependence of the $\cos \theta$ distribution and the impact of the
$p_T$ thrown function on the $\cos \theta$ acceptance was also studied. The 
$\cos \theta$ acceptance was calculated by weighting the $p_T$ distribution 
according to the real data. Another calculation was performed without weighting to
the real data. The $<p_{T}>$ with and without weighting differed by 5$\%$, which 
is compatible with the uncertainties in extracting the p0 parameter in the $p_T$ 
thrown formula. It was then found that within this variation, the $\lambda$ values
moved by $\pm 0.06$.

\subsection{Systematical Errors from Peak Fitting}

Other contributions to the systematic errors come from the $J/\psi$ peak fitting
process. The $\lambda$ values were found to change slightly with different choices
of the fitting limits. With the same background function, changing the 
fitting limit was equivalent to changing the continuum shape within the 
uncertainties. It was found that this contributed a $\pm 0.03$ uncertainty to the
systematic errors. 

The selection of the background function form also produced systematic errors. 
Three different functions, equations 4.1, 4.2, 4.3, were used to fit the
continuum distribution. The uncertainty from different background functions on  
$\lambda$ is $\pm 0.02$ for large $x_F$ and is about $\pm 0.05$ for small $x_F$.

Table \ref{systemE} gives a summary of the systematic errors from all the 
contributions.

\begin{table}[tbp]
\caption{Summary of the systematic errors from all the sources.}
\label{systemE}
%\begin{small}
\begin{center}
\begin{tabular}{ccccccccc}
\hline \hline  

$x_F$ & SM12 & SM3 & angle &centroid &$<p_T>$& fit limit& background & overall\\
\hline
 0.3 &  0.01 & 0.01  & 0.04   &   0.02  &  0.06  &   0.05   &   0.06   &  0.109\\  
 0.4 &  0.01 & 0.01  & 0.04   &   0.02  &  0.06  &   0.04   &   0.05   &  0.099\\
 0.5 &  0.01 & 0.01  & 0.02   &   0.02  &  0.06  &   0.04   &   0.03   &  0.084\\
 0.6 &  0.01 & 0.01  & 0.02   &   0.02  &  0.06  &   0.03   &   0.02   &  0.077\\
 0.7 &  0.01 & 0.01  & 0.02   &   0.02  &  0.06  &   0.03   &   0.02   &  0.077\\
 0.8 &  0.01 & 0.01  & 0.02   &   0.02  &  0.06  &   0.03   &   0.02   &  0.077\\
 0.9 &  0.01 & 0.01  & 0.02   &   0.02  &  0.06  &   0.03   &   0.02   &  0.077\\
\hline \hline
\end{tabular}
\end{center}
%\end{small}
\end{table}

\chapter{RESULTS}

The angular distributions of the $J/\psi$ decay in the dimuon channel have 
been measured for the process $p + Cu \rightarrow J/\psi + X$ using an 800 GeV proton 
beam. In this
chapter the results are presented, along with comparison to the theoretical
predictions and the results from other experiments.

\section{Drell-Yan Angular Distribution}

In the same data sample used in this study, about 200K dimuon pairs with mass
ranging from 4.0 GeV to 7.5 GeV were also recovered, of which most are 
Drell-Yan events. This data sample is of interest because the target was copper
and no angular-distribution measurements had ever been published for the 
proton-induced Drell-Yan process in this mass range. Though the amount of data 
sample was not enough
to study the polarization as a function of $x_F$, it was still useful to 
examine the overall polarization parameter $\lambda$, which was expected to be
equal to unity based on the standard Drell-Yan production mechanism, as a 
cross check for the $J/\psi$ angular distribution results. Below the procedures
are described in more detail, and the results are presented

\subsection{Random Background}

Random muon pairs were the most significant background contamination in the 
mass range of interest. The definition of a random pair is that two opposite-sign 
muons, which were produced independently by pion decay or other processes,
coincidentally fired the trigger system and appeared to be a valid target 
dimuon event. The random pair distribution could not be measured directly,
because the pairs were indistinguishable from the real Drell-Yan dimuons in the 
spectrometer; they were however simulated from the like-sign event distribution
by changing the sign of the Y-momentum of one of the like-sign tracks to calculate
other kinematic variables of the pair, based
on the assumption that the probability to form a random pair is the same as to 
form a like-sign pair given the first muon track. Single-muon trigger
rates were used to normalize the ratio of like-sign pairs and the randoms; the
like-sign pairs were expected to have the same single-muon trigger rate as
the random pair would have. Figure \ref{Fig:rawrdn} shows some kinematic-variable 
distributions of the random pairs. 

\begin{figure}
\begin{center}
\mbox{\epsfxsize=5.7in\epsfysize=6.5in\epsffile{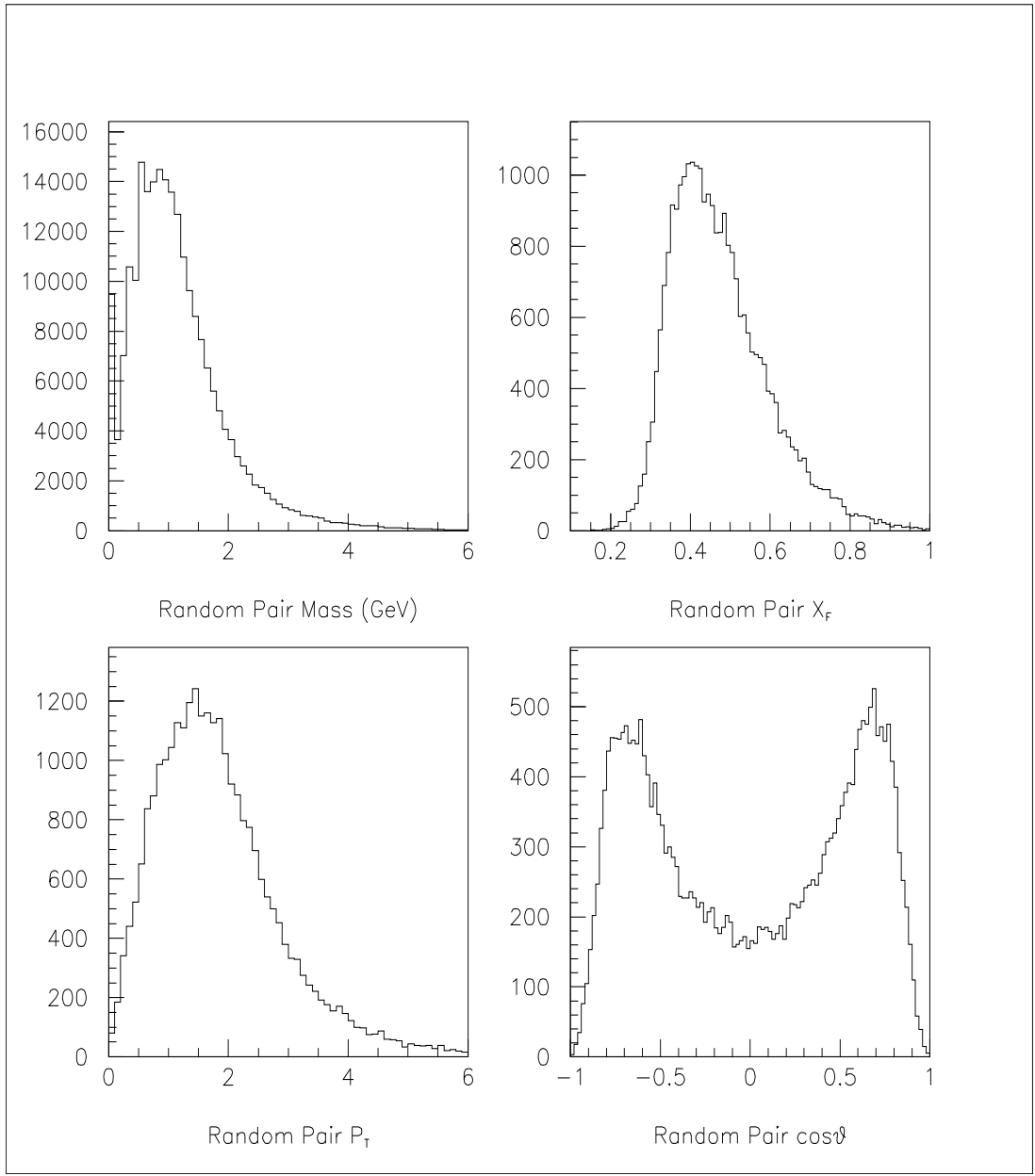}}
\end{center}
\vspace*{+0.25in}
\caption[Some kinematic variables of the random pairs]
{Some kinematic variables of the random pairs. The mass range of the 
$x_F$, $p_T$, and $\cos \theta$ plots is above 2 GeV.}
\label{Fig:rawrdn}
\end{figure}

%-- Figure 5.1  Raw spectrum of random pairs --

\begin{figure}
\begin{center}
\mbox{\epsfxsize=5.7in\epsfysize=6.5in\epsffile{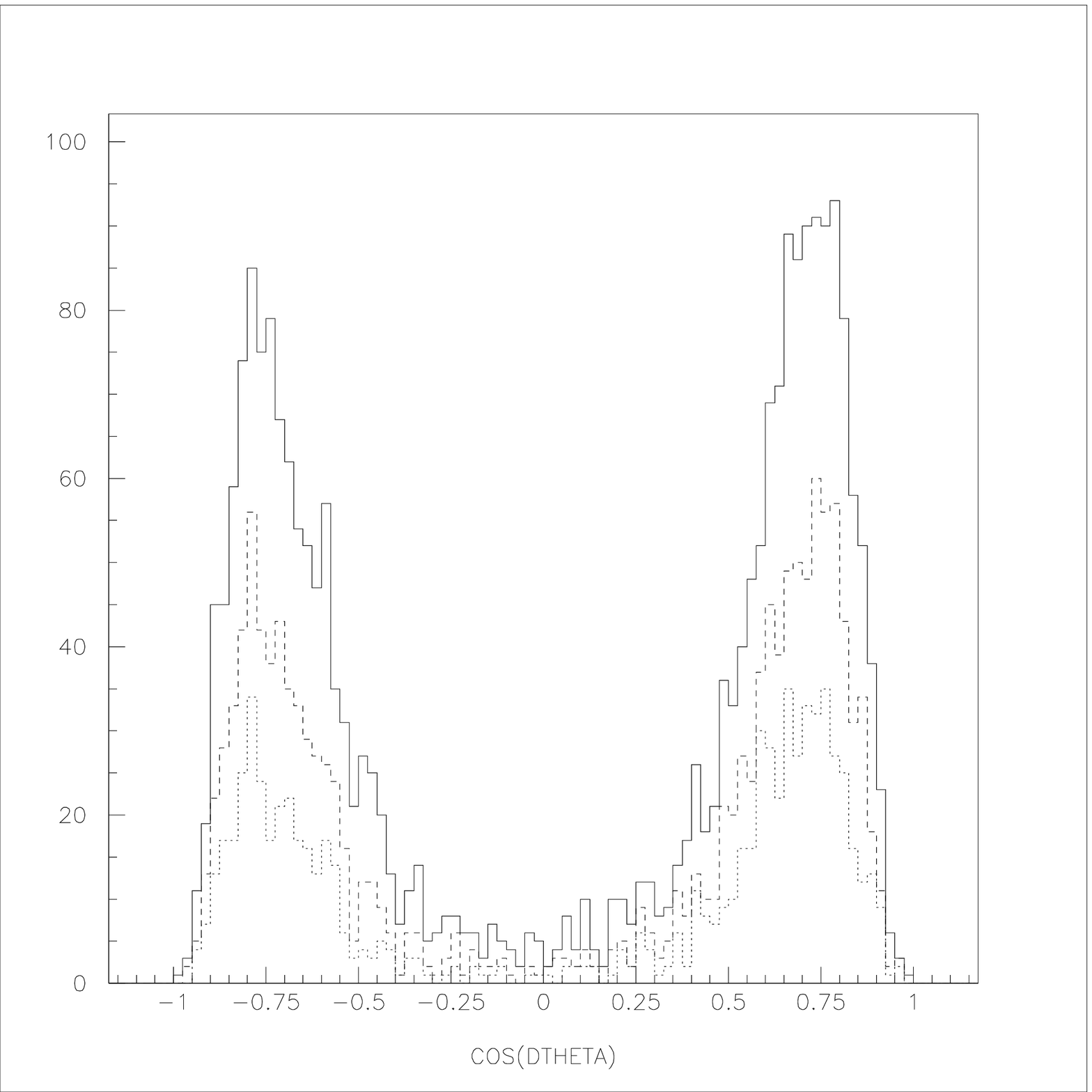}}
\end{center}
\vspace*{+0.25in}
\caption[$\cos \theta$ of the random pairs]
{$\cos \theta$ of the random pairs. Solid line: pair mass $>$ 4 GeV. 
Dashed line: pair mass $>$ 4.5 GeV. Dotted line: pair mass $>$ 5 GeV.}
\label{Fig:rdntheta}
\end{figure}

%-- Figure 5.2  random $\cos \theta$ cut on various mass range --

\subsection{Random Subtraction and Results}

From Figure \ref{Fig:rdntheta} it was understood that the random subtraction 
was important in order to
extract the Drell-Yan angular distribution, even for pairs of mass greater
than 5.0 GeV. The normalization factor for the random events was determined by
matching the number of like-sign pairs from the random ntuple file and from the
data, since the random events were generated according to the amount of 
like-sign pairs of the data. Then the angular acceptance of the Drell-Yan events 
was obtained by a dedicated Monte Carlo run. The $\cos \theta$ distribution of the 
Drell-Yan data, after subtracting out the randoms, was corrected for 
acceptance and the angular distribution was obtained. The angular distribution
of Drell-Yan events in the mass range of 4.0 GeV to 7.0 GeV is shown in 
Figure \ref{Fig:randomsub}. The rise at the edges is understood as a resolution
problem and the same effect has been reproduced by broadening the $\cos \theta$ 
resolution. Also, the random pairs show a very strong rise at large
$\cos \theta$; a slight mismatch in the normalization can result in the same 
effect. For these reasons, for the Drell-Yan data the angular distribution was 
fitted in the range of $-0.7<\cos \theta <0.7$ in which the systematic uncertainties
are best handled. A result of $\lambda = 0.98 \pm 0.04$ was obtained. This is 
consistent with 1.0 as predicted. This provided a confidence check for the 
$J/\psi$ angular distribution presented in next section.   

\begin{figure}
\begin{center}
\mbox{\epsfxsize=5.7in\epsfysize=6.5in\epsffile{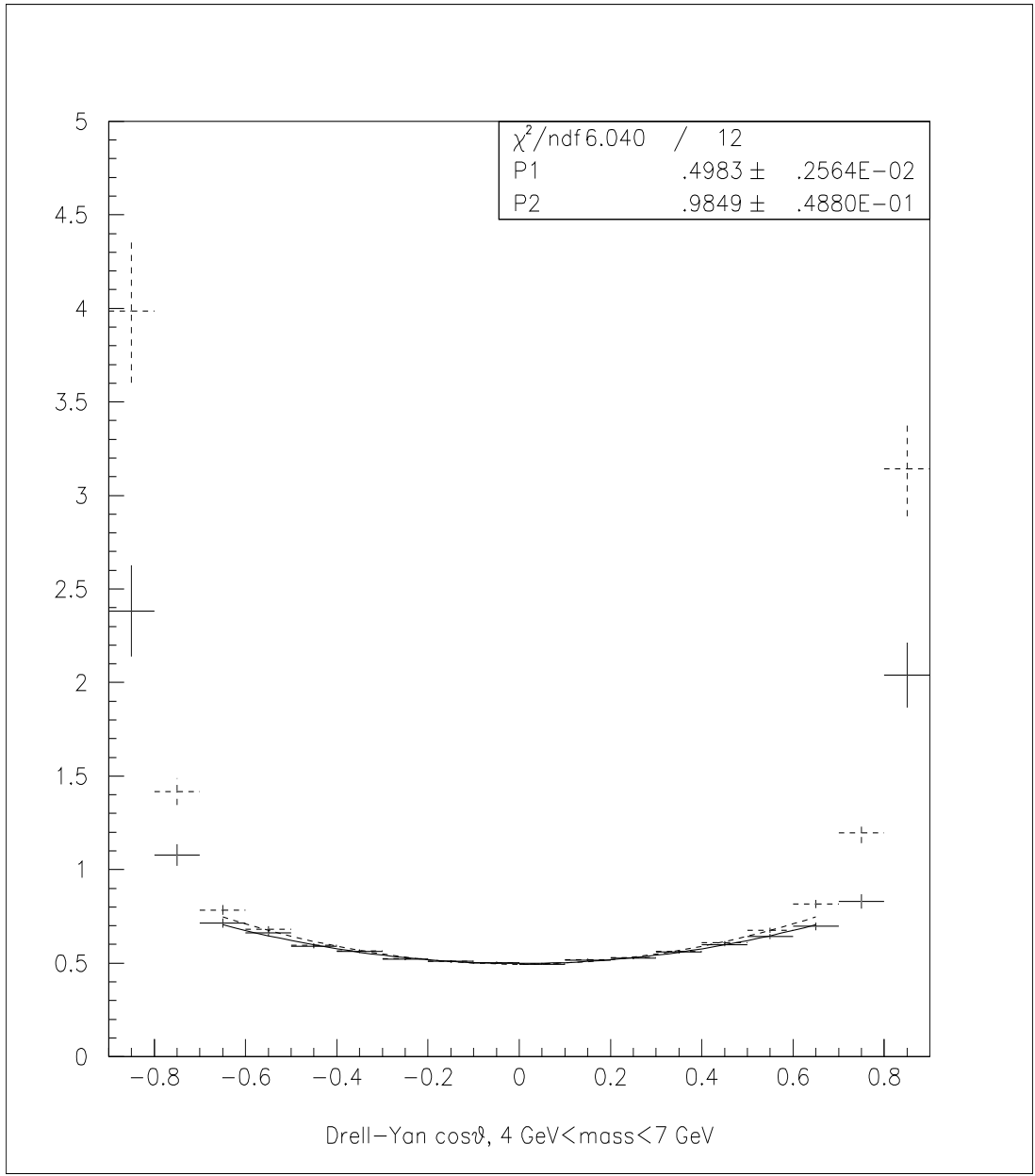}}
\end{center}
\vspace*{+0.25in}
\caption[$\cos \theta$ of the Drell-Yan pairs]
{$\cos \theta$ of the Drell-Yan pairs with mass ranging between 4 GeV
and 7 GeV. Solid line: After random subtraction. Dashed line: Before random
subtraction. A $\lambda$ value of 0.98 $\pm$ 0.04 is obtained after 
correcting for the random pairs. The rise at the edges is due to 
resolution effects.}
\label{Fig:randomsub}
\end{figure}

%-- Fig 5.3 Drell-Yan angular distribution after random subtraction --

\section{$J/\psi$ Angular Distribution Results}

In this section the $J/\psi$ angular distribution results are presented. The
measurements were performed under two different magnet configurations, therefore
different acceptances, in order to minimize the systematic bias. They are 
effectively two independent measurements.
The polarization parameter $\lambda$ in bins of $p_T$ and $x_F$ is presented
in Tables \ref{tblxfpt2040} and \ref{tblxfpt2800}, and plotted in Figure 
\ref{Fig:lambdaxfpt}. In the figures only the statistical errors are shown. The
results from the two measurements are in agreement with each other. This provides
a confirmation that the results are not affected by the specific apparatus 
settings. 

\begin{table}[tbp]
\caption[$\lambda$ in $x_F$ and $p_T$ bins for the ``SM12 = 2040'' data]
{$\lambda$ in $x_F$ and $p_T$ bins with statistical errors only. This 
table is for the SM12 = 2040 data.}
\label{tblxfpt2040}
\begin{center}
\begin{tabular}{ccccc}
\hline \hline
$x_F$ & $0<p_{T}<1$ & $1<p_{T}<2$ & $2<p_{T}<3$ & $3<p_{T}$\\
\hline
0.25 - 0.35 & $0.153\pm0.037$ & $0.057\pm0.024$  & $0.093\pm0.026$ & $0.124\pm0.049$\\
0.35 - 0.45 & $0.218\pm0.031$ & $0.015\pm0.019$  & $0.095\pm0.026$ & $0.141\pm0.056$\\
0.45 - 0.55 & $0.146\pm0.023$ & $0.035\pm0.017$  & $0.101\pm0.025$ & $-0.052\pm0.049$\\
0.55 - 0.65 & $0.151\pm0.039$ & $-0.013\pm0.027$ & $0.072\pm0.041$ & $-0.06\pm0.08$\\
0.65 - 0.75 & $0.111\pm0.070$ & $-0.211\pm0.046$ & $0.023\pm0.093$ & $-0.43\pm0.13$\\
0.75 - 0.85 & $-0.17\pm0.15$  & $-0.22\pm0.09$   & $-0.14\pm0.30$  &   -  \\
$> 0.85$    & $-0.44\pm0.42$  &   -              &      -          &     -  \\
\hline \hline
\end{tabular}
\end{center}
\end{table}

\begin{table}[tbp]
\caption[$\lambda$ in $x_F$ and $p_T$ bins for the ``SM12 = 2800'' data]
{$\lambda$ in $x_F$ and $p_T$ bins with statistical errors only. This 
table is for the SM12 = 2800 data.}
\label{tblxfpt2800}
\begin{center}
\begin{tabular}{ccccc}
\hline \hline
$x_F$ & $0<p_{T}<1$ & $1<p_{T}<2$ & $2<p_{T}<3$ & $3<p_{T}$\\
\hline
0.25 - 0.35 & $0.189\pm0.044$  & $0.135\pm0.023$  & $0.120\pm0.034$  & $0.060\pm0.064$\\
0.35 - 0.45 & $0.162\pm0.029$  & $0.095\pm0.015$  & $0.170\pm0.021$  & $0.170\pm0.045$\\
0.45 - 0.55 & $0.115\pm0.022$  & $0.051\pm0.013$  & $0.153\pm0.020$  & $0.057\pm0.040$\\
0.55 - 0.65 & $0.018\pm0.028$  & $-0.053\pm0.020$ & $-0.026\pm0.028$ & $0.01\pm0.06$\\
0.65 - 0.75 & $-0.032\pm0.049$ & $-0.174\pm0.033$ & $-0.167\pm0.059$ & $-0.09\pm0.12$\\
0.75 - 0.85 & $-0.25\pm0.10$   & $-0.09\pm0.10$   & $-0.21\pm0.19$   &   -  \\
$> 0.85$    & $-0.51\pm0.54$   &   -              &      -           &     -  \\
\hline \hline
\end{tabular}
\end{center}
\end{table}

\begin{figure}
\begin{center}
\mbox{\epsfxsize=5.7in\epsfysize=6.5in\epsffile{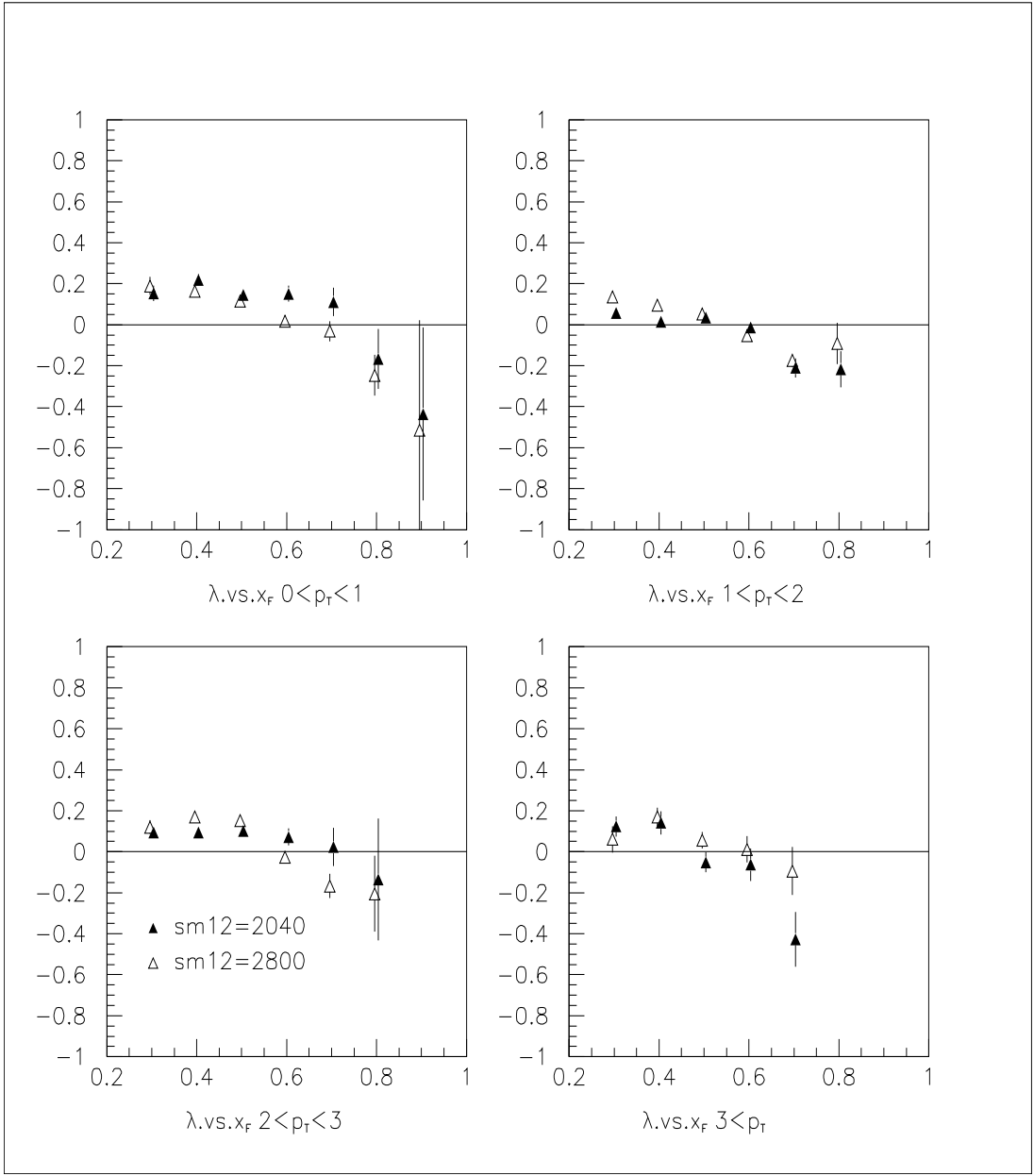}}
\end{center}
\vspace*{+0.25in}
\caption[$J/\psi$ polarization parameter $\lambda$ in $x_F$ and $p_T$ bins]
{$J/\psi$ polarization parameter $\lambda$ in $x_F$ and $p_T$ bins. The
errors shown here are statistical only.}
\label{Fig:lambdaxfpt}
\end{figure}

%-- Fig 5.4  $J/\psi$ polarization parameter $\lambda$ in $x_F$ and $p_T$ bins --

\section{Comparison}

In order to compare with other experiments, $\lambda$ values in $x_F$ and $p_T$
bins from 
E866 measurements were combined to obtain $\lambda$ in $x_F$ bins and the overall
integrated $\lambda$ using the following the relations:

\begin{center}
\begin{eqnarray}
\bar A = \frac{\sum _{i}w_{i}A_{i}}{\sum _{i} w_{i}} \\
w_{i} = 1/\sigma_{i}^{2}
\end{eqnarray}
\end{center}
\noindent
where $\sigma_{i}$ is the statistical error of some measurement $A_{i}$, and 
$\bar A$ is the average value of $A_{i}$. The results are shown in 
Table \ref{tblxf}. One can see from the table that the $J/\psi$ starts slightly
transversely polarized at small $x_F$, then eventually becomes partially 
longitudinally polarized as $x_F$ increases toward unity. The 
$\lambda$ values from both magnet settings versus $x_F$ are plotted in 
Figure \ref{Fig:lambdaxf}.  

\begin{table}[tbp]
\caption{$\lambda$ in $x_F$ bins with statistical errors only.}
\label{tblxf}
\begin{center}
\begin{tabular}{cccc}
\hline \hline
$x_F$ & $\lambda(x_F)$(SM12 = 2040) & $\lambda(x_F)$(SM12 = 2800) & Combined\\
\hline
0.25 - 0.35 & $0.092 \pm 0.015$  & $0.134 \pm 0.017$  & $0.110 \pm 0.011$  \\
0.35 - 0.45 & $0.081 \pm 0.013$  & $0.129 \pm 0.011$  & $0.109 \pm 0.008$  \\
0.45 - 0.55 & $0.073 \pm 0.012$  & $0.086 \pm 0.009$  & $0.081 \pm 0.007$  \\
0.55 - 0.65 & $0.041 \pm 0.019$  & $-0.026 \pm 0.014$ & $-0.002 \pm 0.011$ \\
0.65 - 0.75 & $-0.116 \pm 0.034$ & $-0.134 \pm 0.024$ & $-0.128 \pm 0.020$ \\
0.75 - 0.85 & $-0.200 \pm 0.073$ & $-0.174 \pm 0.066$ & $-0.186 \pm 0.049$ \\
$> 0.85$    & $-0.44 \pm 0.42$   & $-0.51 \pm 0.54$   & $-0.47 \pm 0.33$   \\
\hline
All data    & $0.065 \pm 0.007$  & $0.070 \pm 0.005$  & $0.069 \pm 0.004$  \\
\hline \hline
\end{tabular}
\end{center}
\end{table}

\begin{figure}
\begin{center}
\mbox{\epsfxsize=5.7in\epsfysize=6.5in\epsffile{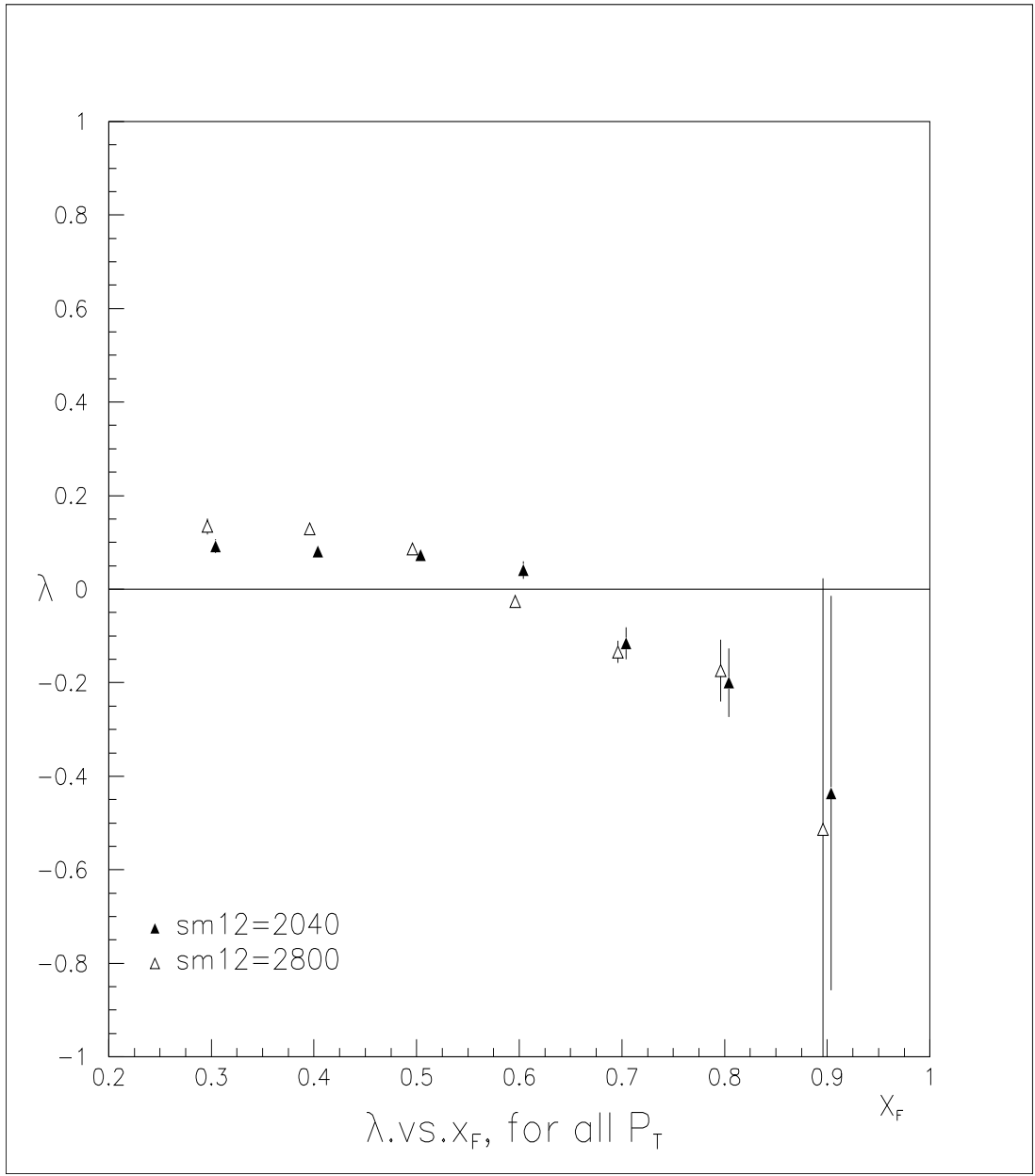}}
\end{center}
\vspace*{+0.25in}
\caption[$J/\psi$ polarization parameter $\lambda$ in $x_F$ bins]
{$J/\psi$ polarization parameter $\lambda$ in $x_F$ bins. The errors are
statistical only.}
\label{Fig:lambdaxf}
\end{figure}

%-- Fig 5.5  $J/\psi$ polarization parameter $\lambda$ in $x_F$ bins --

The E866 results are compared with the results of CIP data \cite{Bii 87}. Figure 
\ref{Fig:lambdacip}
shows the E866 results and data published by the CIP group. Recall that the CIP 
experiment was fixed-target $\pi$N collisions. At large $x_F$, both experiments
observe longitudinal polarizations. At smaller $x_F$, E866 sees small transverse
polarization while the CIP group saw no polarization. Since the dominant Feynman
diagrams are different for pN (mainly g-g fusion) and $\pi$N (it has significant
$q \bar q$ contributions) at small $x_F$, the differences in the polarization are
not unexpected. However if the 0.1 systematic errors are included, the E866 
results at $x_{F} < 0.5$ are marginally in agreement with no polarization.
  
\begin{figure}
\begin{center}
\mbox{\epsfxsize=5.7in\epsfysize=6.5in\epsffile{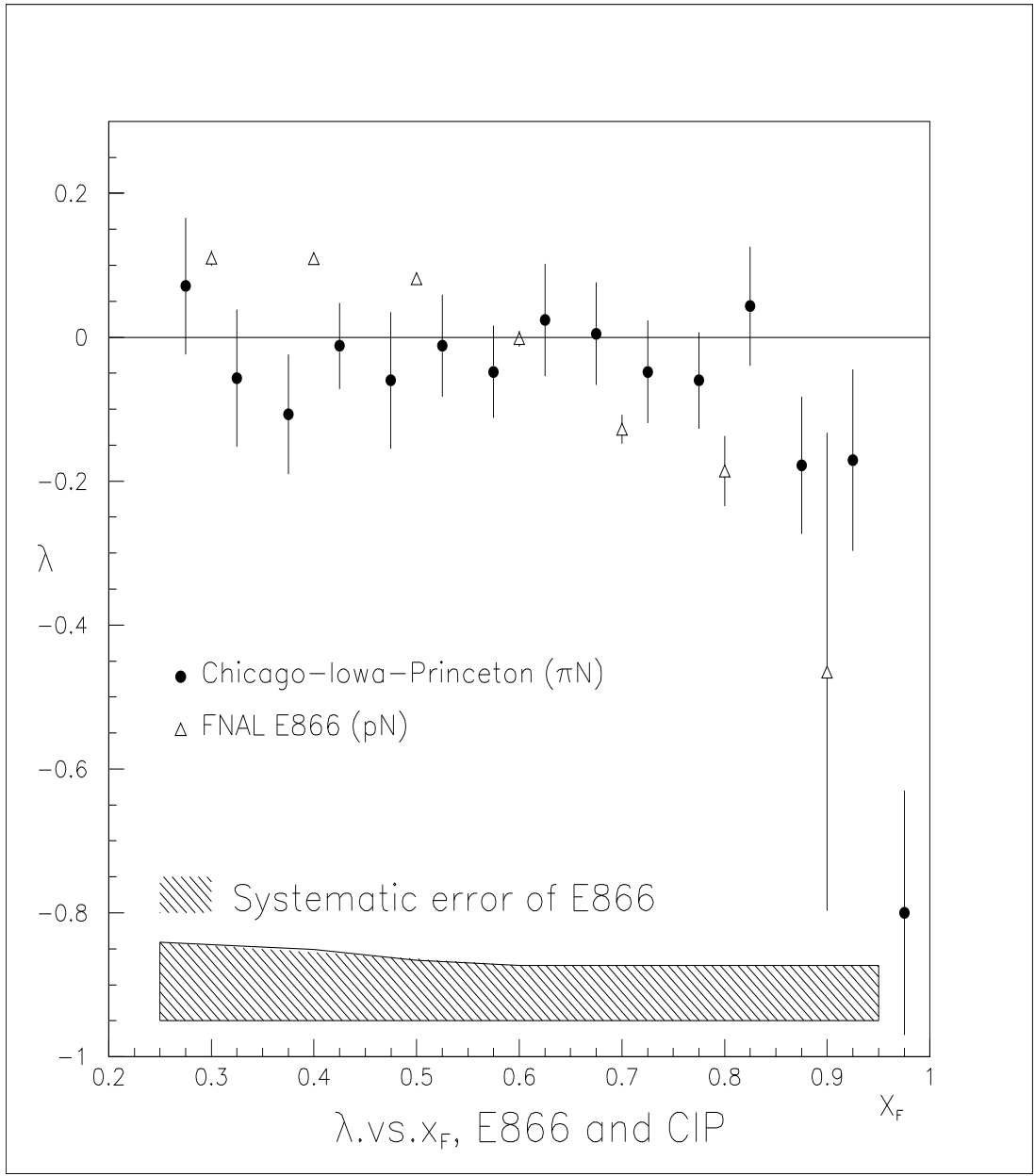}}
\end{center}
\vspace*{+0.25in}
\caption[$\lambda(x_{F})$ from FNAL E866 and from CIP group]
{$J/\psi$ polarization parameter $\lambda$ in $x_F$ bins from the FNAL 
 E866 and from Chicago-Iowa-Princeton collaboration. The error bars on E866 data
 are statistical only; the systematic error is shown in the shadowed band 
 below.}
\label{Fig:lambdacip}
\end{figure}

The integrated polarization parameter $\lambda$ obtained by E866 and other 
previous experiments are presented in Table \ref{compare} for comparison.
Recall that E866 uses the Collin-Soper frame and the other experiments have used
the Gottfried-Jackson frame as their reference frame. However the $p_T$ in 
fixed-target experiments is low enough that the direct comparison is still 
sensible. 
E866 gives $\lambda = 0.069 \pm 0.004$ integrated over all available data. 
If the systematic error is included, the E866 result shows no polarization. 
The transverse polarization at small $x_F$ is partially cancelled by the 
longitudinal polarization at large $x_F$. The overall result is in agreement with 
other experiments, and in contrast to the non-relativistic QCD calculation.

\begin{table}[tbp]
\caption{Overall $\lambda$ values from other fixed-target experiments and E866.}
\label{compare}
\begin{flushleft}
\begin{tabular}{ccccc}
\hline \hline
Experiment & reaction & $\sqrt{S}$ & $x_F$ range & $\lambda$ \\
\hline
E537      & $\overline{p}$ + W & 15.3 GeV & $x_F > 0$ & $-0.115 \pm 0.061$\\
E537      & $\pi^-$ + W        & 15.3 GeV & $x_F > 0$ & $0.028 \pm 0.004$\\
E672/706  & $\pi^-$ + Be       & 31.5 GeV & $0.1<x_F<0.8$ & $-0.01 \pm 0.08$\\
E771      & p + Si             & 38.8 GeV & $-0.05<x_F<0.25$ & $-0.09 \pm 0.12$\\
E866      & p + Cu             & 38.8 GeV & $0.25<x_F<1.0$ & $0.069 \pm 0.004 \pm syst.$\\
\hline \hline
\end{tabular}
\end{flushleft}
\end{table}

%\begin{table}[tbp]
%\begin{flushleft}
%\begin{tabular}{ccc}
%\hline \hline
%Theory & $x_F$ range & $\lambda$\\
%\hline
%CSM &    $x_F > 0$ &   $\sim$ 0.25\\
%CEM &    $x_F > 0$ &   0\\             
%NRQCD &  $x_F > 0$ &   $0.31<\lambda<0.63$\\
%\hline \hline
%\end{tabular}
%\end{flushleft}
%\end{table}

The $\lambda$'s were also integrated in 1-GeV $p_T$ bins for $x_{F}<0.45$ and 
$x_{F}>0.45$ to study the $p_T$ dependence, as shown in Figure \ref{Fig:2xf}. No
$p_T$ dependence was identified for either $x_F$ region. This suggests that
nuclear effects are probably not responsible for the polarization observed, since
one important cause of the broadening in the $p_T$ distribution in nuclear targets
is the multiple scattering of the incoming and outgoing partons with the nuclear
media. If some of the nuclear effects, e.g. energy loss inside nucleus, are 
important, one would expect to see significant $p_T$ dependence on the 
polarization.

\begin{figure}
\begin{center}
\mbox{\epsfxsize=5.7in\epsfysize=6.5in\epsffile{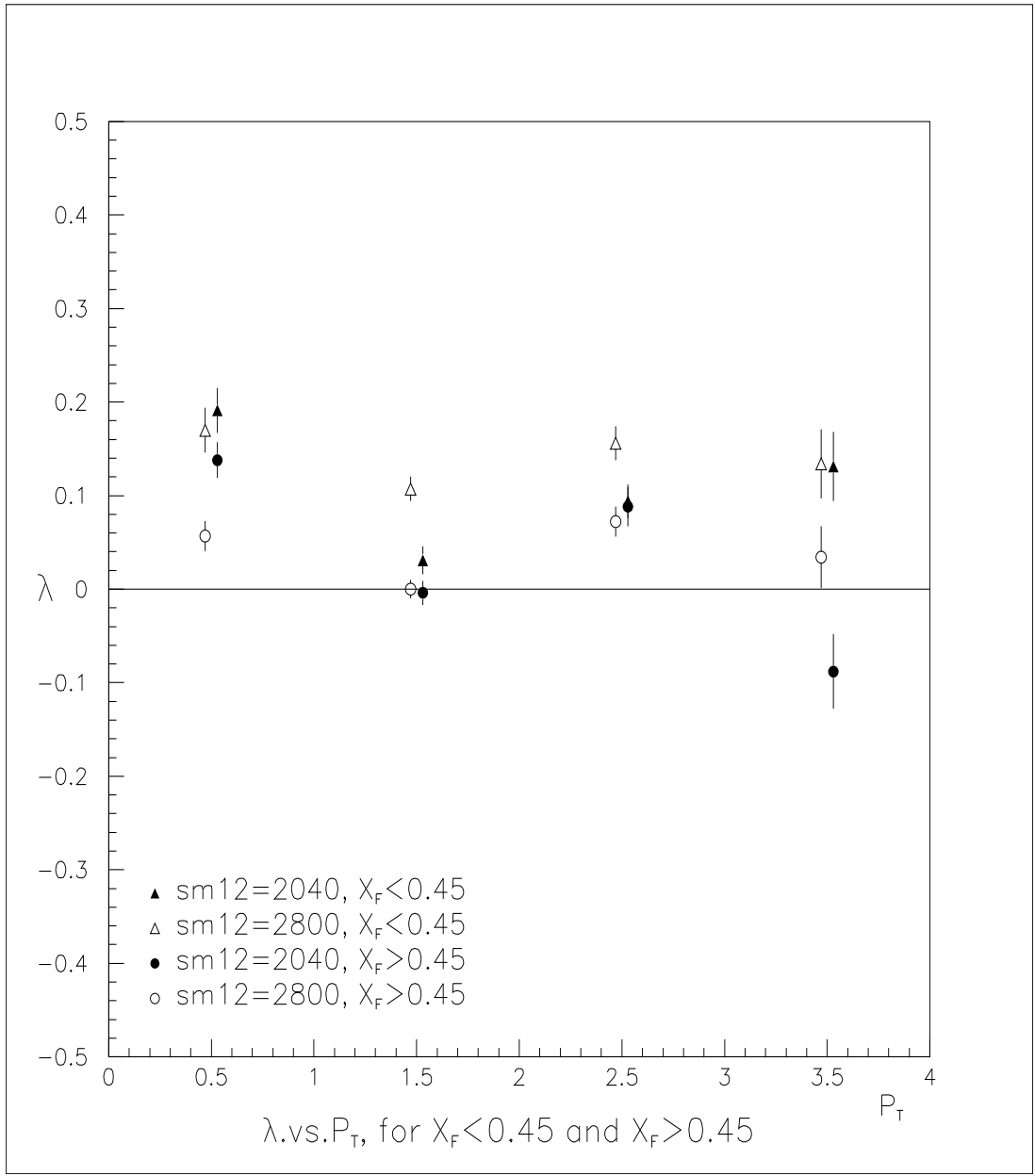}}
\end{center}
\vspace*{+0.25in}
\caption[$J/\psi$ polarization parameter $\lambda$ in 1-GeV $p_T$ bins]
{$J/\psi$ polarization parameter $\lambda$ in 1-GeV $p_T$ bins for
$x_{F}<0.45$ and $x_{F}>0.45$. The errors are statistical only.}
\label{Fig:2xf}
\end{figure}

\chapter{CONCLUSIONS AND FUTURE PROSPECTS}

The angular distribution of $J/\psi$ decays in the $\mu^{+} \mu^{-}$ channel 
produced in 800
GeV proton-copper collisions has been measured for $x_{F} > 0.25$. The 
polarization parameter $\lambda$ is extracted in $p_T$ and 
$x_F$ bins for two magnet configurations with different acceptances. The 
data indicate that the $J/\psi$'s are produced with a slight transverse 
polarization at $x_{F} < 0.6$, which turns to longitudinal at $x_{F} > 0.6$. 
This suggests that gluon-gluon fusion, which dominates at small $x_F$, and 
quark-antiquark annihilation,
which dominates at large $x_F$, leave $J/\psi$'s in different polarization states.
Another fixed-target experiment \cite{Bii 87}, using pion beams, also 
showed longitudinal polarization at $x_{F} \rightarrow 1$. However at smaller 
$x_F$ the uncertainties are large and no evidence of polarization is seen
in Biino's paper. The difference of the results from E866 will provide
interesting information on how the production mechanism affects the polarization,
because in $\pi$N interactions the production is dominated by quark-antiquark 
annihilation while in the pN case the production is dominated by gluon-gluon 
fusion in the range of $x_{F} < 0.6$.  
   
It should be mentioned that the $J/\psi$ samples collected in this study do not
purely come from pN interactions. A significant amount of pions were generated at
the dump by hadronic interactions, and those pions can further interact with the
beam dump to produce $J/\psi$'s. A calculation \cite{Mue 99} 
shows that about $10\%$ of the $J/\psi$'s come from pion interactions at small 
$x_F$. This has to be taken into account when comparing with the theoretical 
calculations. Table \ref{secondary} shows the estimate of the ratios of the
$J/\psi$'s produced by the secondary pions to those produced by the primary 
proton beam in various $x_F$ ranges.

\begin{table}[tbp]
\caption[Ratio of primary proton and secondary pion induced-$J/\psi$]
{Ratio of primary proton and secondary pion induced-$J/\psi$ \cite{Mue 99}.}
\label{secondary}
\begin{center}
\begin{tabular}{cc}
\hline \hline
$x_F$  &  ratio($\%$)\\
\hline
0.2 & 15.1\\
0.3 &  9.3\\
0.4 &  6.2\\
0.5 &  4.4\\
0.6 &  3.4\\
0.7 &  3.0\\
0.8 &  2.9\\
\hline \hline
\end{tabular}
\end{center}
\end{table}

It is also important to keep in mind that a substantial fraction of $J/\psi$'s 
come from
decays of the $\chi _{c}$ states and $\psi ^\prime$ decays in addition to direct
$J/\psi$'s. All the processes contribute different amounts of polarization to
$J/\psi$. Thus one needs to know the relative production cross sections 
of the various charmonium states to interpret the results properly. So far, only
production ratios for pion-produced charmonium states are available. It is also 
necessary to know the polarization of $J/\psi$'s from each process to extract the 
polarization of direct $J/\psi$ decays. A theoretical calculation has been 
done for $\pi$N collisions using the Color-Singlet Model \cite{Van 95}, but the 
results do not agree with the pion data. It would be interesting to see the 
predictions of similar calculations for pN collisions. A measurement of 
$\psi ^\prime$ polarization would be very interesting since the 
$\chi_{c}$ state contribution is absent. In this experiment the mass resolution 
was sacrificed to gain the yield rate and angular coverage; the data sample 
contains only about 1 $\%$ of $\psi ^\prime$'s and they are not resolved from the 
$J/\psi$ peak.  

It is interesting to notice that if the $J/\psi$'s are integrated over the 
entire $x_F$ range, the transverse polarization at small $x_F$ partially cancels 
the longitudinal polarization at large $x_F$, and the overall effect appears to 
be no polarization if the systematic uncertainty is included. Unpolarized 
$J/\psi$'s were also observed in other fixed-target experiments, using either 
proton or pion beams.

Nuclear effects may also affect the $J/\psi$ polarization, since the $J/\psi$
may collide with other nucleons before it can escape the nucleus. The original
polarization may thus be supressed or smeared out. To eliminate such an effect, 
a hydrogen target is preferable, at the price of smaller production rate however.
It would also be interesting to study the nuclear dependence of $\lambda$
to understand the nuclear effects on the polarization patterns. 
 
The large-$x_F$ behavior is of interest and yet remains mysterious. The 
polarization is changed to longitudinal. Similar behavior was observed also in 
the pion data and a possible explanation is higher-twist effects 
\cite{Van 95}. It is not clear however how this mechanism applies to 
proton-induced data. In this 
study the $x_F$ coverage is only up to 0.95. The statistics are too poor to 
produce sensible results for $x_{F} >0.95$. Even for the $0.85<x_{F}<0.95$ bin it 
is desirable to reduce the statistical uncertainty. It would be interesting to 
know whether $\lambda$ actually drops to $-$1 when $x_{F}$ approaches 1.0 in
pN interactions. This might give us better understandings of the higher-twist
effects.

\appendix
%\noindent {\bf APPENDIX A}

%\noindent {\bf $\cos \theta$ Acceptance in $x_F$ and $p_T$ Bins}\\ 
\chapter{$\cos \theta$ Acceptance in $x_F$ and $p_T$ Bins}

The $\cos \theta$ acceptance in $x_F$ and $p_T$ bins for both magnet settings
is presented in this appendix. In Figures A1 to A6 the $\cos \theta$ acceptance
for the ``SM12=2040'' data is plotted, and from Figure A7 to A12 the 
$\cos \theta$ acceptance for the ``SM12=2800'' data is plotted.  

\begin{figure}
\begin{center}
\mbox{\epsfxsize=5.7in\epsfysize=6.5in\epsffile{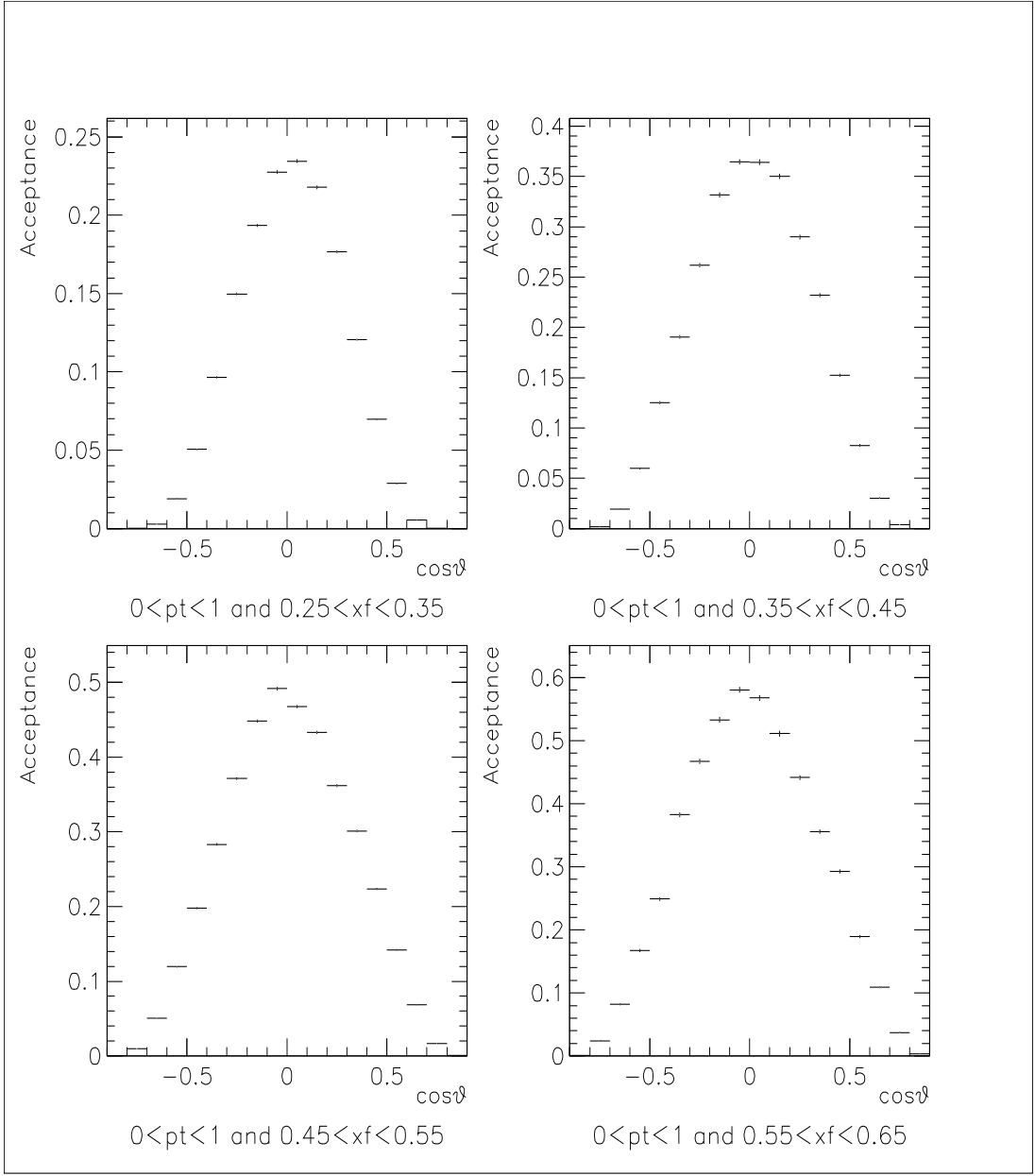}}
\end{center}
\vspace*{+0.25in}
\caption{$\cos \theta$ acceptance in $x_F$ and $p_T$ bins of the 2040Amp data.}
\label{Fig:acpt2040_1}
\end{figure}

\begin{figure}
\begin{center}
\mbox{\epsfxsize=5.7in\epsfysize=6.5in\epsffile{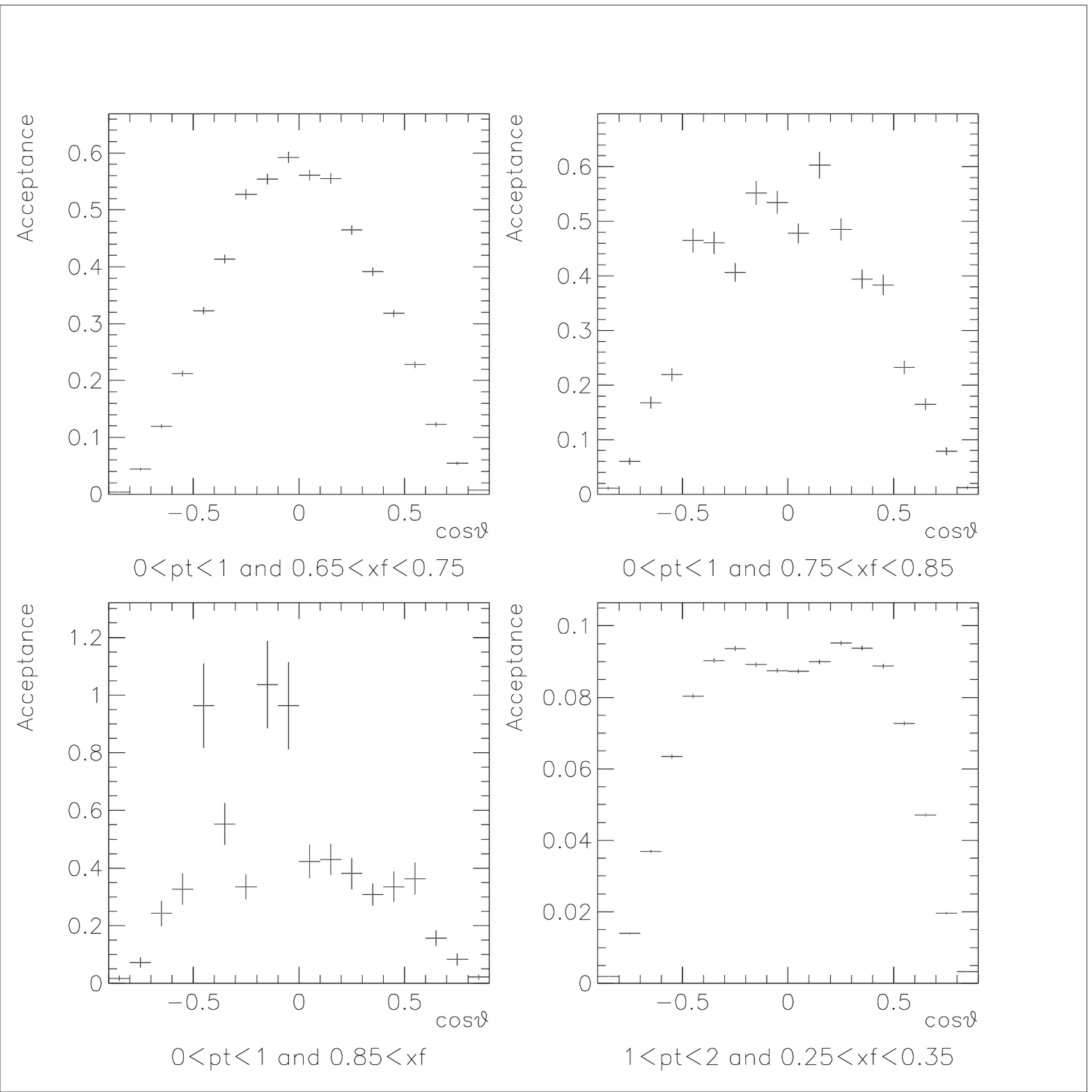}}
\end{center}
\vspace*{+0.25in}
\caption{$\cos \theta$ acceptance in $x_F$ and $p_T$ bins of the 2040Amp data.}
\label{Fig:acpt2040_2}
\end{figure}

\begin{figure}
\begin{center}
\mbox{\epsfxsize=5.7in\epsfysize=6.5in\epsffile{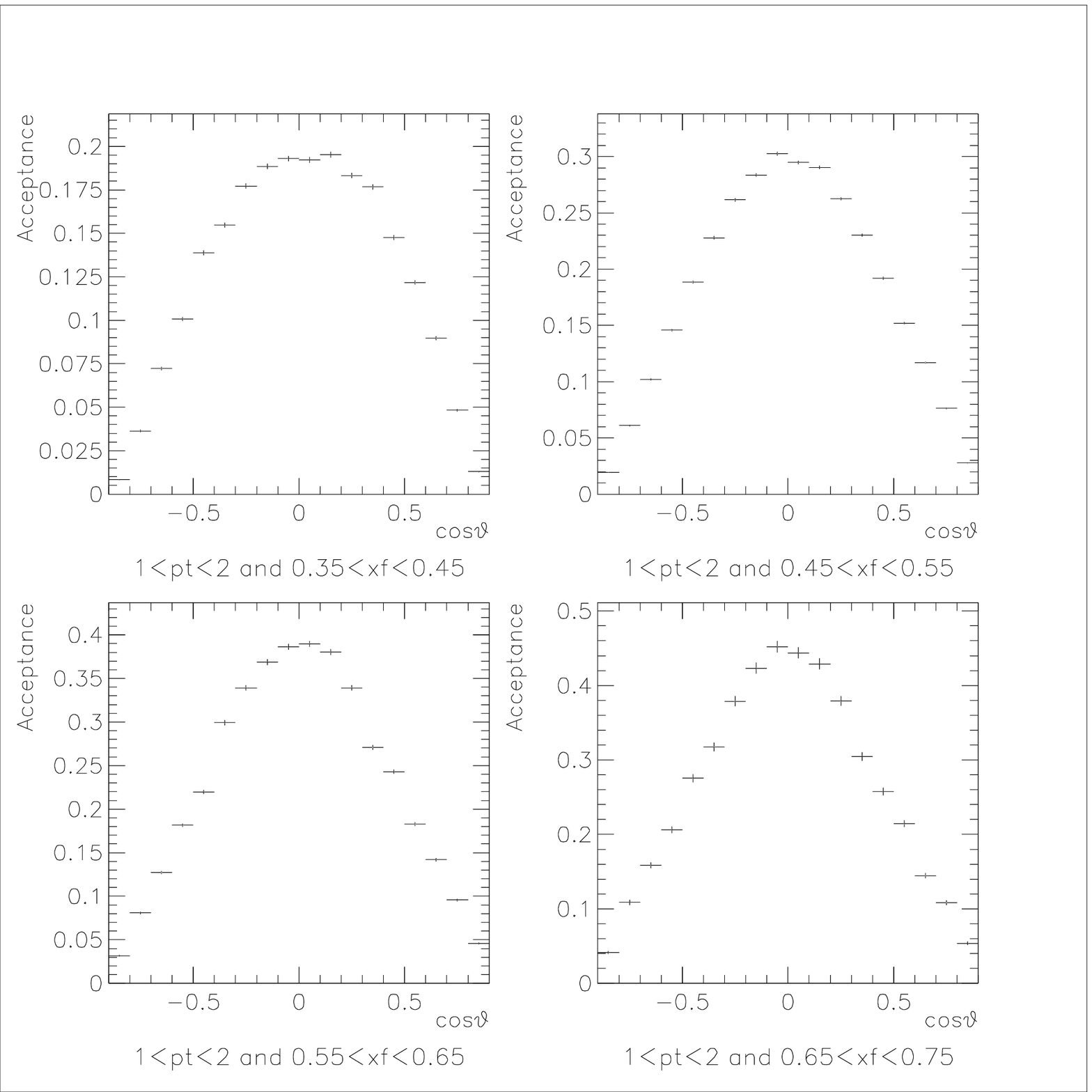}}
\end{center}
\vspace*{+0.25in}
\caption{$\cos \theta$ acceptance in $x_F$ and $p_T$ bins of the 2040Amp data.}
\label{Fig:acpt2040_3}
\end{figure}

\begin{figure}
\begin{center}
\mbox{\epsfxsize=5.7in\epsfysize=6.5in\epsffile{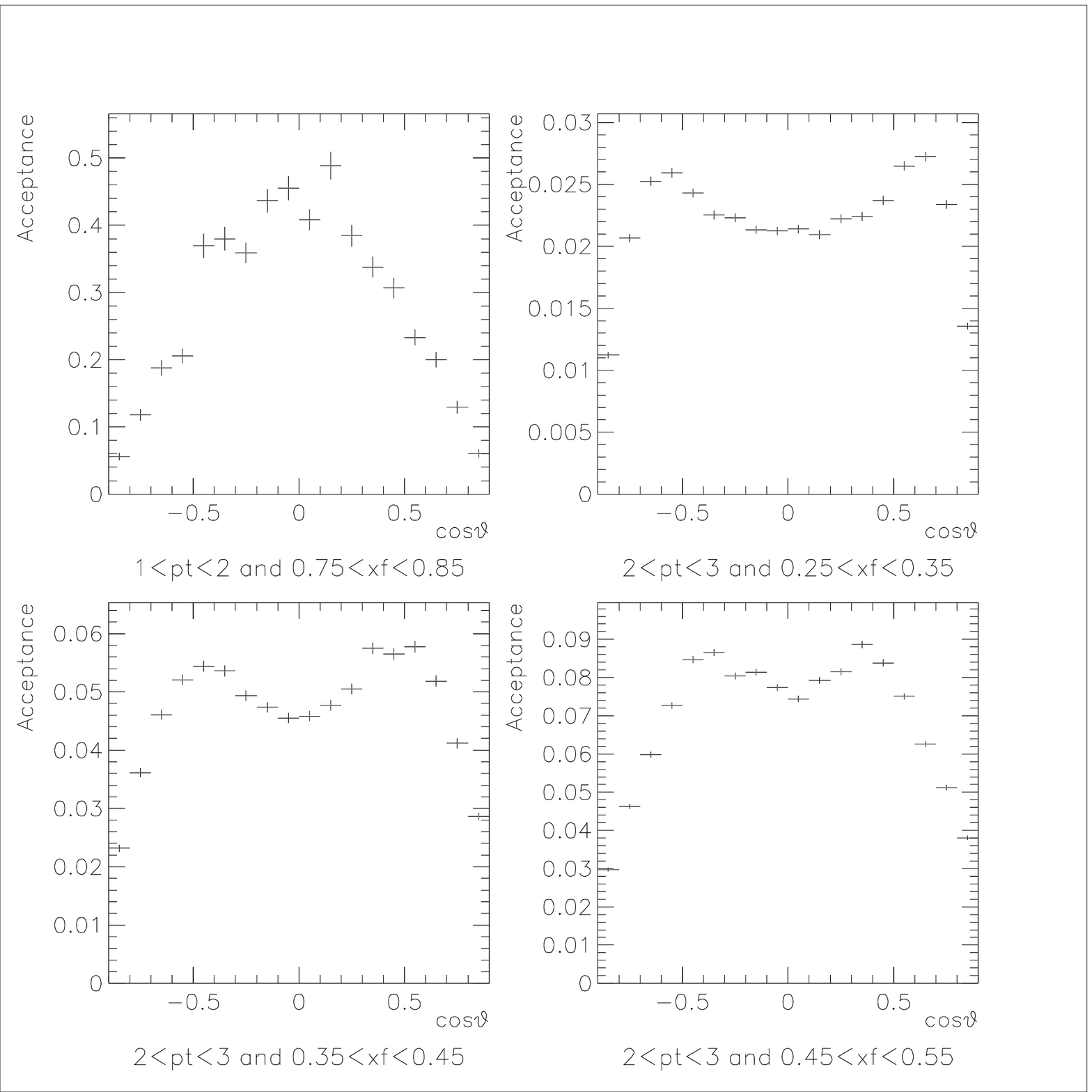}}
\end{center}
\vspace*{+0.25in}
\caption{$\cos \theta$ acceptance in $x_F$ and $p_T$ bins of the 2040Amp data.}
\label{Fig:acpt2040_4}
\end{figure}

\begin{figure}
\begin{center}
\mbox{\epsfxsize=5.7in\epsfysize=6.5in\epsffile{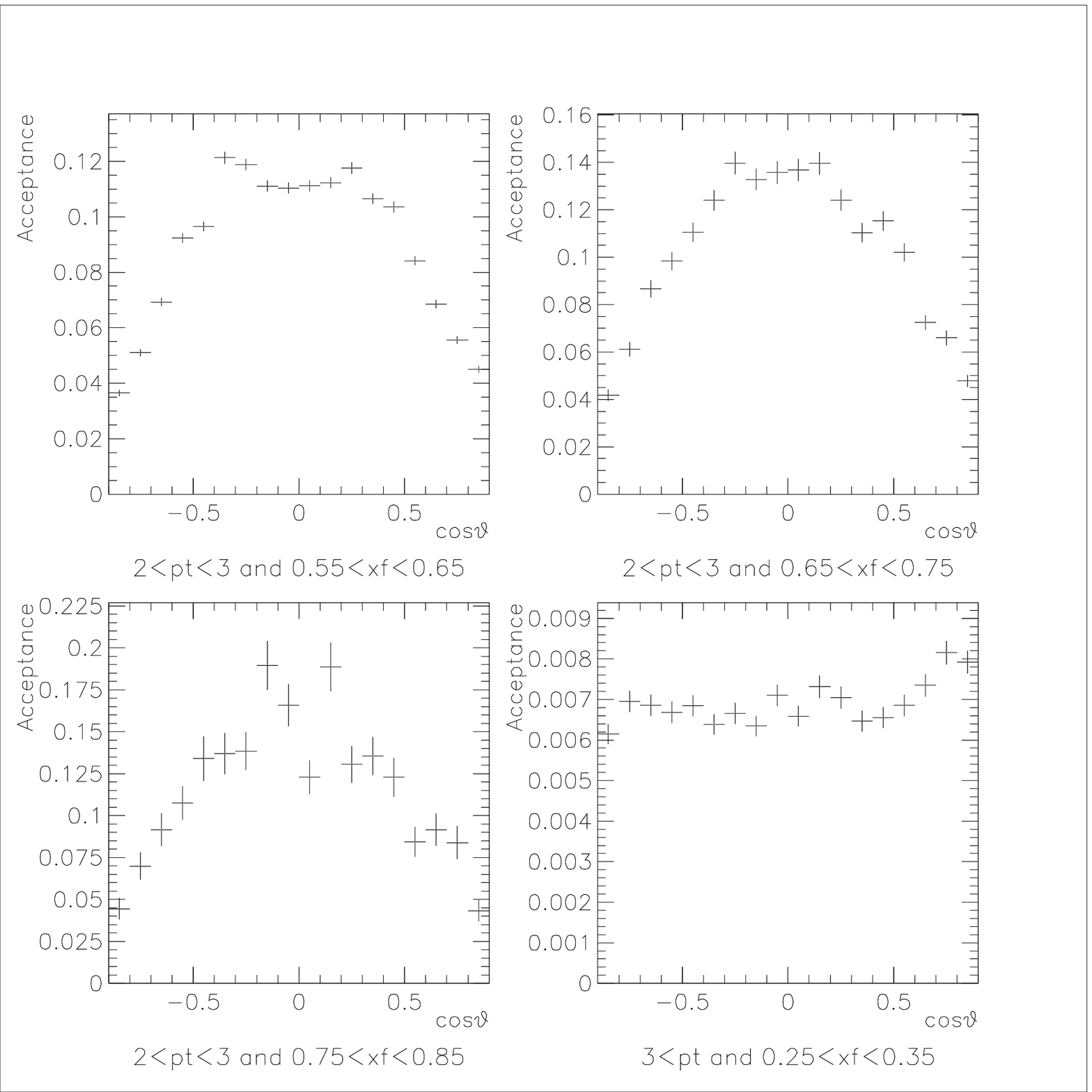}}
\end{center}
\vspace*{+0.25in}
\caption{$\cos \theta$ acceptance in $x_F$ and $p_T$ bins of the 2040Amp data.}
\label{Fig:acpt2040_5}
\end{figure}

\begin{figure}
\begin{center}
\mbox{\epsfxsize=5.7in\epsfysize=6.5in\epsffile{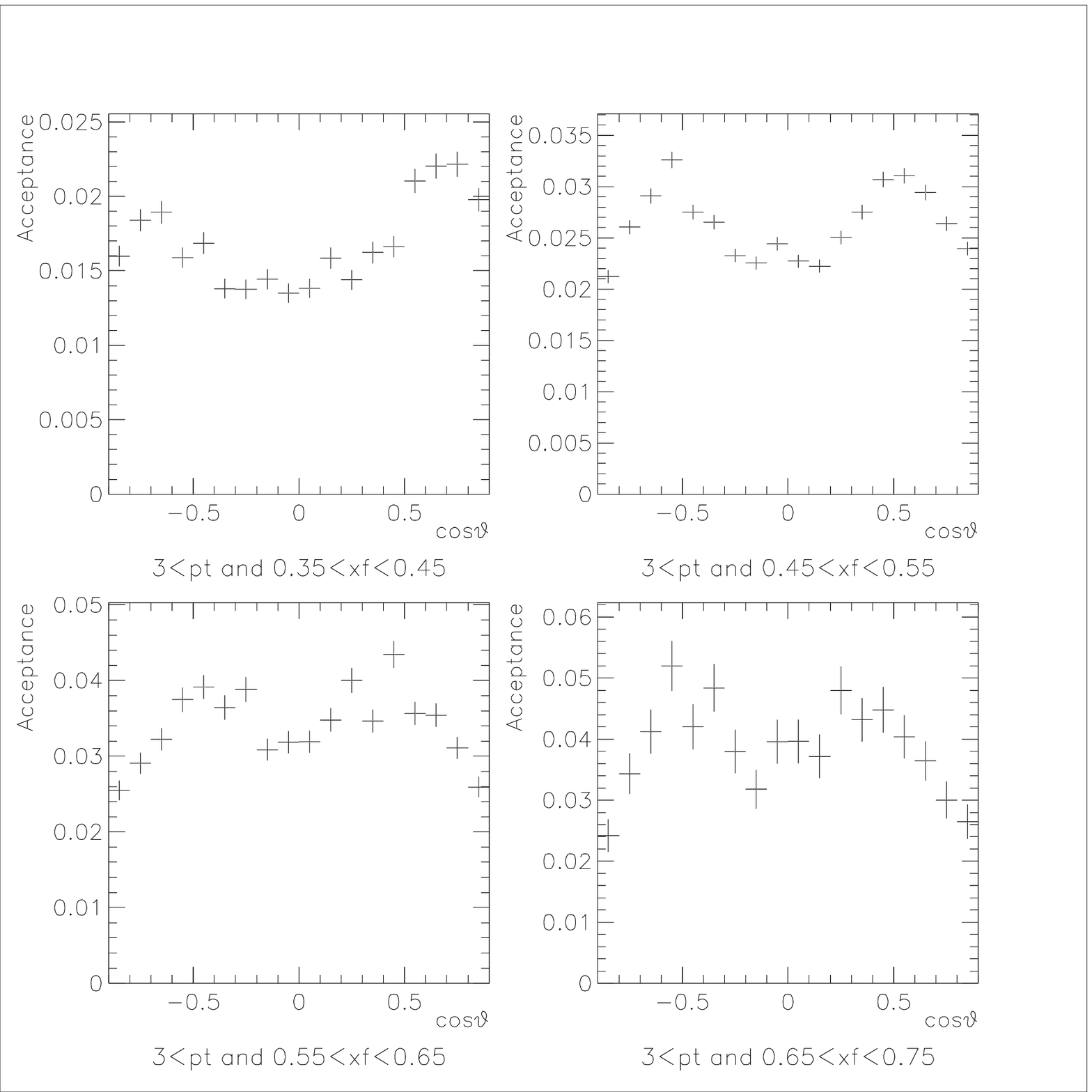}}
\end{center}
\vspace*{+0.25in}
\caption{$\cos \theta$ acceptance in $x_F$ and $p_T$ bins of the 2040Amp data.}
\label{Fig:acpt2040_6}
\end{figure}

\begin{figure}
\begin{center}
\mbox{\epsfxsize=5.7in\epsfysize=6.5in\epsffile{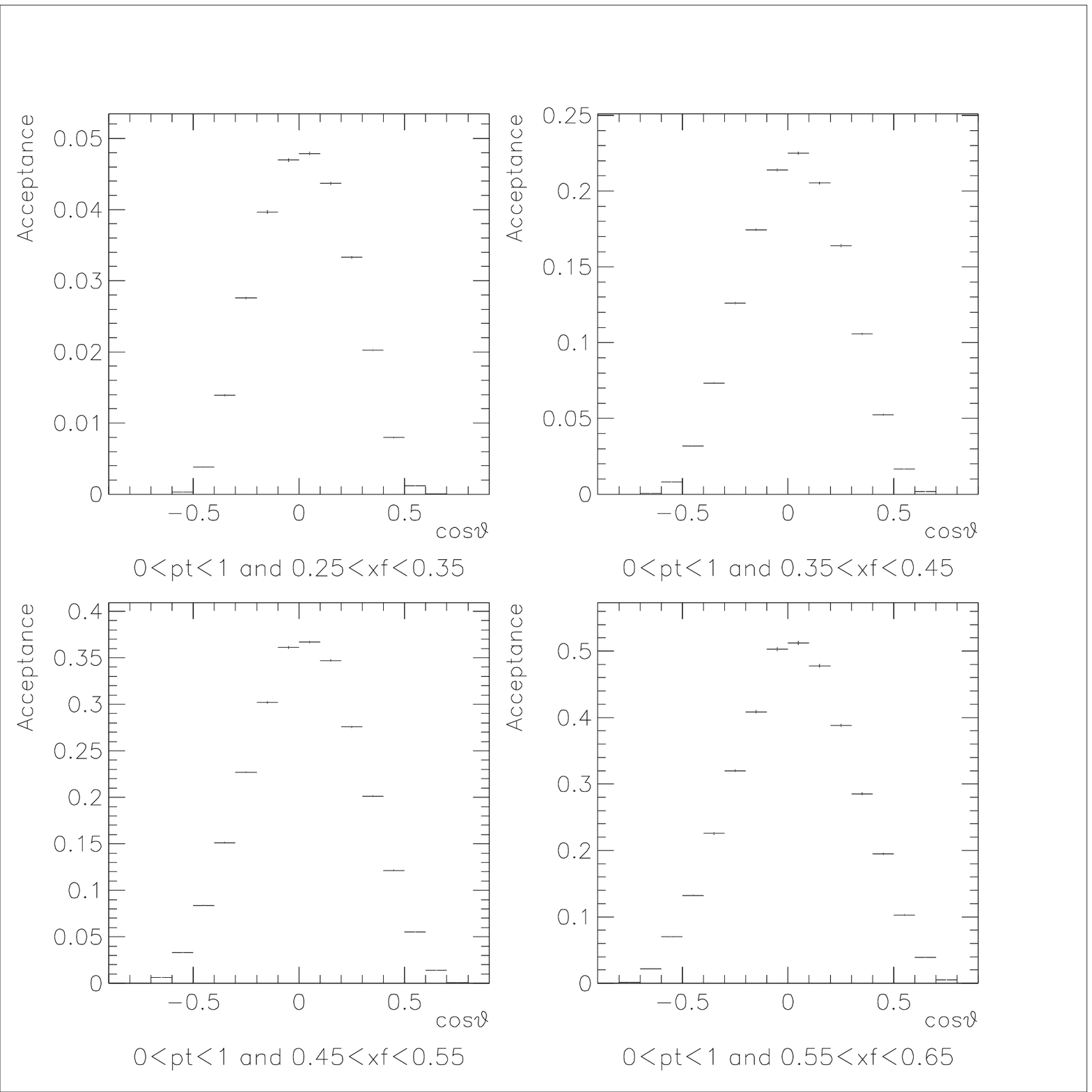}}
\end{center}
\vspace*{+0.25in}
\caption{$\cos \theta$ acceptance in $x_F$ and $p_T$ bins of the 2800Amp data.}
\label{Fig:acpt2800_1}
\end{figure}

\begin{figure}
\begin{center}
\mbox{\epsfxsize=5.7in\epsfysize=6.5in\epsffile{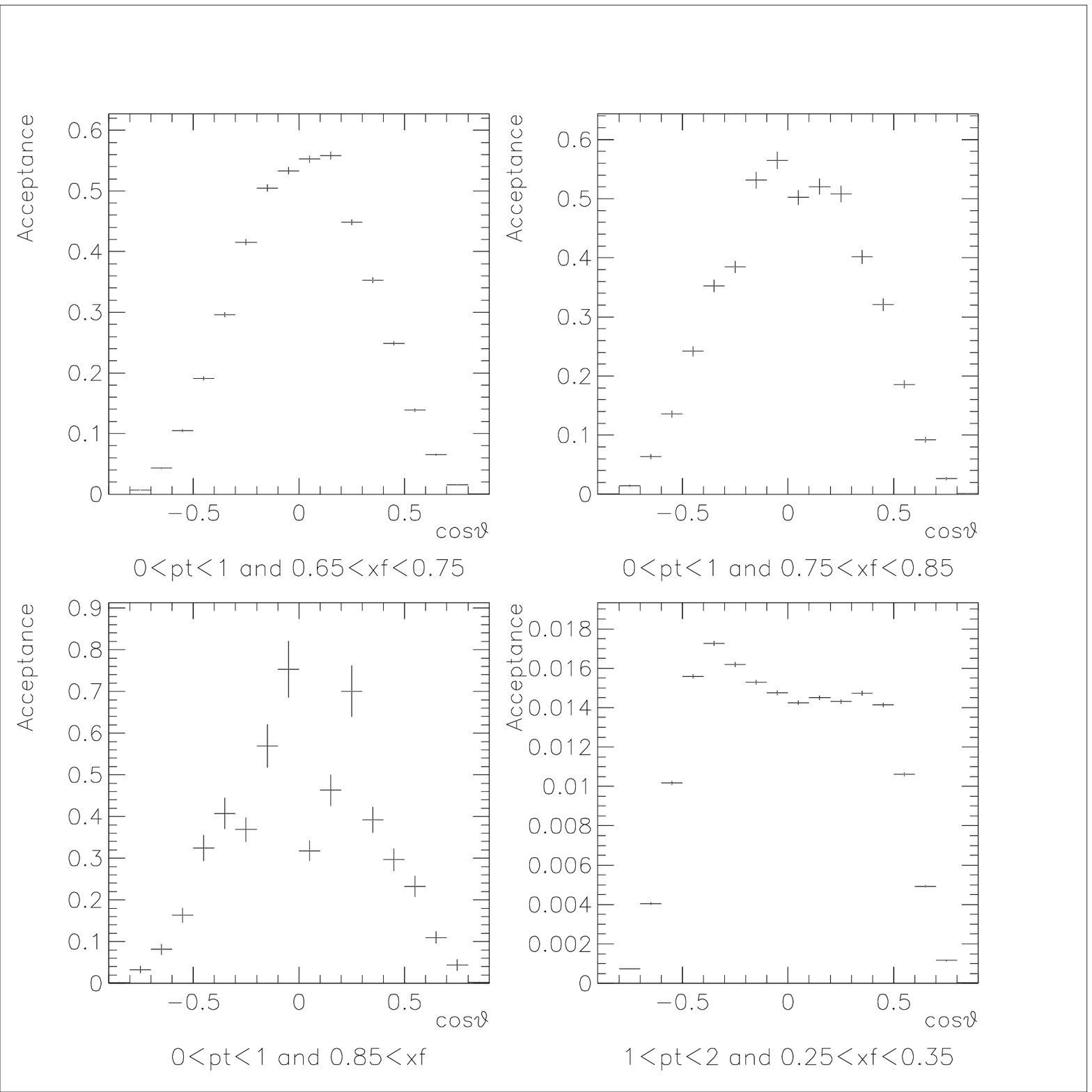}}
\end{center}
\vspace*{+0.25in}
\caption{$\cos \theta$ acceptance in $x_F$ and $p_T$ bins of the 2800Amp data.}
\label{Fig:acpt2800_2}
\end{figure}

\begin{figure}
\begin{center}
\mbox{\epsfxsize=5.7in\epsfysize=6.5in\epsffile{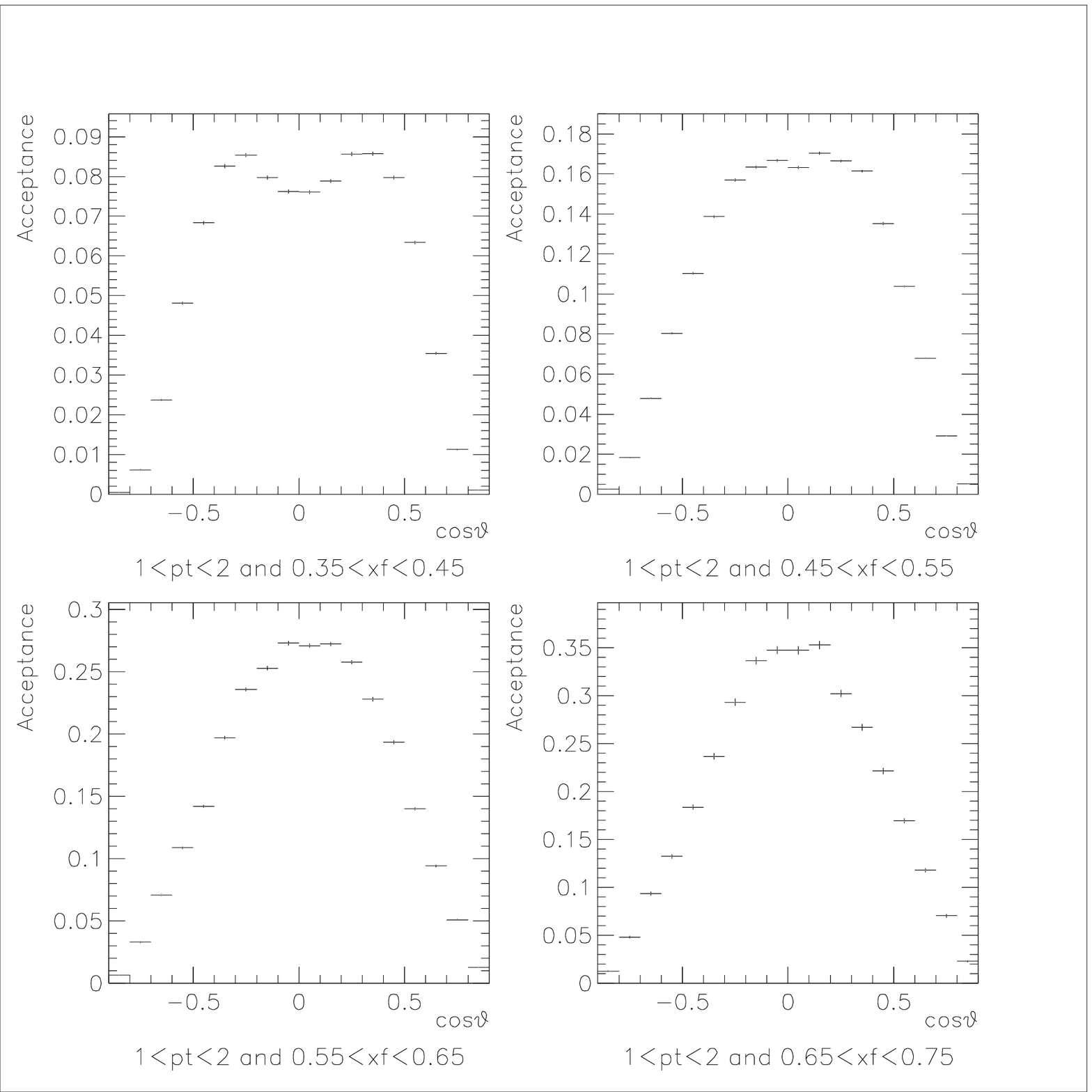}}
\end{center}
\vspace*{+0.25in}
\caption{$\cos \theta$ acceptance in $x_F$ and $p_T$ bins of the 2800Amp data.}
\label{Fig:acpt2800_3}
\end{figure}

\begin{figure}
\begin{center}
\mbox{\epsfxsize=5.7in\epsfysize=6.5in\epsffile{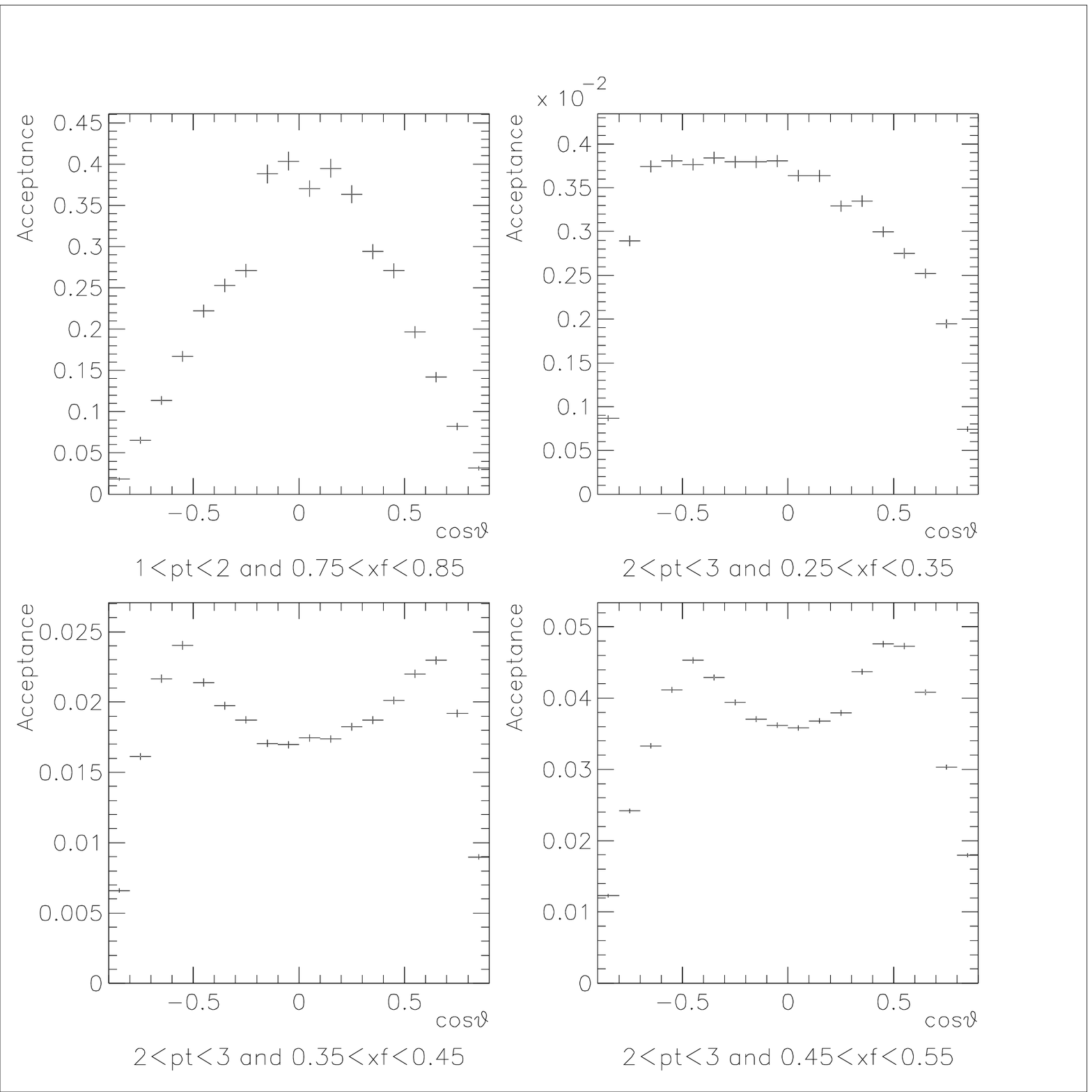}}
\end{center}
\vspace*{+0.25in}
\caption{$\cos \theta$ acceptance in $x_F$ and $p_T$ bins of the 2800Amp data.}
\label{Fig:acpt2800_4}
\end{figure}

\begin{figure}
\begin{center}
\mbox{\epsfxsize=5.7in\epsfysize=6.5in\epsffile{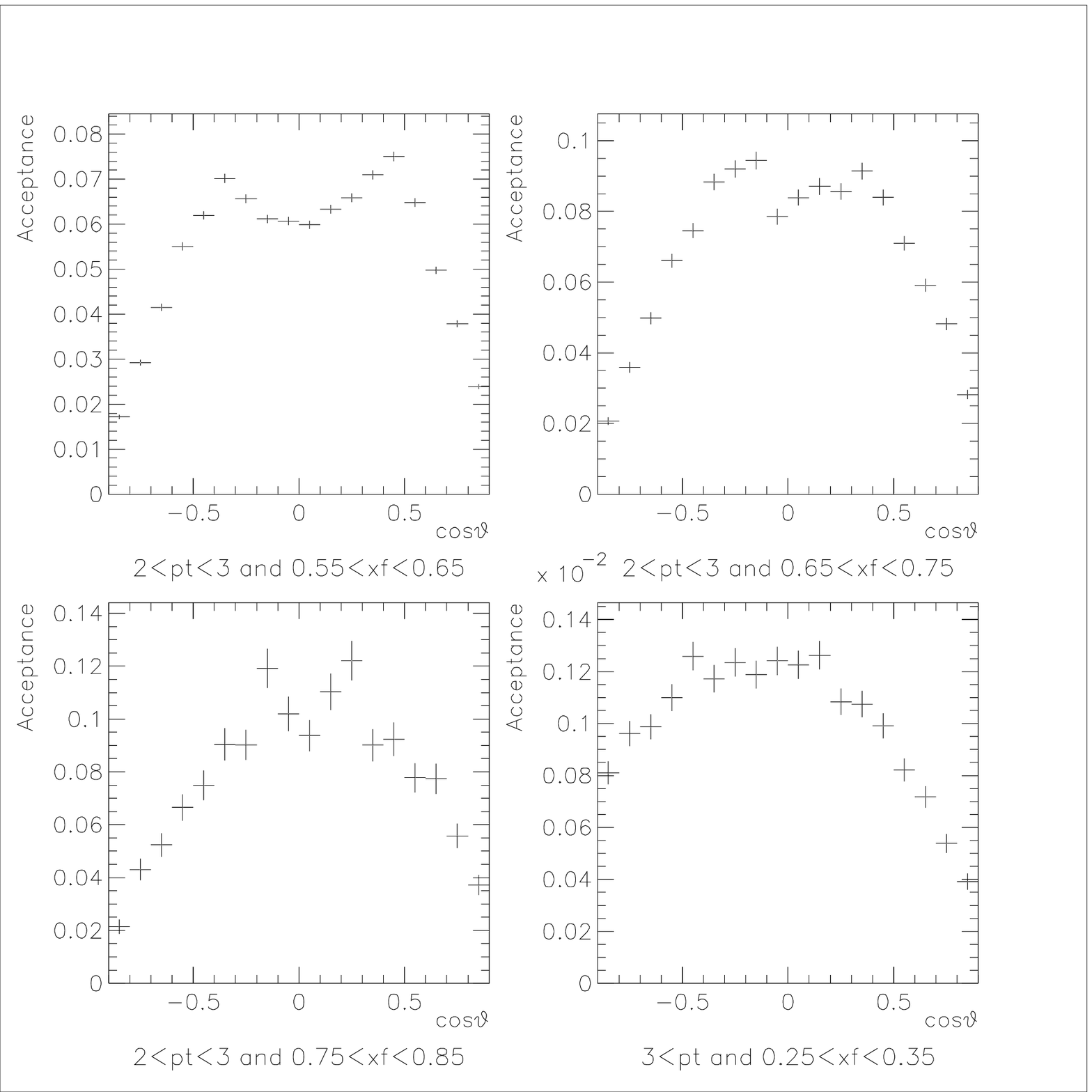}}
\end{center}
\vspace*{+0.25in}
\caption{$\cos \theta$ acceptance in $x_F$ and $p_T$ bins of the 2800Amp data.}
\label{Fig:acpt2800_5}
\end{figure}

\begin{figure}
\begin{center}
\mbox{\epsfxsize=5.7in\epsfysize=6.5in\epsffile{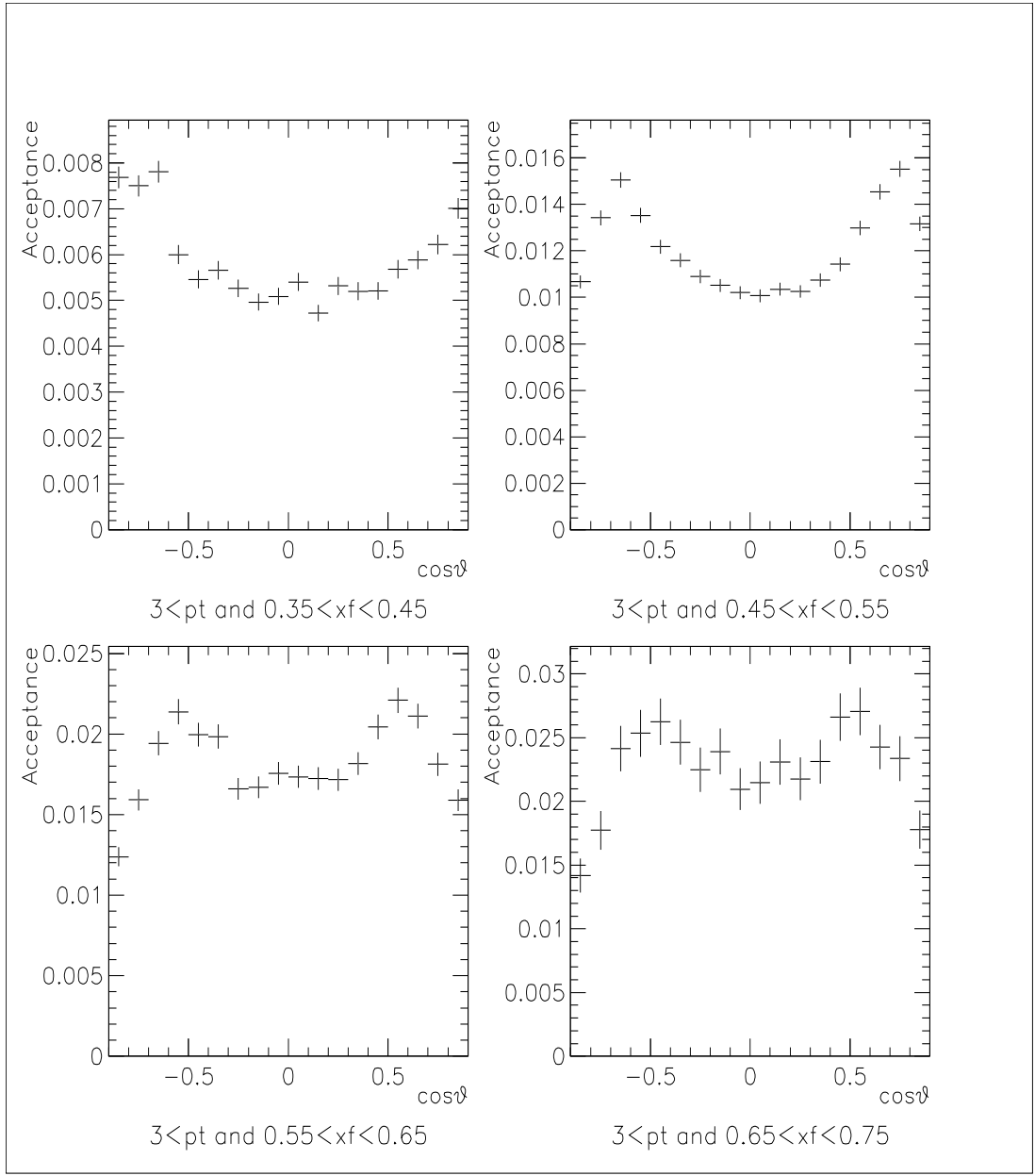}}
\end{center}
\vspace*{+0.25in}
\caption{$\cos \theta$ acceptance in $x_F$ and $p_T$ bins of the 2800Amp data.}
\label{Fig:acpt2800_6}
\end{figure}

\chapter{Fitting the $J/\psi$ Peaks}
%\noindent {\bf APPENDIX B}

%\noindent {\bf Fitting the $J/\psi$ Peaks}\\

In this appendix the fitting of the $J/\psi$ peaks is presented. The di-muon
mass spectrum from each bin of $\cos \theta$, $p_T$, and $x_F$, are fitted to
a Gaussian plus a background function with $J/\psi$'s fitted to a Gaussian 
shape. The count of $J/\psi$'s in each bin was then calculated according to 
the output parameters of the Gaussian fit. Figure B1 to B42 show the fittings
from the ``SM12 = 2040 Amp'' data set, and Figure B43 to B84 show the fittings
from the ``SM12 = 2800 Amp'' data set.

\begin{em}
These plots have been removed from this version of the thesis.  A
complete copy of this Appendix is available from the author, or by
contacting Department of Physics, New Mexico State University, Box 3D,
Las Cruces, NM 88003.
\end{em}

\newpage
\setcounter{page}{200}

\chapter{Fitting the $\lambda$'s}

%\noindent {\bf APPENDIX C}

%\noindent {\bf Fitting the $\lambda$'s}\\

After the counts of the $J/\psi$ peaks for each bin of $\cos \theta$, $x_F$,
and $p_T$, were determined, the number of counts was plotted versus $\cos \theta$.
This gave us the accepted $\cos \theta$ distributions for $J/\psi$ in bins of 
$x_F$ and $p_T$. Then those accepted $\cos \theta$ distributions were divided
by the acceptance curves shown in Appendix A to obtain the true distributions.
The corrected $\cos \theta$ distributions were then fitted to $1 + \lambda \cos^{2}
\theta$ times an arbitrary normalization constant. Those distributions and the 
fits are presented in this appendix. In Figures C1 to C6 are the curves for
the ``SM12=2040'' data set, and from Figures C7 to C12 are the curves for the
``SM12=2800'' data set.

\begin{figure}
\begin{center}
\mbox{\epsfxsize=5.7in\epsfysize=6.5in\epsffile{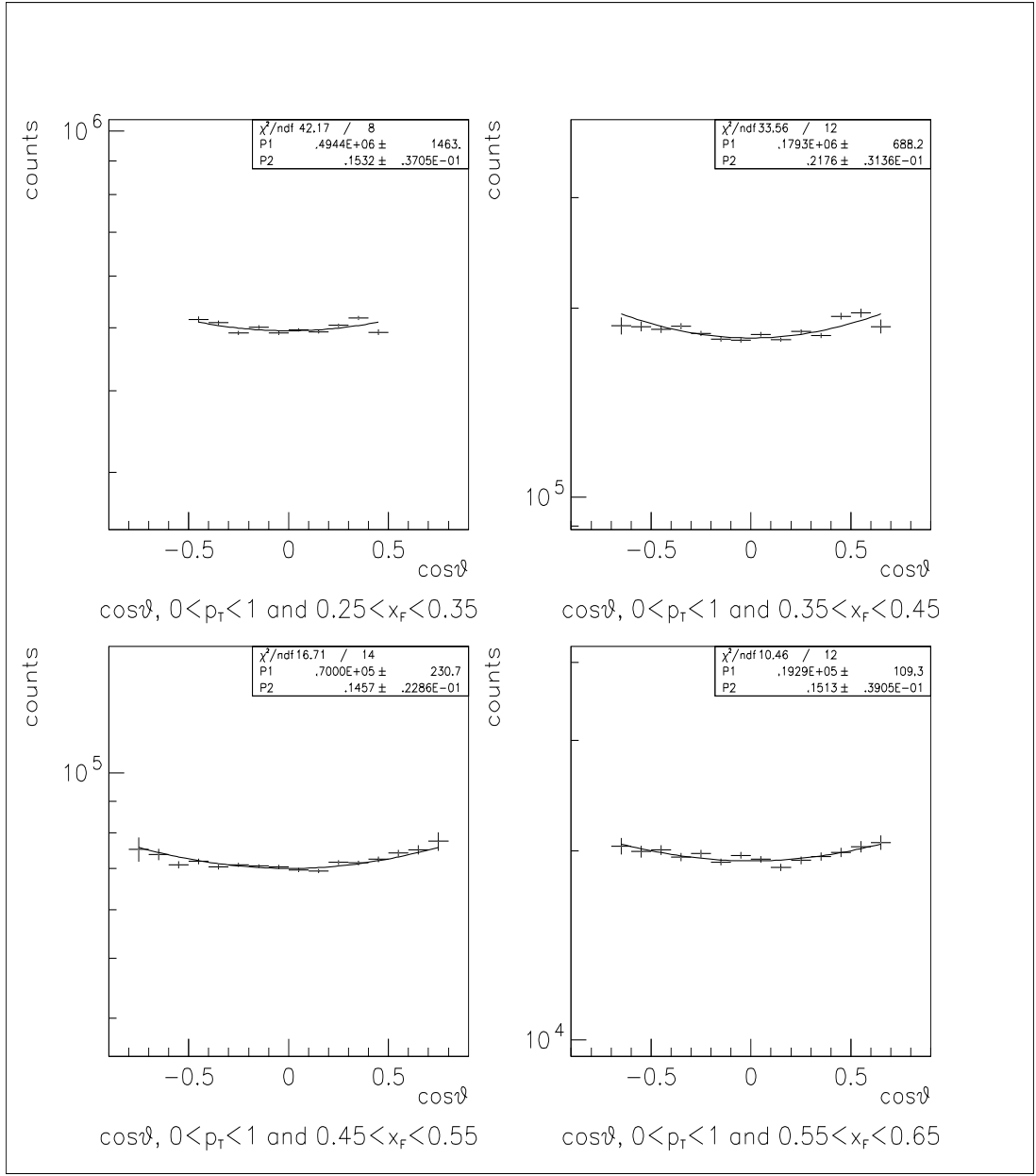}}
\end{center}
\vspace*{+0.25in}
\caption[The corrected $\cos \theta$ distributions and the polarization parameter $\lambda$]
{The corrected $\cos \theta$ distributions and the polarization 
parameter $\lambda$. Upper left: $0<p_{T}<1$ and $0.25<x_{F}<0.35$. Upper right:
$0<p_{T}<1$ and $0.35<x_{F}<0.45$. Lower left: $0<p_{T}<1$ and 
$0.45<x_{F}<0.55$. Lower right: $0<p_{T}<1$ and $0.55<x_{F}<0.65$. SM12=2040.}
\label{Fig:lfit2040_1}
\end{figure}

\begin{figure}
\begin{center}
\mbox{\epsfxsize=5.7in\epsfysize=6.5in\epsffile{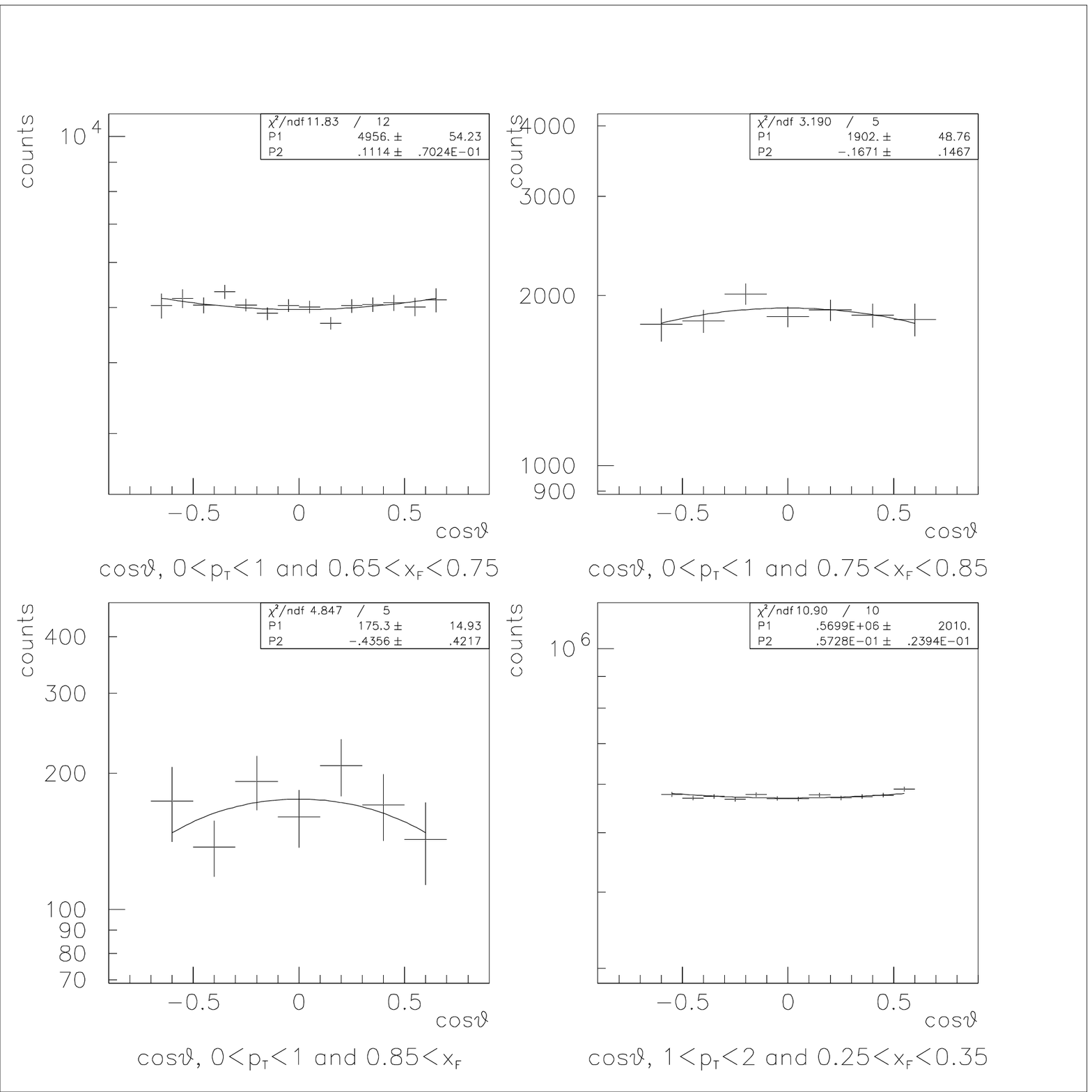}}
\end{center}
\vspace*{+0.25in}
\caption[The corrected $\cos \theta$ distributions and the polarization parameter $\lambda$]
{The corrected $\cos \theta$ distributions and the polarization 
parameter $\lambda$. Upper left: $0<p_{T}<1$ and $0.65<x_{F}<0.75$. Upper right:
$0<p_{T}<1$ and $0.75<x_{F}<0.85$. Lower left: $0<p_{T}<1$ and 
$0.85<x_{F}$. Lower right: $1<p_{T}<2$ and $0.25<x_{F}<0.35$. SM12=2040.}
\label{Fig:lfit2040_2}
\end{figure}

\begin{figure}
\begin{center}
\mbox{\epsfxsize=5.7in\epsfysize=6.5in\epsffile{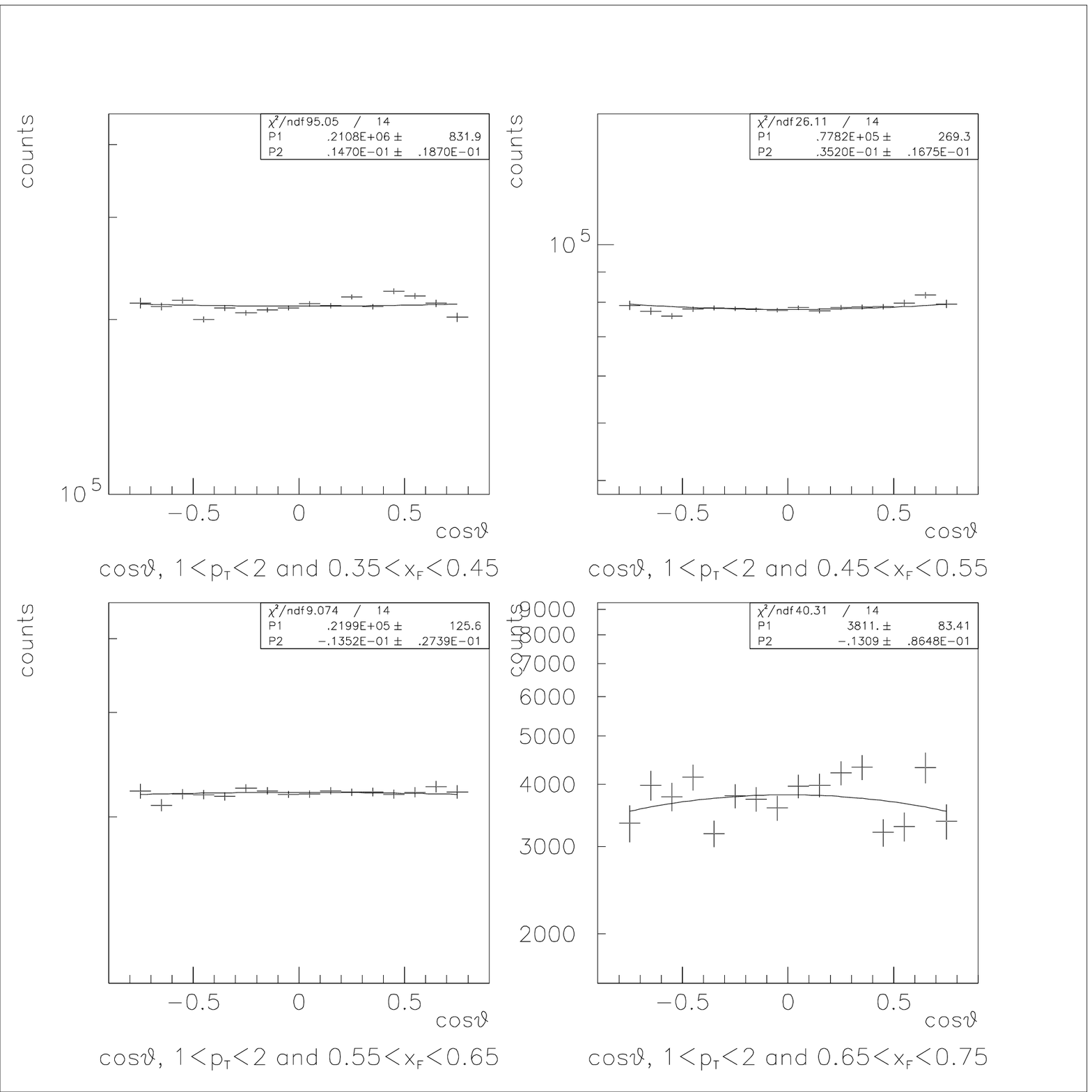}}
\end{center}
\vspace*{+0.25in}
\caption[The corrected $\cos \theta$ distributions and the polarization parameter $\lambda$]
{The corrected $\cos \theta$ distributions and the polarization 
parameter $\lambda$. Upper left: $1<p_{T}<2$ and $0.35<x_{F}<0.45$. Upper right:
$1<p_{T}<2$ and $0.45<x_{F}<0.55$. Lower left: $1<p_{T}<2$ and 
$0.55<x_{F}<0.65$. Lower right: $1<p_{T}<2$ and $0.65<x_{F}<0.75$. SM12=2040.}
\label{Fig:lfit2040_3}
\end{figure}

\begin{figure}
\begin{center}
\mbox{\epsfxsize=5.7in\epsfysize=6.5in\epsffile{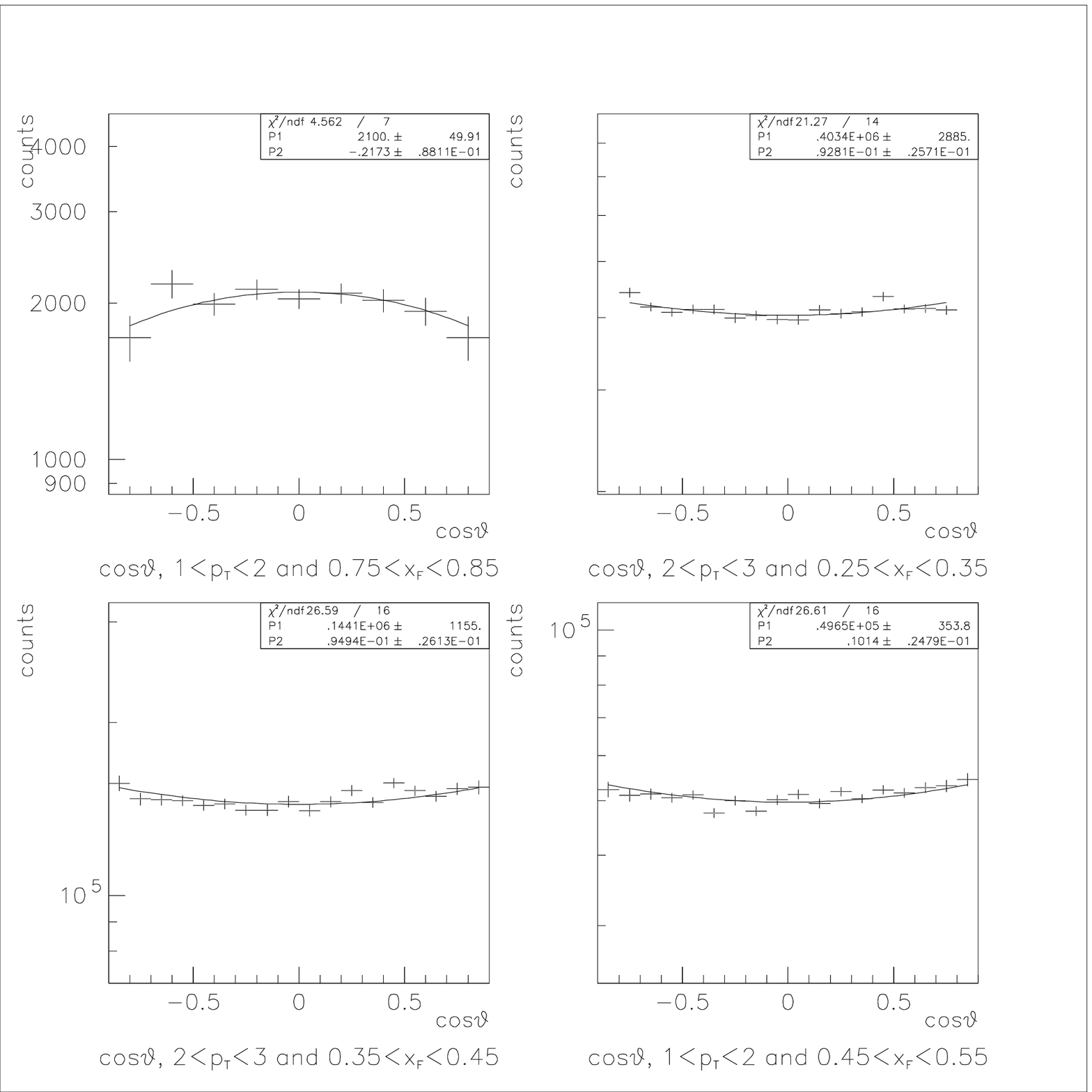}}
\end{center}
\vspace*{+0.25in}
\caption[The corrected $\cos \theta$ distributions and the polarization parameter $\lambda$]
{The corrected $\cos \theta$ distributions and the polarization 
parameter $\lambda$. Upper left: $1<p_{T}<2$ and $0.75<x_{F}<0.85$. Upper right:
$2<p_{T}<3$ and $0.25<x_{F}<0.35$. Lower left: $2<p_{T}<3$ and 
$0.35<x_{F}<0.45$. Lower right: $2<p_{T}<3$ and $0.45<x_{F}<0.55$. SM12=2040.}
\label{Fig:lfit2040_4}
\end{figure}

\begin{figure}
\begin{center}
\mbox{\epsfxsize=5.7in\epsfysize=6.5in\epsffile{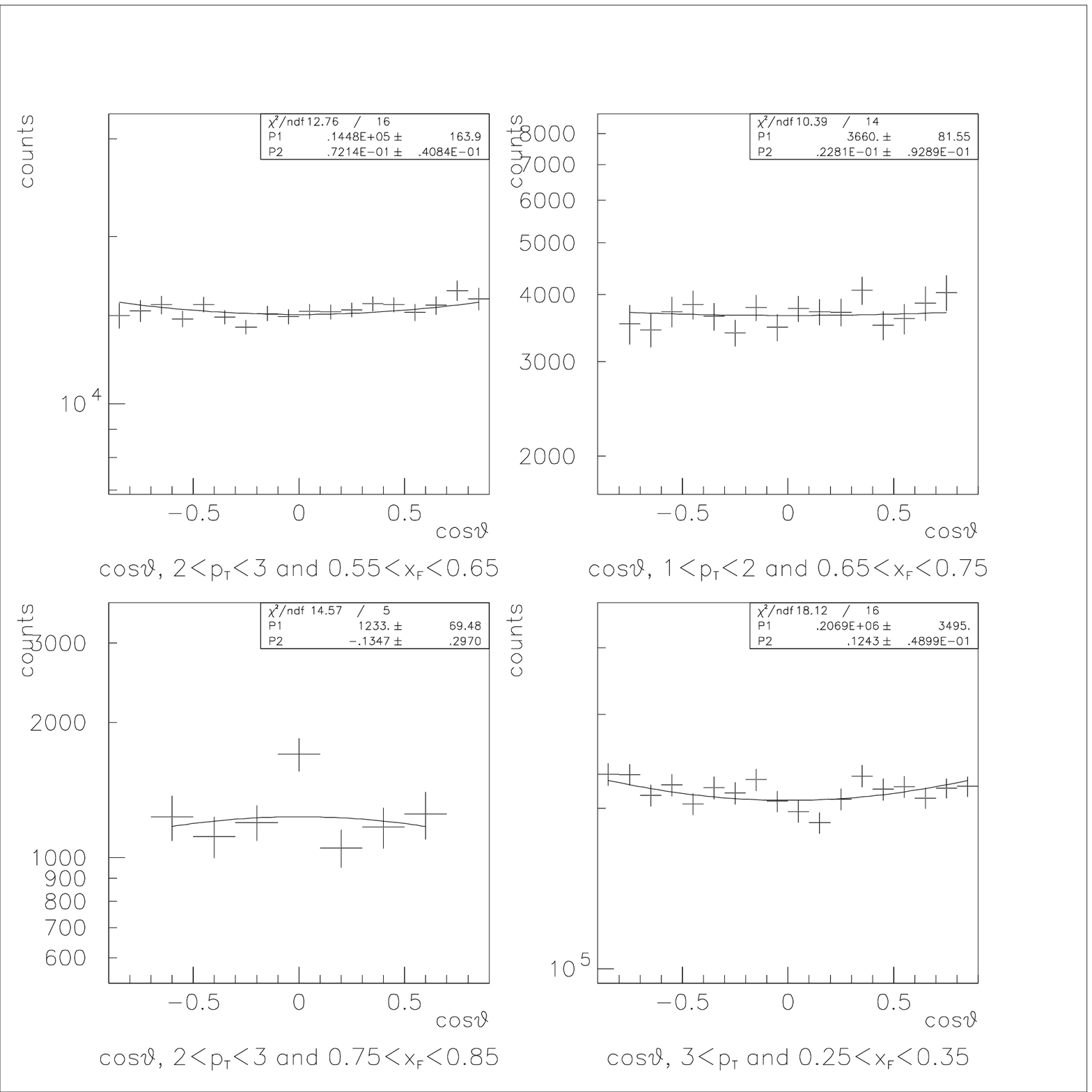}}
\end{center}
\vspace*{+0.25in}
\caption[The corrected $\cos \theta$ distributions and the polarization parameter $\lambda$]
{The corrected $\cos \theta$ distributions and the polarization 
parameter $\lambda$. Upper left: $2<p_{T}<3$ and $0.55<x_{F}<0.65$. Upper right:
$2<p_{T}<3$ and $0.65<x_{F}<0.75$. Lower left: $2<p_{T}<3$ and 
$0.75<x_{F}<0.85$. Lower right: $3<p_{T}$ and $0.25<x_{F}<0.35$. SM12=2040.}
\label{Fig:lfit2040_5}
\end{figure}

\begin{figure}
\begin{center}
\mbox{\epsfxsize=5.7in\epsfysize=6.5in\epsffile{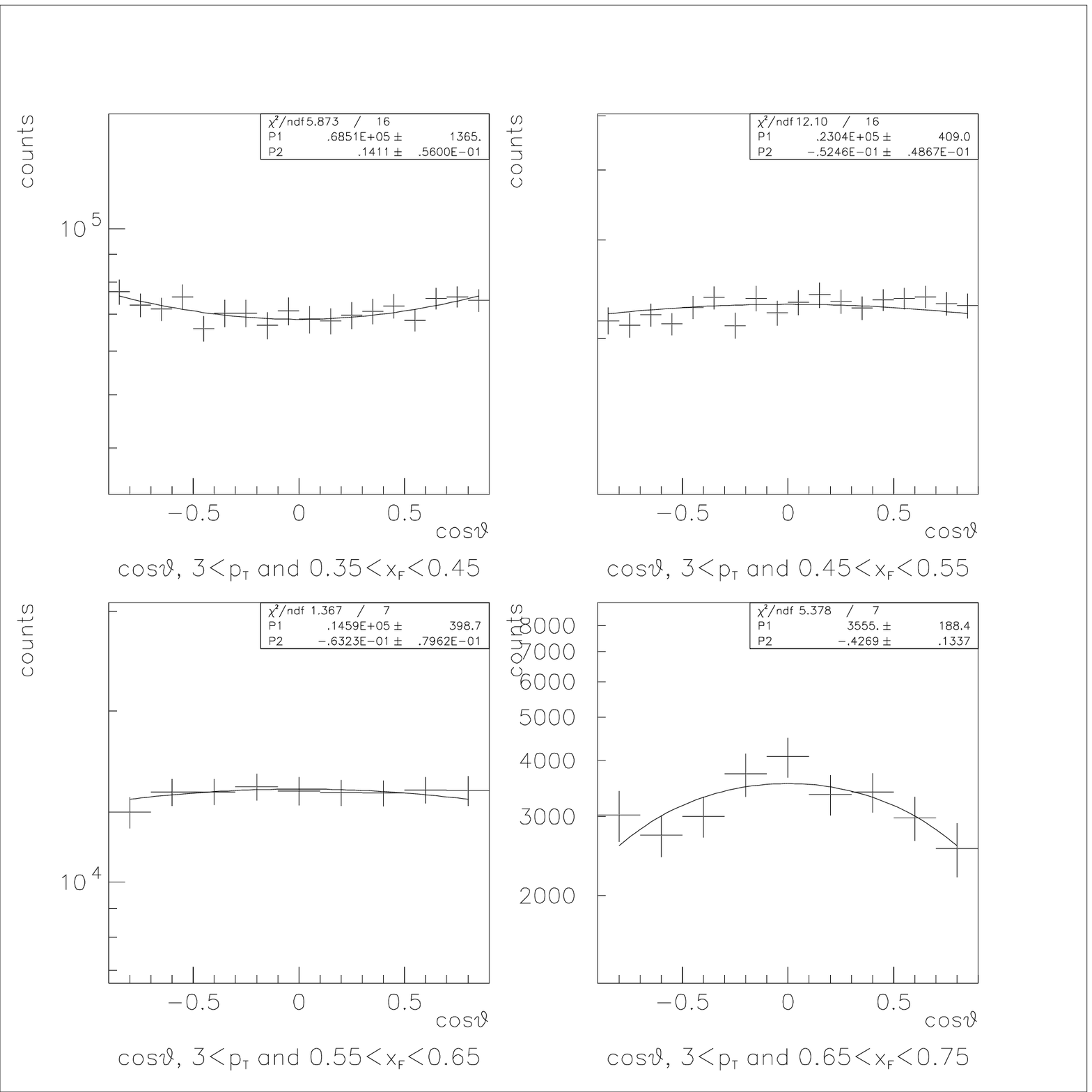}}
\end{center}
\vspace*{+0.25in}
\caption[The corrected $\cos \theta$ distributions and the polarization parameter $\lambda$]
{The corrected $\cos \theta$ distributions and the polarization 
parameter $\lambda$. Upper left: $3<p_{T}$ and $0.35<x_{F}<0.45$. Upper right:
$3<p_{T}$ and $0.45<x_{F}<0.55$. Lower left: $3<p_{T}$ and 
$0.55<x_{F}<0.65$. Lower right: $3<p_{T}$ and $0.65<x_{F}<0.75$. SM12=2040.}
\label{Fig:lfit2040_6}
\end{figure}

\begin{figure}
\begin{center}
\mbox{\epsfxsize=5.7in\epsfysize=6.5in\epsffile{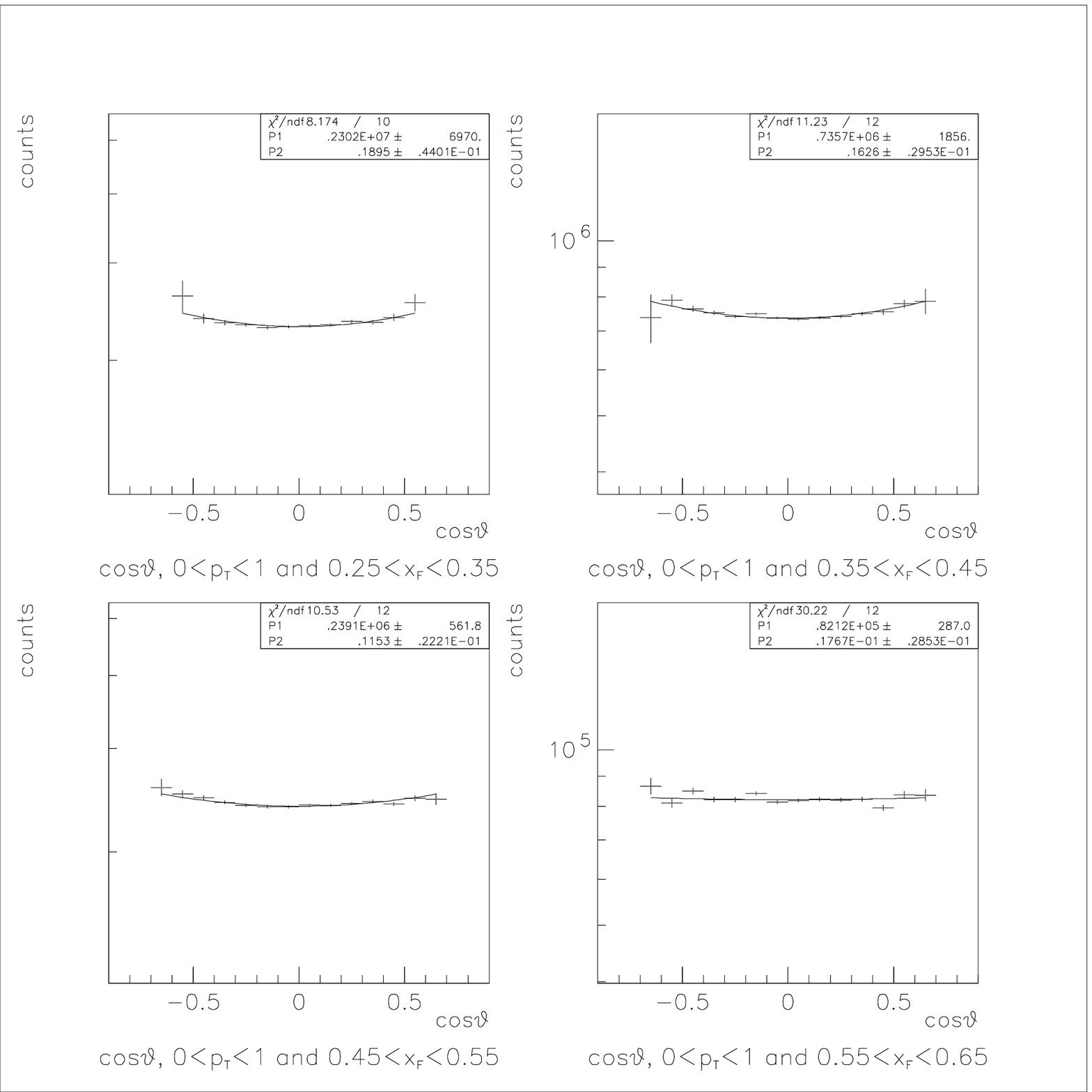}}
\end{center}
\vspace*{+0.25in}
\caption[The corrected $\cos \theta$ distributions and the polarization parameter $\lambda$]
{The corrected $\cos \theta$ distributions and the polarization 
parameter $\lambda$. Upper left: $0<p_{T}<1$ and $0.25<x_{F}<0.35$. Upper right:
$0<p_{T}<1$ and $0.35<x_{F}<0.45$. Lower left: $0<p_{T}<1$ and 
$0.45<x_{F}<0.55$. Lower right: $0<p_{T}<1$ and $0.55<x_{F}<0.65$. SM12=2800.}
\label{Fig:lfit2800_1}
\end{figure}

\begin{figure}
\begin{center}
\mbox{\epsfxsize=5.7in\epsfysize=6.5in\epsffile{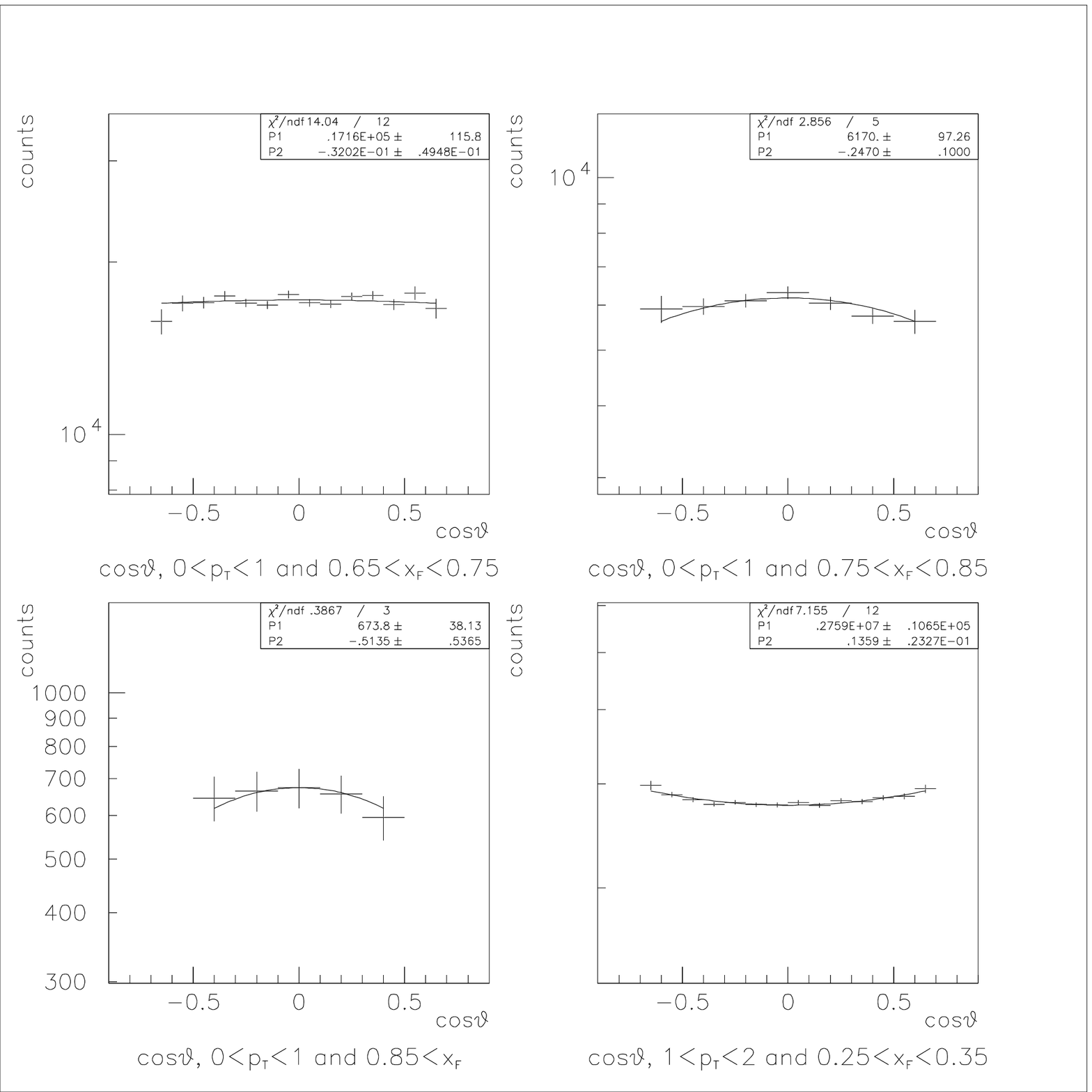}}
\end{center}
\vspace*{+0.25in}
\caption[The corrected $\cos \theta$ distributions and the polarization parameter $\lambda$]
{The corrected $\cos \theta$ distributions and the polarization 
parameter $\lambda$. Upper left: $0<p_{T}<1$ and $0.65<x_{F}<0.75$. Upper right:
$0<p_{T}<1$ and $0.75<x_{F}<0.85$. Lower left: $0<p_{T}<1$ and 
$0.85<x_{F}$. Lower right: $1<p_{T}<2$ and $0.25<x_{F}<0.35$. SM12=2800.}
\label{Fig:lfit2800_2}
\end{figure}

\begin{figure}
\begin{center}
\mbox{\epsfxsize=5.7in\epsfysize=6.5in\epsffile{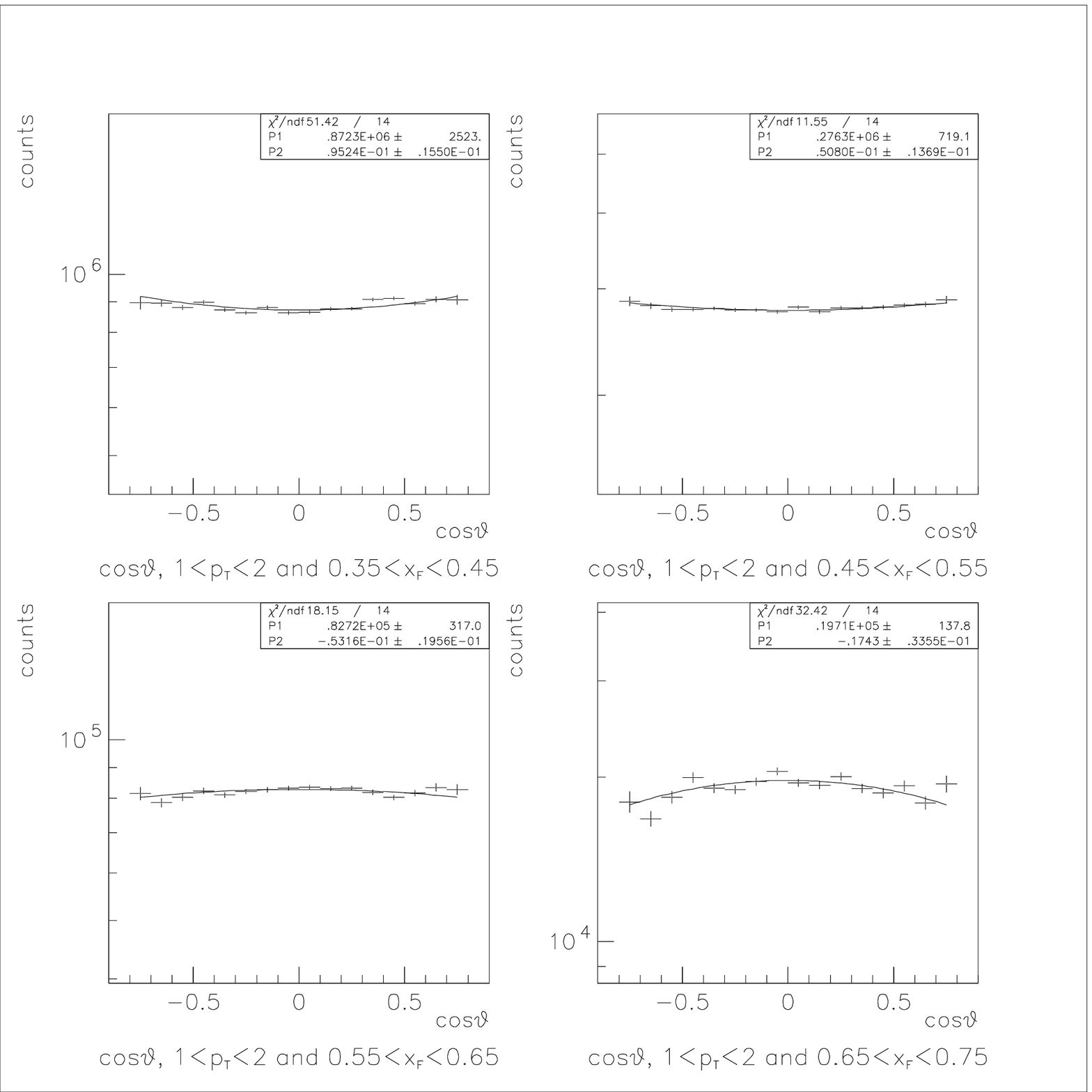}}
\end{center}
\vspace*{+0.25in}
\caption[The corrected $\cos \theta$ distributions and the polarization parameter $\lambda$]
{The corrected $\cos \theta$ distributions and the polarization 
parameter $\lambda$. Upper left: $1<p_{T}<2$ and $0.35<x_{F}<0.45$. Upper right:
$1<p_{T}<2$ and $0.45<x_{F}<0.55$. Lower left: $1<p_{T}<2$ and 
$0.55<x_{F}<0.65$. Lower right: $1<p_{T}<2$ and $0.65<x_{F}<0.75$. SM12=2800.}
\label{Fig:lfit2800_3}
\end{figure}

\begin{figure}
\begin{center}
\mbox{\epsfxsize=5.7in\epsfysize=6.5in\epsffile{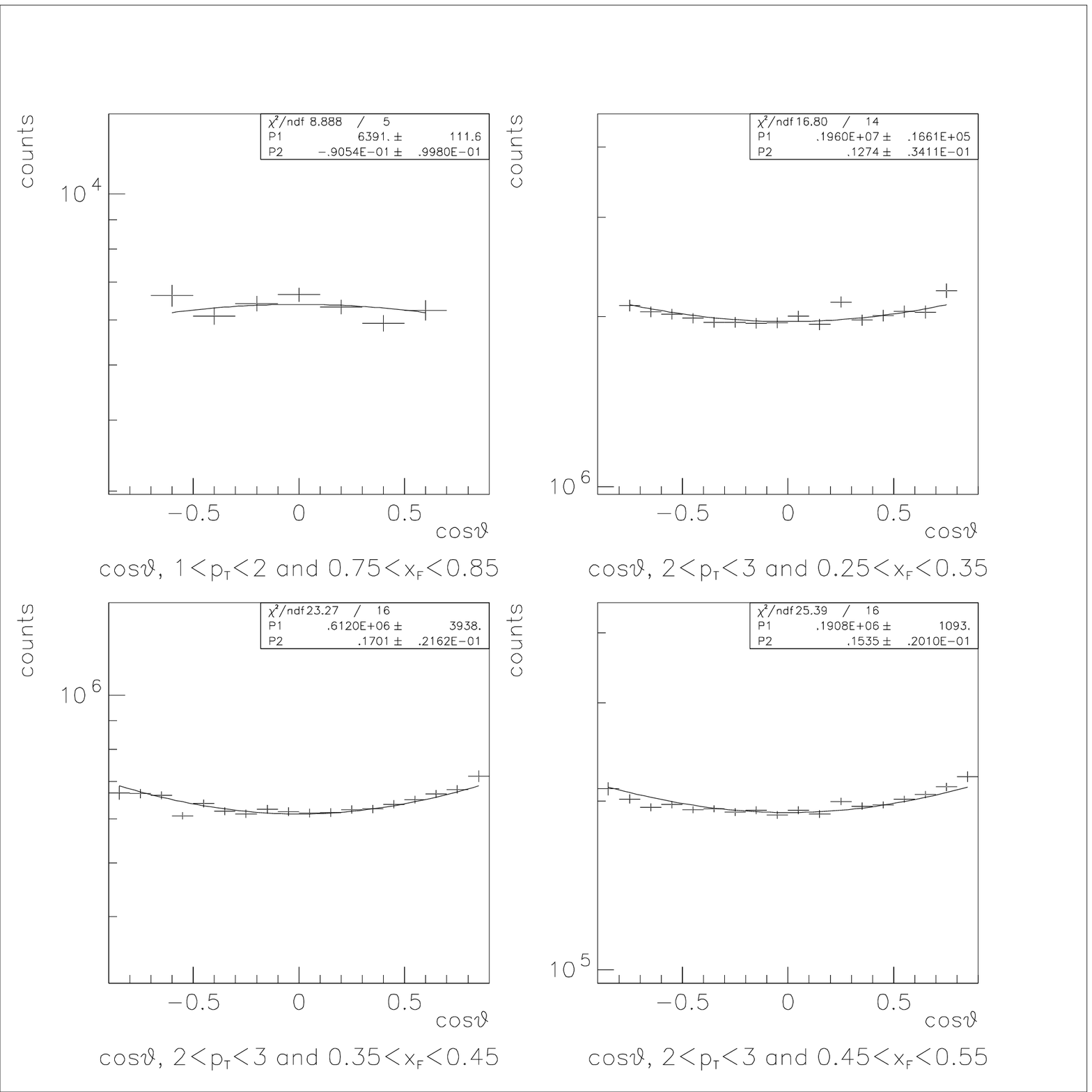}}
\end{center}
\vspace*{+0.25in}
\caption[The corrected $\cos \theta$ distributions and the polarization parameter $\lambda$]
{The corrected $\cos \theta$ distributions and the polarization 
parameter $\lambda$. Upper left: $1<p_{T}<2$ and $0.75<x_{F}<0.85$. Upper right:
$2<p_{T}<3$ and $0.25<x_{F}<0.35$. Lower left: $2<p_{T}<3$ and 
$0.35<x_{F}<0.45$. Lower right: $2<p_{T}<3$ and $0.45<x_{F}<0.55$. SM12=2800.}
\label{Fig:lfit2800_4}
\end{figure}

\begin{figure}
\begin{center}
\mbox{\epsfxsize=5.7in\epsfysize=6.5in\epsffile{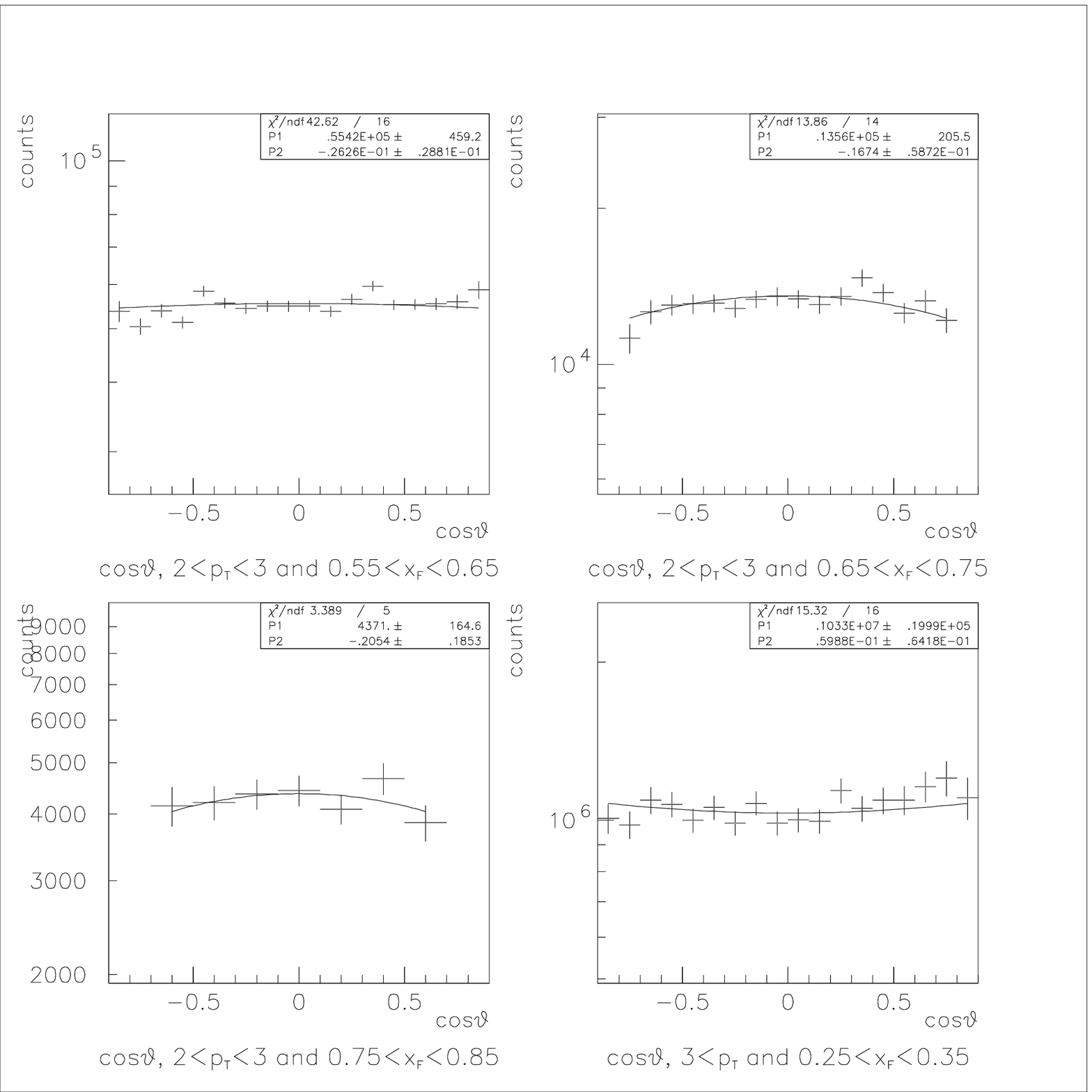}}
\end{center}
\vspace*{+0.25in}
\caption[The corrected $\cos \theta$ distributions and the polarization parameter $\lambda$]
{The corrected $\cos \theta$ distributions and the polarization 
parameter $\lambda$. Upper left: $2<p_{T}<3$ and $0.55<x_{F}<0.65$. Upper right:
$2<p_{T}<3$ and $0.65<x_{F}<0.75$. Lower left: $2<p_{T}<3$ and 
$0.75<x_{F}<0.85$. Lower right: $3<p_{T}$ and $0.25<x_{F}<0.35$. SM12=2800.}
\label{Fig:lfit2800_5}
\end{figure}

\begin{figure}
\begin{center}
\mbox{\epsfxsize=5.7in\epsfysize=6.5in\epsffile{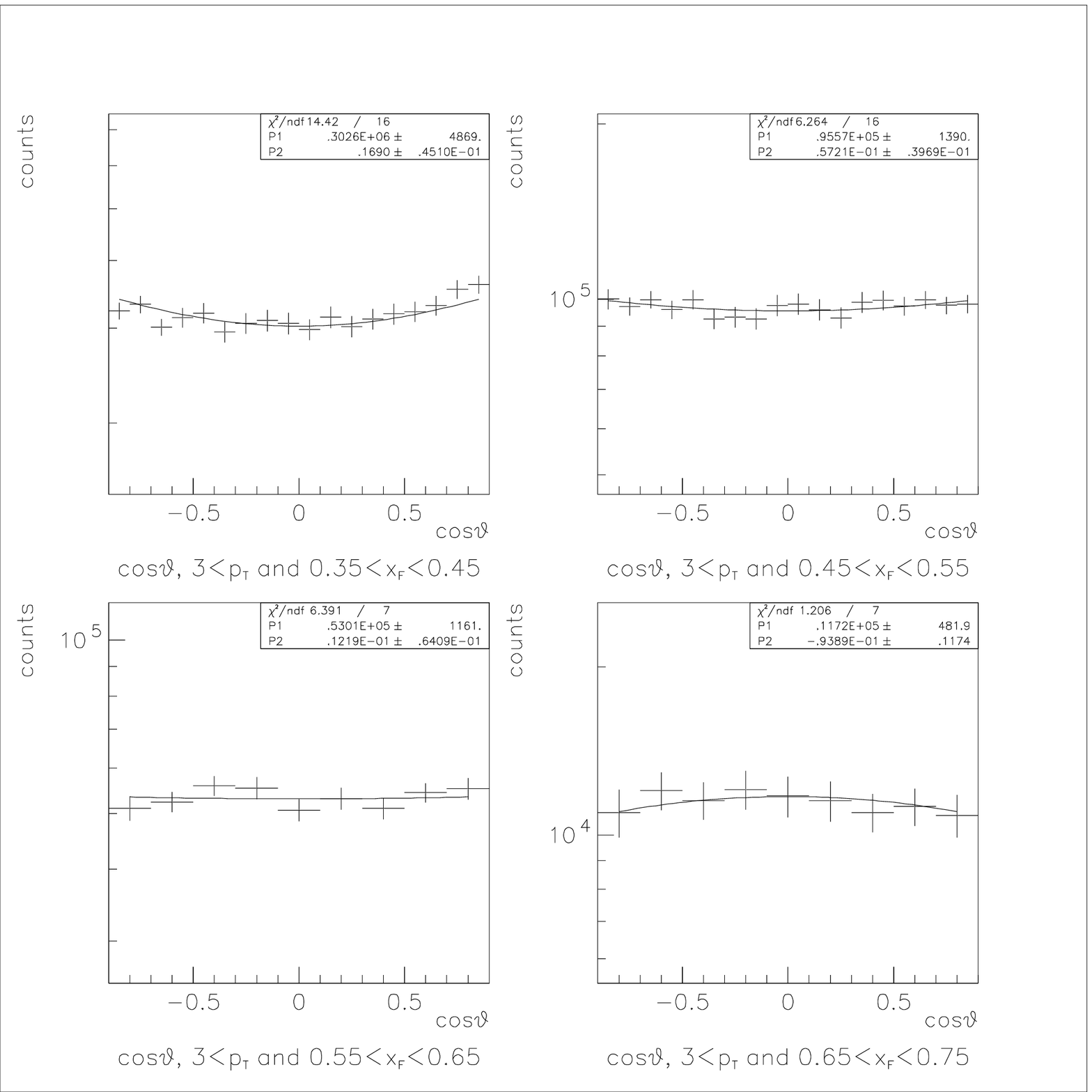}}
\end{center}
\vspace*{+0.25in}
\caption[The corrected $\cos \theta$ distributions and the polarization parameter $\lambda$]
{The corrected $\cos \theta$ distributions and the polarization 
parameter $\lambda$. Upper left: $3<p_{T}$ and $0.35<x_{F}<0.45$. Upper right:
$3<p_{T}$ and $0.45<x_{F}<0.55$. Lower left: $3<p_{T}$ and 
$0.55<x_{F}<0.65$. Lower right: $3<p_{T}$ and $0.65<x_{F}<0.75$. SM12=2800.}
\label{Fig:lfit2800_6}
\end{figure}

\end{document}